\documentclass[11pt]{article}

\usepackage{latexsym}
\usepackage{amsmath}
\usepackage{amssymb}
\usepackage{float}
\usepackage{graphicx}

\def\beq{\begin{eqnarray}}
\def\eeq{\end{eqnarray}}
\def\fsp{\rule{0pt}{10pt}}

\def\barW{{\overline W}}
\def\shft{{\leftarrow}}

\def\pro{\Pr\nolimits_o}
\def\prs{\Pr\nolimits_s}
\def\and{\,\&\,}
\def\giv{\,|\,}

\def\upst{^{\!\!\!\begin{array}{l}\scriptstyle\shft t\\[-9pt]
                                  \scriptstyle\shft s\end{array}\!\!\!}}

\def\QED{\rule[-0.5pt]{5pt}{10pt}}

\newfloat{algorithm}{tbph}{log} \floatname{algorithm}{Algorithm}

\begin{document}

\fontsize{11}{14.5pt}\selectfont


\begin{center} \Large \bf 
 When to Take One Box, When to Take Two: \\[3pt]
 A Concrete Analysis of Newcomb's Problem
\end{center}

\vspace*{8pt}

\begin{center} 
  {\Large Radford M. Neal\,\footnote{\fsp
  University of Toronto, Dept.\ of Statistical Sciences,
  \texttt{https://glizen.com/radfordneal},
  \texttt{radford@utstat.utoronto.ca}. See
  \texttt{gitlab.com/radfordneal/newcomb} for supplemental information,
          including R code for generating the figures.}}

  \vspace{4pt}

  {\large 22 August 2026~~}
\end{center}

\vspace{10pt}

\noindent {\bf Abstract.}  Fifty-seven years of debate have not led to
a consensus on Newcomb's problem. I argue that this failure comes from
not viewing the problem in concrete terms, for which common-sense
intuitions are clearly applicable.  If near-perfect predictions are
not required --- only predictions good enough that Evidential Decision
Theory (EDT) recommends taking only one box --- Newcomb scenarios
without any fantastic elements are actually feasible.  I give a
detailed account of such concrete scenarios, in which the
decision-maker has varying amounts of information about the prediction
process. I argue that in these scenarios it is rational to take both
boxes, as Causal Decision Theory (CDT) recommends. To make this more
convincing, I introduce a simpler impersonal Newcomb problem, also with a
concrete realization, to help bridge causal intuitions from everyday
decisions to the standard Newcomb scenario.  I also argue that Newcomb's
problem with nearly perfect predictions can only be realized by a
(currently fantastic) process in which the Predictor accurately
simulates the thoughts of a Participant.  In a functionalist view of
consciousness, this simulation will be a conscious entity, whose
experience will not be distinguishable from that of a ``real'' person.
In this scenario, CDT recommends taking one box. I conclude that when
Newcomb's problem is properly analysed in concrete terms, CDT gives
the correct answer in all scenarios. Some problems such as Prisoner's
Dilemma and voting decisions have been seen as analogous to Newcomb's
problem.  I argue that these are more closely analogous to the
impersonal Newcomb problem, and that for them CDT gives the correct
decision, analogous to two-boxing. Finally, I consider the ``Newcomb
insurance'' variant, and argue that in contrast to the standard
Newcomb problem, the insurance variant cannot be realized by
non-fantastic methods.  In a fantastic Newcomb insurance scenario
using thought simulation, applying CDT potentially results in
decision instability, but reasonable ways of resolving this lead to a 
sensible decision.

\vspace{4pt}

\section{\hspace*{-8pt} Introduction}\vspace{-10pt}

A facility exists where Participants enter, are examined by
a Predictor, and then are shown two boxes, one transparent and one
opaque. Each Participant then chooses one of two options:\ \ Take only
the opaque box, which contains either nothing or one million dollars
(\$1M), or take both this opaque box and the transparent box, which
can be seen to contain one thousand dollars (\$1K).  Participants keep
the money in the box or boxes they take. The Predictor predicts, based
on an examination of a Participant, and perhaps knowledge of
the Participant's history, whether the Participant will take one box
or two, and puts nothing in the opaque box if the prediction is that
the Participant will take both boxes, but puts \$1M in the opaque box
if the prediction is that the Participant will take only this box.
The prediction and filling of the boxes is done before the Participant
chooses, but Participants believe that the Predictor is likely to
predict their choice correctly.  If you are a Participant, what is the
rational choice for you to make --- to take the opaque box only, or to
take both boxes?

This is Newcomb's problem, first discussed in print by Nozick (1969).

An argument for taking only the opaque box is that if you do,
the Predictor will likely have predicted that you will take only one box,
and will have put \$1M in it, but if you take both boxes, the
Predictor will likely have predicted \textit{that}, and put nothing in
the opaque box, so you obtain only the \$1K in the transparent box.

An argument for taking both boxes is that the contents of the opaque
box are already determined when you make your choice --- what you do
does not change what is in the opaque box --- so taking both boxes
gets you \$1K more than taking just the opaque box.

The one box argument is often seen as employing Evidential Decision
Theory (EDT) --- discussed, for example, by Ahmed (2014).  The two box
argument can be seen as employing Causal Decision Theory (CDT) ---
discussed, for example, by Lewis (1981) and Joyce (1999).  Newcomb's
problem thus helps us assess which (if either) of these
frameworks for decision theory is correct, with significance beyond
a mere amusing puzzle.

Later, I will look at other arguments for one-boxing or two-boxing that
don't relate so directly to EDT or CDT, but I consider the two arguments
above to be central to what defines Newcomb's problem.

My approach to resolving the problem will be to first consider
\textit{concrete} Newcomb scenarios that could \textit{actually
happen}, or that can be imagined as happening without making any
fantastic assumptions, and in which the Predictor's accuracy is high
enough that the recommendations of EDT and CDT conflict (even if the
predictions do not have near-perfect accuracy).  In these
realistic scenarios, I hope it will be clear to everyone that the
correct decision is to take both boxes.

In presenting such scenarios, I depart from those who contend that
Newcomb's problem is unrealizable.  For example, Ahmed (2014, p.~81) says
(of a version of the problem in which the Predictor is 
90\% accurate):\vspace{-8pt}
\begin{quotation}\noindent
Newcomb's problem involves a science-fictional situation that never
has faced and never will face any real person.\vspace{-8pt}
\end{quotation}
Mackie (1977) says
of Newcomb's problem (with the Predictor being highly accurate):\vspace{-8pt}
\begin{quotation}\noindent
There is no conceivable kind of situation that satisfies at once the
whole of what it is natural to take as the intended specification of
the paradox.\vspace{-8pt}
\end{quotation}
Berm\'udez (2018), while not convinced
that Mackie has covered all conceivable possibilities, still thinks
that\vspace{-8pt}
\begin{quotation}\noindent
\ldots nobody who is skeptical about NP’s utility in settling disputes about the
nature of rational decision-making is likely to think it proﬁtable to think up
increasingly baroque explanations of the extraordinary powers of a mysterious 
and possibly supernatural entity.\vspace{-8pt}
\end{quotation}
Accordingly, Berm\'udez instead considers proposals that some real-life
situations have the character of a Newcomb problem (not necessarily with a
near-perfect Predictor), and argues that
none of the situations so far proposed is an actual Newcomb problem,
leading him to say:\vspace{-8pt}
\begin{quotation}\noindent
\ldots I end with a challenge to those who think that
NP really does force a rethinking of evidential approaches to decision theory.
Please provide a fully described decision problem in which conditions 
[ for Newcomb's problem ] hold, and where we have a convincing explanation of
how and why a rational decision-maker can have the conditional probabilities 
required for those conditions to be met.\vspace{-8pt}
\end{quotation}
My presentation in this paper of realistic Newcomb scenarios
can be seen as taking up this challenge.

The detailed scenarios I present for how a Predictor could have good
(but not near-perfect) accuracy could occur in real life. They
do not involve the Predictor being a supernatural entity, or
possessing technology beyond that of the present day.  Nor do they
rely on some tricky redefinition of the problem.  They simply exploit
statistical patterns that are very likely to exist in reality. These
scenarios are unlikely to ever actually be realized, for practical, ethical,
and financial reasons, but they can readily be imagined as being
actual. I will present a detailed analysis of how a Predictor could
achieve significant accuracy in these scenarios, and of what a
Participant should believe regarding the Predictor's accuracy,
verifying that it is plausible for Participants to face a decision with 
the characteristics of Newcomb's problem.

I will argue that in these concrete scenarios, the correct decision is
to take both boxes. To help persuade readers of this, I introduce an
impersonal Newcomb problem, also with a concrete realization, for
which I hope causal intuitions are more clearly applicable. This
serves as a bridge to the use of causal decision theory for Newcomb's
problem itself. Looking at concrete mechanisms for such problems also
mitigates the possibility of uncertainty itself inclining people
towards ``one-box'' choices, as discussed by Shafir and Tversky (1992).

I then consider scenarios in which the Predictor is almost
perfectly accurate, which I argue is only possible using a method
in which the Predictor simulates Participants' thoughts --- though
this is beyond present human technology, and may be beyond any possible
technology.  I again consider the \textit{actual} implications of
such a Predictor for what it is rational to do.  If one takes a
functionalist view of consciousness, someone deciding to take
one box or two will not know whether they are the ``real'' Participant
or the simulated Participant.  One might think that this violates
the conditions of Newcomb's problem, but if one treats the scenario as
legitimate, I argue that taking one box is the correct decision,
according to both EDT and CDT.  If simulations are somewhat
unreliable, and the Predictor therefore does several simulations to
obtain a better prediction, I show that a Participant may be best off
randomly choosing whether to take one box or two.

I go on to consider two decision problems that have been seen as
analogous to Newcomb's problem:\ \ the Prisoner's Dilemma, and deciding
whether to vote in a large election.  I argue that these problems are
actually more like an impersonal Newcomb problem than a
standard Newcomb problem.  Given the assumptions commonly made (which
rule out most people's actual motives) the appropriate decision, found
using CDT, is analogous to taking two boxes:\ \ not cooperating in a
Prisoner's Dilemma, and not voting.

Finally, I consider the ``Newcomb insurance'' variant of the problem
(Ahmed 2014, Section~7.4.3), which involves sequential decisions. I
argue that \textit{this} version of Newcomb's problem \textit{cannot}
be realized in a non-fantastic manner, and that in a
fantastic realization using thought simulation CDT gives the right answer,
albeit only after resolving potential decision instability.

\section{\hspace*{-8pt}
  Clarifying Newcomb's problem}\vspace{-10pt}\label{sec-cond}

Before considering concrete realizations, I will clarify 
the exact conditions required for a situation to be a Newcomb
problem, starting with clarifications that I believe most of those who
have considered the problem would assent to (noting some exceptions),
and then considering two aspects on which there is less agreement.

\subsection{\hspace*{-8pt}
  Uncontroversial clarifications}\label{sec-uncont}\vspace{-6pt}

I take the following clarifications of Newcomb's problem to be
uncontroversial (except as noted), though some of the points
below are seldom (or ever) stated explicitly:\footnote{\fsp
Joyce (2018) lists five conditions for a valid Newcomb problem (p.~139).
His conditions $N\!P_1$ and $N\!P_2$ refer to the Predictor examining a
Participant's ``brain state'', which I consider an undue restriction on
how the Predictor operates (and also not sufficient on its own for 
highly reliable
prediction). Condition $N\!P_3$ of Joyce corresponds to my condition~(e), 
condition $N\!P_4$ to my conditions~(i) and~(k), and condition $N\!P_5$ 
to my condition~(g).
Berm\'udez (2018) also lists conditions required for a legitimate
Newcomb-like problem, generalizing beyond Newcomb's problem itself,
which are compatible with mine, though not explicit about some matters.
He emphasizes that to be of interest the problem must lead to conflicting 
recommendations for EDT and CDT (at least as these theories are commonly 
understood), which is also a focus of my conditions. Ahmed (2014, pp.~109,
footnote~30) also sees such a conflict as essential.\vspace{-25pt}
}\vspace{-4pt}

\begin{itemize}
\item[a)] The Participants are human beings. Later, I will briefly consider
          the possibility of Participants who are Artificial Intelligences
          running on a digital computer, but Participants are not AIs in 
          the standard Newcomb problem.
\item[b)] Participants are sampled from some subset of human beings, which
          may restrict them to, say, being between 15 and 75 years of age,
          or to having read a book on decision theory.
          However, such restrictions must
          not eliminate all people who end up making one of the two possible
          choices, and also must not eliminate all people who are inclined
          to think carefully, since either of these 
          restrictions would render the problem uninteresting.  
          (A fifty-seven-year-long
          academic inquiry into what you should do if you aren't going to think 
          very much would be ridiculous.)
\item[c)] The Predictor may or may not be a human being (or beings), but
          is in any case part of the physical world.  In particular, the 
          Predictor is not some omniscient being who can directly see 
          the future.\footnote{\fsp
            Nozick (1969, p.~130 and endnote~14) explicitly
            states that he will not consider influence backwards in time.
            When the Predictor is perfectly accurate, McKay (2004) wonders
            whether ``\ldots presented with what is apparently a Newcomb case,
            you might still decide that there is reason to believe your
            actions now can have a causal influence on the apparently past
            action of the predictor'', but in the end decides that
            backwards causation ``brings in a whole host of new 
            problems'', and instead considers the possibility of cheating.
            Schmidt (1998) argues that a Newcomb problem with a highly
            accurate Predictor is best viewed as embodying backward causation,
            even though his scenario has no backward causation in
            what I take to be the ordinary sense.\label{foot-cause}
          }
\item[d)] When Participants enter the facility, they are for a short while in
          an examination room, where the Predictor may examine them by 
          means that cause no non-negligible damage, and in particular do
          not alter the Participants' thought processes in other than 
          mundane ways (such as boredom from waiting for the examination to 
          conclude). For example, the Predictor might measure each Participant's
          height, weight, head size, and perhaps more detailed information
          about the Participant's brain, if this can be done without 
          significantly affecting the Participant.
\item[e)] After a Participant leaves the examination room, the Predictor
          makes a prediction, based on results of the examination, and
          perhaps on knowledge of the Participant's history (for example: age,
          grades in courses taken, location of birth). 
          The opaque 
          box is then filled with \$1M if the prediction is that the Participant
          will take only the opaque box, or left empty if the Participant
          is predicted to take both boxes.
          The transparent box is filled with \$1K. 
          The contents of the boxes cannot be changed later.
\item[f)] After the boxes are filled, the Participant enters a deliberation
          room, and is afforded ample time and resources (such as
          paper and pencil) to consider whether to take one box or 
          two.\footnote{\fsp Allowing ample time for thought after
          the Predictor has irrevocably set up the boxes is crucial for 
          Newcomb's problem to produce a contrast in the recommendations 
          of EDT and CDT, 
          since deliberation before that point might causally
          influence the prediction. Schmidt (1998) sets a time limit of
          only one minute for deciding whether to take one box or two, 
          apparently to make the rest of his story work well, but it seems
          he could easily have allowed more time for thought. If such a
          small limit is actually necessary for his argument, his paper is 
          self-refuting, since it clearly took much more than a minute to write.
          Slezak (2013) suggests that a neurological phenomenon in which
          decisions are detectable by EEG 350 to 400 milliseconds before
          conscious awareness could be used to realize a Predictor, but
          allowing less than half a second between prediction and choice
          is ridiculous.}
          Since the boxes have already been filled, the Participant's thoughts
          while deliberating can have no causal effect on the content of the
          opaque box.\footnote{\fsp
          Spohn (2011) views the decision to take one or
          two boxes as having been made before the prediction, even if
          the Participant has not even considered the question until after
          the prediction is made and the boxes filled, on the grounds that
          consideration of the question will simply reveal your already
          latent view of the matter. Viger \textit{et al.}\ (2019) argue
          similarly that the final choice is the result of a process
          extending to before the prediction. Both these positions seem to
          contradict the common sense notion that one can in fact reconsider
          a previously-held view and come to a quite different conclusion
          from before.
          }
\item[g)] Participants may have thought about whether to take one box or two
          before entering the deliberation room, but are free to reconsider
          the question, possibly reaching a
          different conclusion. Here, ``free'' is meant in the ordinary
          sense, not necessarily implying indeterminism.\footnote{\fsp
          Mackie (1977) considers the question of what to do in a Newcomb
          situation to be ``idle'' if one's choice is determined by one's 
          previous character, but softens this position somewhat when the 
          choice is not rigidly determined (and hence the Predictor is not 
          perfectly accurate). Joyce (2018) elaborates on how
          freedom of choice is compatible with the conditions of 
          Newcomb's problem.}
\item[h)] After deliberating, a Participant proceeds to either take 
          just the opaque box or to take both boxes, and receives whatever 
          money the boxes taken contain.
\item[i)] The Predictor predicts correctly with some probability, $R$, the
          same for Participants who take one box and those who take 
          two.

          Writing $A_1$ and $A_2$ for the events that a Participant takes 
          one box or two boxes, and $S_1$ and $S_2$ for the events that
          the Predictor predicts that one or two boxes will be taken, 
          this can be expressed as\vspace{-2pt}
\beq \pro(S_1\giv A_1)\ =\ \pro(S_2\giv A_2) \ =\ R
\label{pred-cond}
\eeq

\vspace{-9pt}

where $\pro$ denotes probabilities held by an outside observer, based on
objective information, such as frequencies in past episodes of prediction.
   Writing $C$ for the event that the Predictor is correct (that is,
   either $S_1\and A_1$ or $S_2\and A_2$), the above is 
   equivalent to\vspace{-4pt}
\beq \pro(C\giv A_1)\ =\ \pro(C\giv A_2) \ =\ R
\label{pred-correct1}
\eeq

\vspace{-9pt}

   Allowing somewhat different accuracies for Participants who take one box and
   those who take two boxes would be possible (and is done by Joyce (2018)), 
   but for simplicity I will usually
   assume equal accuracy.\footnote{\fsp Instead of 
          condition~(1), Levi (1975) interprets the requirement that
          the Predictor be highly accurate as asserting that 
          $\pro(C)$ is close to 1,
          which needn't imply condition~(\ref{pred-cond}) with $R$ 
          being close to 1, if the fractions of Participants
          who chose one box or two are very different. Levi argues that this
          makes the solution indeterminate, since condition (\ref{pred-cond})
          is what is necessary for applying EDT. However, it is natural to 
          simply require that condition~(\ref{pred-cond}) hold, which focuses 
          on the skill of the Predictor, regardless of the frequency of one-box
          and two-box Participants. (Berm\'udez (2018) comments similarly on
          Mackie.)  Wolpert and Benford (2011) also claim that Newcomb's problem
          is incompletely specified, with one interpretation corresponding to
          condition~(\ref{pred-cond}) and the other corresponding to an
          assumption that the prediction and the actual choice are
          probabilistically independent. But this second interpretation
          discards the defining feature of Newcomb's problem ---
          that the Predictor has some skill at predicting the choices of
          Participants. That the prediction is 
          probabilistically dependent on the actual choice is compatible
          with the prediction not being \textit{causally} dependent
          on the actual choice, and hence with not conditioning on the 
          actual choice when assessing the value of choices in CDT --- 
          as Lewis (1981) says, of actions $A$,
``It is essential to define utility as we did using the unconditional credences
$C(K)$ of dependency hypotheses, not their conditional credence 
$C(K|A)$''.
}
\item[j)] When deliberating, Participants believe they know everything 
          about the setup, as stated here,
          including the accuracy, $R$, of the Predictor, and possibly including
          some information on the method that the Predictor uses to make
          predictions, but of course excluding the actual prediction made.
          These beliefs are supported by evidence they consider 
          sufficiently strong that when deliberating they ignore the 
          possibility that they could be
          mistaken regarding the setup, or the Predictor's accuracy, $R$.

\item[k)] During deliberations, nothing a Participant knows should rationally 
          affect their
          subjective probability, denoted by $\prs$, that the prediction 
          made by the Predictor 
          is correct, conditional on the action chosen. That is, if you are 
          a Participant,\vspace{-4pt}
\beq
 \prs(C\, |\, A_1\and B) \ =\ 
 \prs(C\, |\, A_2\and B) \ =\ R 
\label{pred-correct2}
\eeq

\vspace{-8pt}

where $B$ denotes everything you know that could potentially influence
your subjective judgement of predictability.
This will include knowledge of $R$ and the problem setup described in 
condition~(j) above, which I will denote by $B_0$.  For example, 
perhaps\vspace{-4pt}
\beq
   B & = & B_0 \ \, \&\ \,
            \mbox{You live in Paris}\ \,\&\ \,\mbox{You know Bayes' Rule}
            \ \,\&\ \,\mbox{You have a headache now}\ \&\ \cdots\ \ 
\eeq

\vspace{-8pt}

Note that $B$ may include private knowledge, such as ``You hated your
decision theory course'', and facts concerning the progress of your
deliberations, such as, ``You feel an initial inclination to take one
box, but will think some more before deciding'', or ``You have now
recalled Nozick's arguments for taking two boxes''.  $B$ will also
include whatever you may know about the Predictor's method of
prediction.

Put another way, a Participant's judgement of the
Predictor's accuracy is conditionally independent of other beliefs
given the choice actually made.\footnote{\fsp This condition is often
implicit, but Ahmed (2014, p.~191) states something similar ---
``Anyone facing Newcomb's problem has \textit{no} evidence that
relevantly distinguishes him now from\ldots all other persons who ever
face this problem.'' Eells (1981) distinguishes ``Type-A'' beliefs,
regarding a randomly selected Participant, from ``Type-B'' beliefs,
regarding the specific decision maker whose deliberations are being
considered, with the latter being what is relevant for decision making. 
This seems analogous to the distinction between the probabilities $\pro$
in equation~(\ref{pred-correct1}) and the probabilities $\prs$ in 
equation~(\ref{pred-correct2})
above. However, as I note later, Eells' subsequent assumption that all
Participants are rational results in probabilities that violate condition~(k).}

\item[l)] A Participant's subjective utilities for the various outcomes
           depend solely on the amount of money obtained.  Furthermore,
           differences in utility for outcomes are proportional to 
           the differences in amount of money for those outcomes.
           This last assumption is just a convenience --- non-linear utility
           of money could be accommodated by adjusting the monetary amounts to 
           produce the same ratios of differences in utilities for the possible 
           outcomes.\footnote{\fsp Newcomb's problem is sometimes presented
           with different payouts, often with the ratio \$M/\$K
           reduced from 1000 to a smaller value of 100 or 10, perhaps to
           make the linearity of utility with money more plausible. This
           increases the accuracy of prediction needed for EDT and CDT
           to conflict, but even for a ratio of 10, any $R$ greater than
           0.55 suffices (see Section~\ref{sec-recommend} below).}

\item[m)] Only the amount of money a Participant obtains in a single
           episode of Newcomb's problem is relevant. Participants 
           have no motive to behave a certain way in order to affect
           what they might be predicted to do in some future episode,
           nor to affect what some other Participant might receive,
           nor are they concerned with the wealth of the Predictor.
           This condition likely requires that no Participant take part 
           in more than one episode of Newcomb's problem.\footnote{\fsp Mackie
           (1977) considers the possibility of repeated play, but regards
           it as not in the spirit of the problem.}
\vspace{-4pt}

\end{itemize}

In order to allow concrete Newcomb scenarios that could actually be
implemented, a slight relaxation of requirements is necessary. The
equating of probabilities to $R$ in (i) and (k) above will be
loosened to a requirement that these probabilities be in the interval
$[R-\epsilon,\,R+\epsilon]$, for some small $\epsilon$. This allows for
a realistic degree of knowledge of the Predictor's accuracy, and a
realistic judgement of near (rather than exact) conditional
independence of the prediction from other beliefs, given the actual
choice.

In the next section, I will consider weakening condition~(k) in the context
of randomized decisions.

Conditions (j) to (m) concern the subjective beliefs and desires of
Participants. As phrased, they are required to be true of all
Participants, but in reality, one cannot rule out the possibility that
some Participant has bizarre desires or delusional beliefs, though the
problem becomes uninteresting if that is typical.  I will therefore
take these requirements as applying to most Participants, and in
particular to ``you'', whose rational deliberation I am trying to
explicate. More specifically, one might consider a Newcomb problem to
be valid even though for a small group of unusual Participants not all
of the conditions above are satisfied, but it cannot be considered
valid if these conditions are not satisfied for Participants who are
following one of the decision procedures whose correctness Newcomb's
problem is intended to test.

One should note that Newcomb's problem definitely does not
require that all Participants be perfectly rational. Perfectly
rational Participants would all make the same choice --- whichever one
is actually rational --- violating condition (b), and making the task
of the Predictor trivial. In a valid Newcomb problem, a big part of
the Predictor's skill will presumably come from being able to predict
(explicitly or implicitly) what decision method (rational or not) each
Participant will employ.

\subsection{\hspace*{-8pt}
  Are randomized decisions allowed?}\label{sec-rand}\vspace{-6pt}

One aspect of Newcomb's problem that might be disputed is whether a
Participant may choose randomly, taking one box with a certain
probability they have decided on. Nozick (1969, endnote 1) specifies
that if the Participant uses randomization, the Predictor will not put
\$1M in the opaque box.

Not allowing randomization seems to me to be an unmotivated
restriction on how Participants may choose. Presumably, everyone would
think it rather odd to consider Newcomb's problem with the proviso
that the Predictor does not put \$1M in the opaque box if, at any
point while deliberating, the Participant engages in mathematical
reasoning that involves logarithms. It seems similarly odd to
stipulate that a decision strategy that is commonly used (for example,
in games) is not allowed. 

However, if Participants may use an effective randomization method,
whose result the Predictor cannot predict, condition~(k) above will
not be satisfied, even if condition~(i) holds. As an illustration,
suppose that all Participants randomize, with half of them choosing
one box with probability 0.8 and the other half choosing one box with
probability 0.2. A Predictor who can perfectly distinguish these two
groups could achieve an overall accuracy of $R=0.8$, by predicting that
Participants in the first group will take one box and those in the
second will take two boxes. Condition~(i) can thus be satisfied.
However, condition~(k) will not be satisfied. Suppose you are a Participant,
and have decided to take one box with probability 0.8. This decision
is part of the knowledge, $B$, referenced in equation~(\ref{pred-correct2}).
At this point, you can deduce that the Predictor must have predicted
that you will take one box, and hence that\vspace{-4pt}
\beq
 \prs(C\, |\, A_1\and B) \ =\ 1\ \ \ \mbox{and}\ \ \ 
 \prs(C\, |\, A_2\and B) \ =\ 0  \\[-20pt]\nonumber
\eeq
contrary to the requirement of condition~(k) that they both equal $R$.
However, it will be the case that\vspace{-4pt}
\beq
 \prs(C\,|\,B) \ =\ \prs(C\, |\, A_1\and B)\prs(A_1\,|\,B)\ +\
                    \prs(C\, |\, A_2\and B)\prs(A_2\,|\,B) \ =\
                    1 \times 0.8\ +\ 0 \times 0.2 \ =\ R
 \\[-20pt]\nonumber
\eeq
So one could consider weakening condition~(k) to\vspace{-8pt}
\begin{itemize}
\item[k$'$)] During deliberations, nothing a Participant knows should rationally
          affect their subjective probability that the prediction 
          made by the Predictor will be correct. That is, if you are 
          a Participant,\vspace{-4pt}
\beq
 \prs(C\, |\, B) \ =\ R  \\[-20pt]\nonumber
\eeq
where $B$ denotes everything you know that could potentially influence
your subjective judgement of predictability, including any decision you
may have made to choose randomly with some probability.\vspace{-8pt}
\end{itemize}
I will consider this more satisfiable version later in scenarios with 
unpredictable randomization.\footnote{\fsp If it can
be observed whether a Participant randomizes, and if so with what
probability, $p$, of one-boxing, one could consider a prediction of
one-boxing to be correct if $p \ge 1/2$, and a prediction of
two-boxing to be correct if $p \le 1/2$. Aaronson (2013, p.~295)
supposes that if the Predictor predicts that you will randomize with
probability $p$, they will put \$1M in the opaque box with probability
$p$. This leads EDT to recommend $p=1$ and CDT to recommend $p=0$,
preserving the point of Newcomb's problem. However, the Predictor's
accuracy can be as low as $1/2$. I will not pursue these variations in
this paper.}

One can conceive of fantastic scenarios in which a Participant seems
to be randomizing, but the results of randomization can actually be
predicted.  For example, in the scenarios I discuss later in which the
Predictor performs a detailed simulation of each Participant's
thoughts, it will be necessary for the Predictor to simulate a
Participant's physical environment as well, perhaps accurately
simulating coins being flipped.  Alternatively, a Participant might
perform ``randomization'' not using some unpredictable random physical
process, but via internal mental operations. For example, you could
choose effectively at random (from your viewpoint) by deciding to take
one box if the sum of the digits in your phone number is even
(something you are unlikely to be aware of before carefully computing
it).\footnote{\fsp The possibility of such mental ``randomization'' is
contemplated for a related problem involving choosing between black
and white doors by Gallow (2021).\vspace{-8pt}}  Such mental randomization might
conceivably also be predictable if detailed thoughts can be simulated
accurately.

For the standard Newcomb problem, Participants typically have no
reason to use randomization.  The usual arguments lead either to
preferring to take one box, or preferring to take two.  Even if you
follow some atypical argument that leads to indifference between these
choices, you can just select arbitrarily (swayed by some unconscious
bias), rather than by any explicitly random method. So it seems that
this issue can be ignored in most discussions of Newcomb's problem ---
at worst, one could just say that the discussion doesn't apply to the
rare Participant who for some bizarre reason chooses to randomize.

As I discuss later, however, randomization will be potentially
advantageous if a Participant believes that the Predictor makes
predictions by performing several somewhat unreliable simulations of a
Participant's thoughts.  Randomization will also be a central issue in
my discussion in Section~\ref{sec-insure} of the Newcomb insurance
variant of the problem, where I will argue that mental randomization
can sometimes be an inevitable result of reasoning using CDT.

Note that I take it as being uncontroversial that the Predictor may
use randomization, which as we will see is sometimes necessary to
achieve a predictive accuracy of $R$ both for Participants who take one
box and for those who take two.

\subsection{\hspace*{-8pt}
  How accurate must the Predictor be?}\vspace{-6pt}

There is no consensus on what the Predictor's accuracy, $R$, may be.

Naive use of EDT favours taking only one box as long as $R$ is greater
than 0.5005 (see Section~\ref{sec-recommend} below), while naive use
of CDT always recommends taking both boxes regardless of $R$. So there
is a conflict between their recommendations whenever $R>0.5005$.

However, Nozick (1969, p.~140) thought opinions would differ only if
$R$ is very close to 1, remarking that ``I presume, if the probability
of the beings predicting correctly were only .6, each of us would
choose to take what is in both boxes''. He also says ``It is crucial
that the predictor is almost certain to be correct''.

As an extreme, one could consider Newcomb's problem with $R=1$ ---
that is, when one believes that the Predictor is certain to be
correct.  According to Wolpert and Benford (2011), Newcomb himself did
not insist that the Predictor be perfect (though they do not report
how low he would let $R$ be). I will not consider Newcomb's problem
with a perfect Predictor, since I do not see how this is possible in
the real world.  Even if one thinks a perfect Predictor is physically
possible, it is hard to see how a Participant could be \textit{absolutely
certain} that the actual Predictor is perfect.

I believe that Nozick (who himself favoured taking two boxes) was
mistaken regarding peoples' behaviour when $R=0.6$.  I think
that many people would take only one box in this circumstance.  And
many people have discussed Newcomb's problem assuming a similarly
modest level of accuracy.

Gallow (2021), for example, describes the Predictor as being 90\%
reliable, as does Arntzenius (2008). Ahmed (2014) also supposes that
the Predictor is 90\% reliable (p.~47), while saying that the
Predictor is ``highly accurate''.  Ahmed (2018) again chooses 90\%
reliability when needing a quantitative level (p.~11), while also
saying (p.~3) that the Predictor has ``almost always'' predicted 
the actual choice of a Participant.

The argument for taking one box is often presented as a calculation of
expected utility given that you take one box versus that you take both
boxes.  Ahmed and Price (2012), for example, say ``The \textit{act}
that EDT recommends in a Newcomb type situation --- namely, one-boxing
--- has a better average return than the act that CDT recommends there
--- namely, two-boxing''.  Though Ahmed and Price then suppose that
the Predictor is 95\% reliable, their argument seems to apply equally
well when the Predictor is 51\% reliable, since that is above the
threshold where EDT calculates that taking one box has the higher
average return.

However, very high reliability does make it easier to believe that
condition~(k) above holds. If the Predictor is only 90\% accurate, you
might reason that although the Predictor has that accuracy for most
people, you yourself have some unusual characteristic (shared by less
than 10\% of the population) that makes predictive accuracy for other
people irrelevant.  If the Predictor is 99.99\% accurate, it's harder
to believe that you are in a special 0.01\% category. I will argue
below that it is possible for condition~(k) to hold even when
predictions are only moderately reliable.

It seems that a reliability of 90\% is high enough that many people
intuitively judge taking one box to be the correct action, and even
lower reliabilities should lead anyone who accepts EDT to take
one box.  Furthermore, if Newcomb's problem is to carry lessons for
more practical decision problems, it is best to not assume extremely
high reliability.  


\section{Recommendations of EDT and CDT}\label{sec-recommend}\vspace{-10pt}

Here, I will review the argument from Evidential Decision Theory for
taking one box in a Newcomb situation, and that from Causal Decision
Theory for taking both boxes.  Apart perhaps from my emphasis on
conditional independence from personal knowledge (condition~(k) in
Section~\ref{sec-uncont}), this is the standard formulation of how the
recommendations of EDT and CDT diverge.

EDT, as presented for example by Ahmed (2014), judges which action is
best by the expected utility (reward, payoff) conditional on that action
having been taken.

In a first attempt at applying EDT to Newcomb's problem, one might
evaluate the expected utility of action $A_1$ (take
one box) or $A_2$ (take both boxes) as follows:
\beq
  V(A_1) & = & \pro(S_1\giv A_1)\,\times\,1000000 
                     \ +\ \pro(S_2\giv A_1)\,\times\,0 \nonumber \\
               & = & R\,\times\,1000000 
                     \ +\ (1-R)\,\times\,0\ =\ R \times 1000000 
              \label{EDT-eq1} \\[4pt]
  V(A_2) & = & \pro(S_1\giv A_2)\,\times\,1001000 
                     \ +\ \pro(S_2\giv A_2)\,\times\,1000 \nonumber \\
               & = & (1-R)\,\times\,1001000 
                     \ +\ R\,\times\,1000\ =\ 1000\ +\ (1-R)\times 1000000
               \label{EDT-eq2}
\eeq
where $S_1$ and $S_2$ are the events that the Predictor predicted that
the Participant would take actions $A_1$ and $A_2$, respectively, and
$V(A_i)$ is the expected utility of action $A_i$ as calculated by EDT.  The
conditional probabilities above are as specified by condition~(i) in 
Section~\ref{sec-uncont}.

If one sets $R=0.9$, for example, these expected utilities 
evaluate to\vspace{-2pt}
\beq
  V(A_1) \ =\ 900000,\ \ \ \ V(A_2) \ =\ 101000
\eeq
and hence EDT recommends taking one box.  In general,
EDT recommends taking one box when $V(A_1)>V(A_2)$,
which is the case when $R\times 1000000\, >\, 1000\,+\,(1\!-\!R)\times 1000000$,
which simplifies to $R\,>\,0.5005$.

However, this approach is overly simple. When deciding which action
EDT recommends, one must use expected utilities based on the
subjective probabilities of a Participant (for example, you) that are
conditional not just on each of the possible actions, but on these
actions together with whatever else the Participant knows.  This is why
condition~(k) in Section~\ref{sec-uncont} needs to hold for a Newcomb
problem --- it justifies why you can use equations like~(\ref{EDT-eq1})
and~(\ref{EDT-eq2}) despite knowing many things (here denoted by $B$)
which might on their own be informative of the Predictor's
prediction. In particular, condition~(k) allows us to conclude
that\vspace{-2pt}
\beq
  \prs(S_1\giv A_1\ \&\ B)
  \ =\ \prs(C\giv A_1\ \&\ B) 
  \ =\ R \ =\ \pro(C\giv A_1) \  =\ \pro(S_1\giv A_1)
\eeq
and similarly for $A_2$.  We should really use a conditional
probability such as that on the left when computing expected utility,
but since it is equal to the final probability on the right, we can
use that instead, as in~(\ref{EDT-eq1}) and~(\ref{EDT-eq2}).  Ensuring
that condition~(k) holds will be a major issue when designing
realistic Newcomb scenarios.

CDT, as presented for example by Gibbard and Harper (1976), Lewis
(1981), and Joyce (1999), judges which action is best by the expected
reward conditional on the \textit{causal effects} of the action, but
not on what having taken the action says about aspects of the world
that it does not causally affect. In Newcomb's problem, taking only
the opaque box has the causal effect that you receive whatever money
is in this box, but do not receive the money in the transparent
box. Taking one box is also evidence that the opaque box contains
\$1M, but it cannot have a causal effect on this, since the box's
contents were determined in the past, and causation cannot go
backwards in time.\footnote{\fsp Of course, there are those who would
dispute this. See footnote~\ref{foot-cause} earlier.}

In Newcomb's problem, the causal effects of action $A_1$ or $A_2$ are
determined by the Predictor's prediction, notated here as either $S_1$
or $S_2$, which is itself causally independent of the action taken.
Given background information $B$, your causal
expectations of utility for
actions $A_1$ and $A_2$ according to CDT, written $U(A_1)$ and $U(A_2)$, 
are as follows:\vspace{-4pt}
\beq
  U(A_1) & = & 
\prs(S_1\giv  B)\,\times\,1000000 
\ +\ \prs(S_2\giv  B)\,\times\,0 
\label{eq-CDT1}\\[4pt]
  U(A_2) & = & 
\prs(S_1\giv  B)\,\times\,1001000 
\ +\ \prs(S_2\giv  B)\,\times\,1000\ \ \ \
\label{eq-CDT2}
\eeq

\vspace{-10pt}

In general, comparing such causal expected utilities requires 
calculating their actual numerical values, but for Newcomb's problem
this is not necessary, since action $A_2$ is \textit{dominant} --- 
it has higher causal expected utility regardless of the probabilities
of $S_1$ and $S_2$. This can be seen by comparing corresponding terms
in equations~(\ref{eq-CDT1}) and~(\ref{eq-CDT2}).\footnote{\fsp
In some related problems, such as the ``Murder Lesion'' problem discussed
by Egan (2007) and Armendt (2019), there is no dominant action,
and the subjective probabilities of hypotheses of causal dependence
such as $S_1$ and $S_2$ would need to be assessed, which can be 
confusing when these probabilities are affected by the action being 
deliberated upon.
But there is no need to consider this issue for Newcomb's problem,
though it does arise for the Newcomb insurance variant I discuss
in Section~\ref{sec-insure}, and in more benign form when CDT is
applied to Newcomb's problem with predictions made by thought simulation
in Section~\ref{sec-simdecision}.
}

We thus see that EDT recommends taking one box when $R>0.5005$, whereas
CDT recommends taking both boxes for any value of $R$. A believable
scenario for Newcomb's problem with $R>0.5005$ will therefore show that
there is a real conflict between the recommendations of EDT and CDT.

\section{Prediction using the category of a 
Participant}\label{sec-prcat}\vspace{-10pt}

A Predictor might operate by first classifying each Participant as being
in some category, based on observation and past history.  For example,
one such category could be Participants with a college education who
have high blood pressure.  If the Predictor has data on how often Participants
in each category have taken one box or two in the past, the Predictor
might predict that a Participant will take whichever action has been
more common in the Participant's category (though as will be
seen in Section~\ref{sec-pr} below, this somewhat simplifies the 
actual procedure necessary for achieving a given accuracy,~$R$).

The following vignette from Viger, Hefer, and Viger (2019) portrays
a Predictor operating in this manner:\vspace{-8pt}
\begin{quotation}\noindent
\ldots Our first contestant is a single working parent of three; the second
is a graduate student in philosophy \ldots you each have been interviewed
by Cassandra, our oracle. Based on your interviews, she has predicted whether
you will pick one or two boxes \ldots Cassandra is correct in her predictions
99\% of the time.\\[5pt]
Contestant number 1, this is your big moment. So what are
you going to choose? \\
Contestant 1: Well I really need the money and most of the people who pick
one box win \$1,000,000 so that's my choice, just box A. \\
During a dramatic pause, Contestant 2's classmates scoff at the irrationality
of her decision \ldots\linebreak{}
[ Contestant~1 leaves with \$1,000,000. ]
\\[5pt]
Host: Now Contestant 2, you're a graduate student in philosophy. Did 
Cassandra ask you about that? \\
Contestant 2: Yes. That was the only question she asked me. \\
$<$Host grimaces$>$ \\
Host: Well, you've seen how it's done. What's your choice? \\
Contestant 2: Since \$1,000,000 is already either in box A or it's not and I
can't change that now, to get as much money as possible I choose both boxes.
Why would I leave \$1,000 on the table? \\
Contestant 2's friends nod approvingly. Box A is raised revealing nothing.
Contestant~2 is heard mumbling, ``Good thing I took both boxes or I'd have
gotten nothing'' while exiting the stage with her \$1,000.\vspace{-8pt}
\end{quotation}
Here Viger \textit{et al.}, who favour taking just one box, seem to be
mocking those who would take both boxes, and in particular their
fellow philosophers of that persuasion. A desire to mock may have
blinded them to the actual implications of the vignette they
have written.

Cassandra in this vignette appears to operate by categorizing
Participants on the basis of answers to questions she asks. One such
category is ``philosophy graduate student''. It seems that Cassandra
is familiar with philosophy graduate students, and knows that the vast
majority of them take both boxes; hence she puts nothing in the opaque
box (box~A in the vignette) for Contestant~2.

But Contestant~2 knows this, and hence this vignette does not describe
a proper Newcomb problem.  Contestant~2 can deduce that box~A is empty
before she makes a choice. She knows that the only question Cassandra
asked her was whether she is a philosophy graduate student. She knows
that philosophy graduate students overwhelmingly favour taking both
boxes (as indicated by their scorn when Contestant~1 takes only one
box). So she can deduce that Cassandra will have put nothing in box~A,
before having made her own choice. This is in contradiction to
condition~(k) of Section~\ref{sec-uncont}.

The vignette therefore portrays what is essentially the variation of
Newcomb's problem in which both boxes are transparent, considered for
example by Arntzenius (2008, Section~7) and Viger, \textit{et al.}
themselves.\footnote{\fsp Gibbard and Harper (1978) discuss a similar
variation in which a Participant first opens the opaque box and takes
whatever money it contains, and then decides whether or not to take
the money in the transparent box as well. These scenarios are not
entirely coherent.  As Viger, \textit{et al.}\ note, if a Participant
takes one box if the second is empty, and both boxes if the second has
\$1M, the Predictor cannot possibly be correct for them. In contrast,
if a Participant takes one box if the second has \$1M and both boxes
if the second is empty, the Predictor will be correct regardless of
the prediction made. Unlike in the standard Newcomb problem, the
Predictor's prediction has a causal effect on the Participant's
choice.}  In this variation, it is widely considered by advocates of
both EDT and CDT that it is rational to take both boxes.\footnote{\fsp
Some who adhere to neither EDT or CDT would take only one box when
both are transparent --- Yudkowsky and Soares (2018), for
example. Viger, \textit{et al.} (2019, p.~419) appear
ambivalent.\vspace{-10pt}\label{foot-yud}} So the vignette of Viger
\textit{et al.}\ fails to describe a situation in which the standard
recommendations of EDT and CDT differ.

We can try to fix this issue by supposing that Contestant~2 has just
begun her graduate studies in philosophy, having previously been a
military intelligence officer, and has not heard the opinions of her
fellow students on Newcomb's problem. We can also suppose that these
students refrain from commenting on the choice of
Contestant~1. It then seems plausible that Contestant~2 does not know
what is in box~A before she makes her choice, since she does not know
the dominant opinion among philosophy graduate students. If she also
considers herself to be a typical philosophy graduate student, she
may take her own choice to be indicative of what most philosophy graduate
students would do, and hence of what Cassandra predicted she would do.

However, this version of the vignette will not be a valid Newcomb
problem if Contestant~2 thinks that her uncommon background in
military intelligence makes her atypical of philosophy graduate
students. She might reason that Cassandra could achieve her 99\%
accuracy standard for the category of philosophy graduate students
even if she has no skill in predicting what a philosophy graduate
student with a military intelligence background will do. So if
Contestant~2 sees herself as atypical, she may not think that if she
takes just the opaque box there is a 99\% probability of it
containing~\$1M, contradicting condition~(k) of Section~2.

To clarify how prediction from the category of a Participant can be
the basis for a realistic Newcomb problem, I will examine in detail
how a Predictor can make predictions based on categories, and what a
Participant should rationally believe, at several levels of knowledge
of how the prediction process works. My goal is not to show that every
scheme involving prediction using categories fulfills the conditions
of Newcomb's problem, but just to show that there are some such schemes
that one can easily imagine being
implemented in practice, given some plausible assumptions, by someone
with ample funds and flexible ethical principles.

\subsection{\hspace*{-8pt}How categories can be used to
predict whether a Participant takes one box or 
two}\label{sec-pr}\vspace{-6pt}

Prediction of one-boxing or two-boxing using categories seems possible
due to the common-sense observation that most things differ between
groups, on average.  There is every reason to think, for example, that
graduate students in philosophy differ on average from the general
population in fondness for popcorn, in skill at playing backgammon, in
frequency of movie attendance, in tendency to tap their toes to music,
and in propensity to take one box rather than two in Newcomb
problems. It would only be by coincidence that the distribution of
\textit{any} of these traits was \textit{not} different for philosophy
graduate students than for other people.

However, to achieve an accuracy significantly greater than $R=0.5$, a
Predictor may need to use more information than just that a
Participant is a graduate student in philosophy, or is in another
similarly-broad category. Such broad categories will often include
significant numbers of both one-boxers and two-boxers, whereas in a
narrower category a single choice is more likely to predominate, allowing
for more accurate prediction.

Narrow categories are easily created as conjunctions of broader
categories. For example, one category a Predictor might use could
consist of Participants who\vspace{-8pt}
\begin{itemize}
\item were born in North America,\vspace{-8pt}
\item have not taken a philosophy course,\vspace{-8pt}
\item are less than 180cm tall, and\vspace{-8pt}
\item have a brain volume greater than 1300$\mbox{cm}^3$.
\end{itemize}
Combinations of these attributes or their negations produce 16
categories, which might be the ones the
Predictor uses, though some of these categories could be subdivided
further as necessary to achieve the desired accuracy.

The attributes defining a category might or might not have a causal
effect on whether the Participant takes one or two boxes. For the
category above, taking a philosophy course or not and having a
brain volume greater or less than 1300$\mbox{cm}^3$ seem causally
relevant, but being born in North America probably has causal influence only
rather indirectly, perhaps through ideas people born there are more
likely to encounter.  Being less than 180cm tall seems
likely to have little causal relevance, but could nevertheless be
predictively relevant --- for example, it makes it more likely that
the Participant is female, which plausibly could affect the choice of
one or two boxes.

The procedure a Predictor should use to achieve some specified
accuracy level, $R$, is somewhat non-obvious. Suppose that the
Predictor will use $K$ mutually-exclusive categories, denoted by
$G_1,\ldots,G_K$.  (For these to be useful, the Predictor must be able
to determine which of these categories each Participant belongs to.)
Suppose also that the Predictor knows that, in the population of
possible Participants, the proportions of these categories are
$w_1,\ldots,w_K$, which sum to one. Finally, the
Predictor has good estimates, $q_1,\ldots,q_K$, of the proportions
of potential Participants in each category who would take one box.

One might think that the Predictor should predict that a Participant
in category $G_i$ will take one box if $q_i>0.5$, and two boxes
otherwise.  But only by coincidence could this achieve some desired
accuracy, $R$, for both one-boxers and two-boxers. In general, 
randomization is necessary, with the Predictor predicting with probability
$p_i$ that a Participant in category $G_i$ will take one box. The
Predictor will achieve accuracy $R$ for both one-boxers and two-boxers
if the randomization probabilities satisfy both of the following:
\beq
  R_1\ =\ 
  {\sum\limits_{i=1}^K p_i q_i w_i \over \sum\limits_{i=1}^K q_i w_i} 
   \ =\ R\ \ \ \ \ \mbox{and}\ \ \ \ \ \,
  R_2\ =\ 
  {\sum\limits_{i=1}^K (1\!-\!p_i) (1\!-\!q_i) w_i \over 
   \sum\limits_{i=1}^K (1\!-\!q_i) w_i} \ =\ R
\label{acc-eq-R}
\eeq
where $R_1$ and $R_2$ are the accuracies for one-boxers and two-boxers.
Here, $\sum_i q_i w_i$ is the fraction of the population who would take
one box if they became a Participant, and $\sum_i (1\!-\!q_i)\, w_i$ the
fraction who would take two boxes. The numerator $\sum_i p_i q_i w_i$
is the fraction of the population who are one-boxers and would be predicted 
to take one box, and $\sum_i (1\!-\!p_i) (1\!-\!q_i)\, w_i$ is the corresponding
fraction for two-boxers.

Taking the $w_i$ and $q_i$ to be fixed and known (as they will be for
the Predictor), the two equations in (\ref{acc-eq-R}) are linear in
$p_1,\ldots,p_K$ and $R$.  Their solutions will be in a subspace of
dimension $K\!-1$, restricted to the region where the $p_i$ are in
$[0,1]$.  If $R$ is fixed to some possible value,
the $p_i$ that achieve this $R$ will be in a subspace of dimension $K\!-2$.
When $K=2$, this will be a single point. 

\begin{figure}[t]

\vspace*{-10pt}

\begin{center}
\includegraphics[scale=0.75]{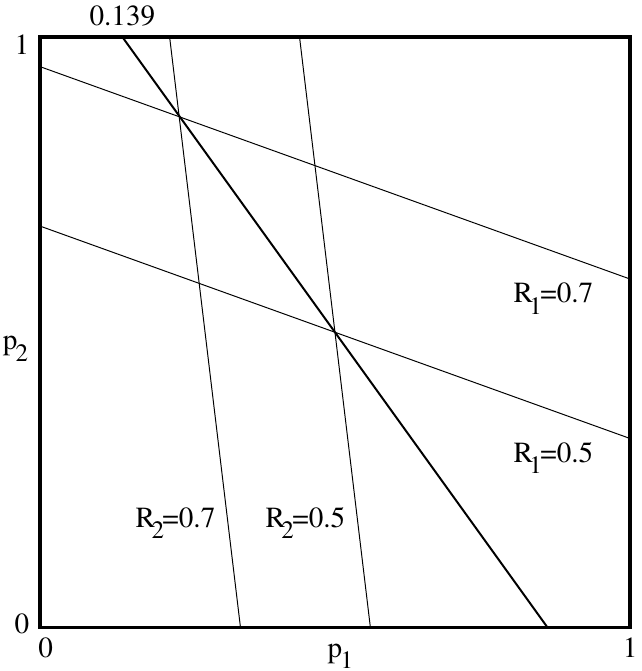}\vspace{-6pt}
\end{center}

\caption{The line of equal accuracy for the example with $w_1=0.8$, $w_2=0.2$,
$q_1=0.05$, and $q_2=0.55$.  Also shown are contours for $R_1$ and $R_2$ at 
values of $0.5$ and $0.7$.  These contours intersect on the line, where 
$R_1=R_2=R$.
The maximum accuracy of $R=R_1=R_2=0.770$ is
achieved at the top left endpoint, where $p_1=0.139$ and $p_2=1$. 
}\label{fig-2dR}

\label{fig1}
\end{figure}

As an example, let $K=2$, with $w_1=0.8$ and $w_2=0.2$, and suppose
that $q_1=0.05$ and $q_2=0.55$, so that
\beq
  \sum\limits_{i=1}^K q_i w_i & = & 0.05\times0.8 \ +\ 0.55\times0.2
                              \ =\  0.04\ +\ 0.11 \ =\ 0.15 \\
  \sum\limits_{i=1}^K (1\!-\!q_i) w_i & = & 0.95\times0.8 \ +\ 0.45\times0.2
                              \ =\  0.76\ +\ 0.09 \ =\ 0.85
\eeq
If the Predictor simply predicts that a participant in category $G_i$
will take one box if $q_i>0.5$, and two boxes otherwise, the accuracy of the
predictions for one-boxers and two-boxers will be
\beq
   R_1\ =\ 0.55\times0.2\ /\ 0.15\ =\ 0.733\ \ \ \ \ \mbox{and}\ \ \ \ \ \,
   R_2\ =\ 0.95\times0.8\ /\ 0.85\ =\ 0.894
\eeq
We'd like for the accuracy to be the same for both one-boxers and two-boxers,
which can be achieved with randomization.  Predicting one-boxing by 
Participants in $G_i$ with probability $p_i$ will lead to $R_1=R_2$ 
if\vspace{-4pt}
\beq
  (1/0.15)\,(0.04p_1+0.11p_2) & = & (1/0.85)\,(0.76(1\!-\!p_1)+0.09(1\!-\!p_2))
  \ =\ R \\[-22pt] \nonumber
\eeq
The line for which this holds is plotted in Figure~\ref{fig1}, along with
contours for $R_1$ and $R_2$ at 0.5 and 0.7.  The maximum possible
value for $R$ is 0.770, achieved when $p_1=0.139$ and $p_2=1$.

\begin{figure}[t]

\vspace*{6pt}

\begin{center}
\includegraphics[scale=0.6]{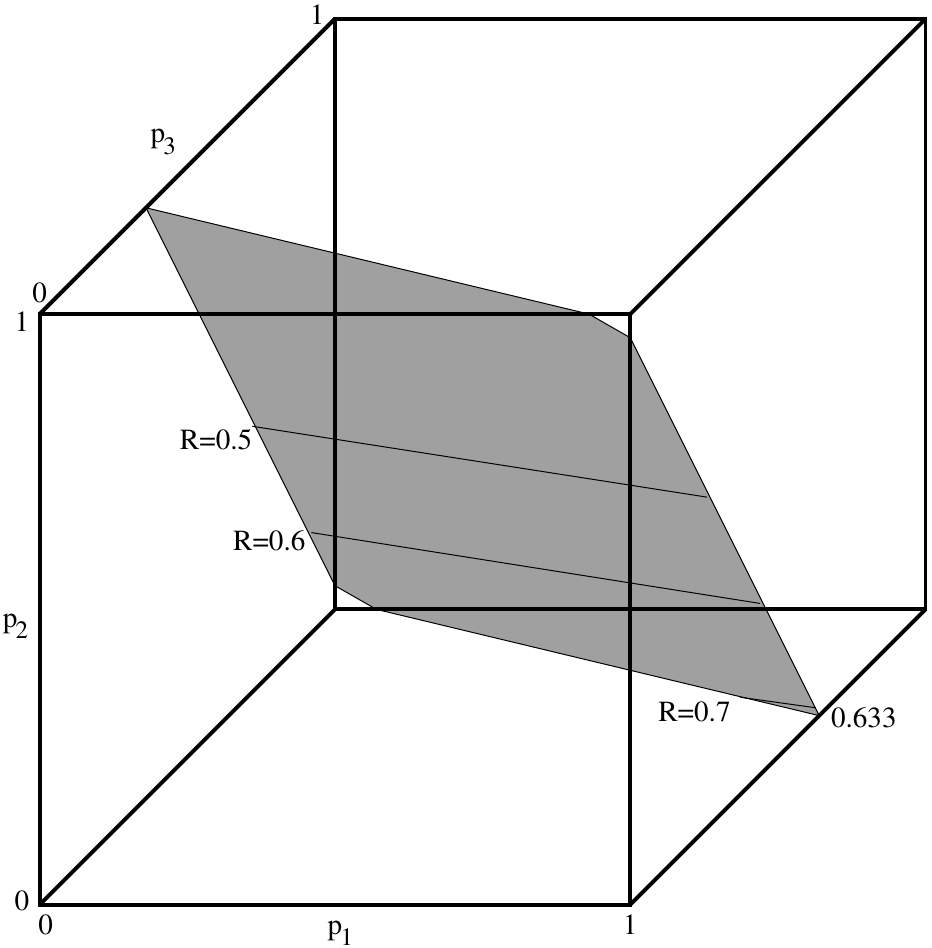}\vspace{-6pt}
\end{center}

\caption{The region of equal accuracy for the example with 
$w_1=0.2$, $w_2=0.3$, $w_3=0.5$, $q_1=0.8$, $q_2=0.1$, and $q_3=0.7$ is
shaded above.
Also shown are contours within this region for $R=R_1=R_2$ at 0.5, 0.6, and 0.7.
The maximum accuracy of $R=R_1=R_2=0.707$ is achieved at the bottom right,
where $p_1=1$, $p_2=0$, and $p_3=0.633$.
}\label{fig-3dR}

\label{fig2}
\end{figure}

When $K=3$, the region in which $R_1=R_2$ will be two-dimensional.
Figure~\ref{fig2} shows this region when $w_1=0.2$, $w_2=0.3$,
$w_3=0.5$, $q_1=0.8$, $q_2=0.1$, and $q_3=0.7$. The maximum possible
accuracy is obtained when $p_1=1$, $p_2=0$, and $p_3=0.633$, for which
$R=R_1=R_2=0.707$. Lower accuracies, such as $R=0.7$, can
be obtained using a range of values for the $p_i$. Indeed, an accuracy
of $R=0.6$ can be obtained using \textit{any} value for $p_1$, along with
suitable values for $p_2$ ranging from 0.176 to 0.287 and 
for $p_3$ ranging from 0.911 to 0.444, as $p_1$ ranges from 0 to 1.

Note that it is possible that achieving a given $R$ will require, for
some category, $G_i$, that the Predictor randomize with a $p_i>0.5$
even though $q_i<0.5$ --- that is, the Predictor usually predicts that
Participants in this category take one box even though over half of
these Participants take two boxes. For example, suppose $K=2$,
$w_1=0.1$, $w_2=0.9$, $q_1=0.4$, and $q_2=0$. All those taking one box
are in category $G_1$, so the accuracy for one-boxers will be
$R_1=p_1$.  Making $R_2$ equal to $R_1$ requires that 
$p_2=(0.96\!-\!1.02p_1)/0.96$, which is in $[0,1]$ only if
$p_1$ is in $[0.0588,0.9412]$. So the highest possible accuracy of
$R=0.9412$ is obtained when $p_1=0.9412$ and $p_2=0$.\footnote{\fsp
The fraction of Participants who take two boxes is
$0.6\times0.1+1\times0.9 =0.96$. The fraction of all Participants who
are correctly predicted to take two boxes will be 
$(1\!-\!p_1)\times0.6\times0.1\ +\ (1\!-\!p_2)\times1\times0.9$, 
and hence accuracy for two-boxers will be
$((1\!-\!p_1)\times0.6\times0.1\ +\ (1\!-\!p_2)\times1\times0.9)\ /\ 0.96=R_2$.
Equating this to $R_1=p_1$ gives $p_2=(0.96\!-\!1.02p_1)/0.96$.\vspace{-12pt}
}

For prediction using categories to be practically feasible, there must
be a way for the Predictor to determine the proportions of the
population in each category, $w_1,\ldots,w_K$, and estimates of the
fractions of people in each category who take one box,
$q_1,\ldots,q_K$.

Finding the $w_i$ should be straightforward, assuming that the category
of a person can easily be determined --- the Predictor can simply sample
people in the same way as actual Participants will be chosen, and count
the fraction in each category. 

Estimating the $q_i$ will require some guile. One cannot expect to
obtain reliable estimates by sampling people and simply
\textit{asking} them whether they would take one box or two in a
hypothetical Newcomb problem in which the Predictor has some specified
reliability, $R$. People's responses to hypothetical questions
will often differ from what they would do in a real situation.

This problem can be overcome by a resourceful and unscrupulous
Predictor. A sufficiently large number of people in each category can
be sampled, told that they are in an \textit{actual} Newcomb
situation, and their actions then observed.  These people must be
convinced that the decision they face is as described in
Section~\ref{sec-cond}.  Since the Predictor does not yet have the
information required to make good predictions, this will require
\textit{faking} evidence that the Predictor has accuracy $R$ ---
perhaps, for example, other people could be bribed to claim that they
have been past Participants, and testify that the predictions of their
choices were accurate.

Carrying out such a scheme would be difficult, but the difficulty is
of an ordinary sort. Common sense tells us that it's not impossible
for someone to succeed at deceiving people about the reliability of
the predictions, and overcoming the other practical difficulties
involved.

Given good estimates for the $q_i$ obtained in this way, the Predictor
can see whether the $w_i$ and $q_i$ found allow for prediction with
the accuracy, $R$, that people were (falsely) told the Predictor had.
If not, a new group of people will need to be sampled, and told that
they are in a Newcomb problem with some smaller~$R$ --- their actions
with this $R$ may, of course, differ from those of the previously
sampled people who were told that the Predictor had higher
reliability. If the categories chosen are actually predictive, a value
of $R$ for the fake Newcomb problems will likely eventually be found
that produces estimated $q_i$ that allow for realizing actual Newcomb
problems with the Predictor having the same reliability
$R$.\footnote{\fsp 
It is conceivable that there is no value of $R$ greater than 0.5005
(the region where EDT and CDT conflict, see
Section~\ref{sec-recommend}) that when purported to be the accuracy of
the Predictor will result in $q_i$ estimates that allow for actual
prediction with accuracy $R$. This is an empirical question (which
could well depend on the particular population from which
Participants are drawn).  It could be, for example, that when told
that the Predictor has accuracy $R=0.7$, Participants behave in such a
way that the maximum achievable accuracy is only 0.65, but when told
that $R=0.65$, the maximum accuracy is 0.6, and so forth, with only
$R=0.5$ being self-consistent. However, common sense leads one to
expect that a self-consistent $R$ greater than 0.5005 is quite likely
to exist. We may expect that Participants who believe the Predictor is
less accurate are more likely to take two boxes, but there seems
little or no reason to think that Participants become less predictable
as $R$ decreases. In any case, all that is really needed for this
paper is that it is easily imaginable that the real situation allows
for an $R$ greater than 0.5005.}

After this, Participants in the real Newcomb problems must be
convinced that their situation is as described in
Section~\ref{sec-cond}. 
Perhaps some guarantees of authenticity could
be offered that overcome any suspicion that the Predictor is still
faking.  Perhaps it will be necessary to nevertheless provide fake
evidence to some initial number of Participants in the real Newcomb
problems, so that the accuracy of predictions for them can be cited as
evidence for later Participants.  If all this is successful, for these
later Participants, the Predictor has accuracy $R$, they believe the
Predictor has accuracy $R$, and this belief is based on true evidence.

This scheme should work well when the number of categories, $K$, is
fairly small, and none of them are very rare (no $w_i$ are near zero),
since it will then be practical to obtain good estimates,
$q_1,\ldots,q_K$, with a feasible sample of Participants. The number
of categories used could easily be very large, however. If the example
above were extended to define categories based on conjunctions
of~$J=20$ attributes or their negations, rather than only four, there
would be $2^J$ categories --- more than a million --- and hundreds of
millions of Participants would need to be observed to get good
estimates of $w_i$ and $q_i$ for every category by simply counting how
many Participants are in each category and how often those in each
category choose one box.

However, one can more efficiently estimate the $w_i$ and $q_i$ when
categories are defined based on $J$ attributes, and $J$ is not small,
if one can suppose with some justification that attributes relate to
each other in a comparatively simple way.  For example, one could use
a log linear model (Agresti 2013, Section~9.4) that assumes that the
attributes, and the decision to take one box or two, have only
pairwise interactions, allowing the $w_i$ and $q_i$ to be jointly
modelled using only $(J\!+\!1)(J\!+\!2)\,/\,2$ parameters rather than the
$2K=2^{J+1}$ needed if no assumptions are made.  The assumptions of
such a model are likely to be at best approximately true, but
adopting them might allow the Predictor
to achieve accuracy within $\epsilon$ of $R$, as discussed in
Section~\ref{sec-cond}.

An alternative, more robust approach would be to use a log linear
model (or some other parsimonious model) to estimate the probability
of one-boxing for each combination of attributes, and then define a
fairly small number of categories based on these estimated
probabilities --- for example, one category for people with a
combination of attributes for which the log linear model predicts
one-boxing with probability between 0 and 0.1, another for those with
estimated one-boxing probability between 0.1 and 0.2, etc. Another
sample of potential participants would then be used to estimate the
one-boxing probabilities, $q_i$, for these categories, which will
produce valid estimates even if the assumptions of the log linear
model are significantly violated. The $w_i$ for these categories would
also be estimated by simple counting.


\subsection{\hspace*{-8pt}Shafir and Tversky's experiment}\vspace{-6pt}

Shafir and Tversky (1992) report on an interesting actual experiment
regarding Newcomb's problem that purported to utilize categories.
They primarily were testing whether, in Prisoner's Dilemma situations,
a player's choice is affected by their uncertainty regarding the other 
player's choice, but after questions regarding
Prisoner's Dilemma situations, they posed to each of the 40 participants the 
following:\vspace{-8pt}
\begin{quotation}\noindent
You now have one more chance to collect additional points. A program
developed recently at MIT was applied during this entire session to
analyze the pattern of your preferences. Based on that analysis, the program
has predicted your preference in this final problem. \\[6pt]
\hspace*{10pt}[ Describes a Newcomb problem, with 20 ``points''
in Box~A and perhaps 250 points in Box~B ]
\\[6pt]
(So far, the program has been remarkably successful: 92\%
of the participants who chose only Box~B found 250 points in it, as opposed
to only 17\% of those who chose both boxes.)

\vspace{6pt}

To insure that the program does not alter its guess after you have indicated
your preference, please indicate to the person in charge whether you 
prefer both boxes or Box B only. After you indicate your preference, press
any key to discover the allocation of points.\vspace{-8pt}
\end{quotation}
They report that 65\% of the Participants chose only one
box, and 35\% chose both boxes.

Since the purported Predictor is said to look only at the
Participant's responses to the previous (several dozen) questions
regarding Prisoner's Dilemma problems, this is an instance of
(purported) prediction based on categories --- the categories being
the different possible patterns of responses to those earlier
questions.

Shafir and Tversky's experiment did not pose an actual Newcomb problem,
however, according to the criteria of Section~\ref{sec-uncont}, for
multiple reasons:\vspace{-6pt}
\begin{itemize}
\item The claimed accuracy of the Predictor was not the same for
      one-boxers (92\%) and two-boxers (83\%), violating condition~(i). 
      This is a minor
      issue, however, since the accuracy was reasonably high for both.
\item The experimenter's claim that the ``program recently developed at MIT''
      had the stated accuracy was a lie. No such program existed.\footnote{\fsp
Shafir and Tversky do not say how they actually filled Box~B. I would
guess that they always put 250 points in it, to avoid controversy.
In any case, what they did at that point would have had no effect
on the results.}
      Although it seems that Participants believed this claim,
      their belief was not well founded.  This belief could be seen as 
      sufficient for condition~(j), though.
\item The monetary stakes were quite small --- around
      one dollar.\footnote{\fsp
      Shafir and Tversky state that the Participants were paid \$6 dollars
      on average.  Presumably, most of the payout was determined by the earlier 
      questions. However, from Shafir and Tversky's description, it may be
      that Participants had no clear idea of the stakes (though they 
      likely realized that they were not very large).}  
      The Participants
      may therefore have been maximizing a utility function that was
      not predominately about monetary gain, violating condition~(l).
\item Participants' uncritical acceptance of the claimed accuracy
      may also have been related to the small stakes, with skeptical
      inquiry not being seen as worth the effort.
\item The low stakes may also be why Participants
      spent little time deliberating. (The entire 
      experiment, including dozens of previous questions, took about 
      40 minutes.) There may also have been a time limit imposed by
      the experimenters --- perhaps implicitly, by the social convention of
      not taking much time to do something when it's not clear that long
      thought is expected. This would violate condition~(f).
\item There is insufficient information to determine whether condition~(k)
      was satisfied --- that is, whether Participants were rationally
      convinced that the
      advertised accuracy of the Predictor applied to them in particular,
      after they had made their choice (but not seen the result).\footnote{\fsp
Ahmed (2014, p.~105) speculates that Shafir and Tversky's
subjects believed that ``The psychological test predicts how I will
act by identifying what causes my initial \textit{inclinations}'', and
argues that if so 
``\ldots the subject's knowledge of his current inclinations
means that Evidential Decision Theory agrees with CDT that the subject
should take both boxes. [~Shafir and Tversky's experiment~] is
then \textit{not} a genuine Newcomb problem.'' 
I agree, given Ahmed's hypothesis, but unlike him,
I don't see much reason to think subjects believed predictions were
based only on initial inclination.\vspace*{-15pt}}
\end{itemize}

Although no prediction program was actually used in their experiment,
Shafir and Tversky report that in retrospect a program that used a
simple rule looking at responses to previous questions would have had
accuracy of 70\% for one-boxers and 71\% for two-boxers. This is
likely an overestimate of the true accuracy of such a program, since
the rule was chosen using the same 40 responses used to assess
accuracy, but it does seem plausible that if they had employed an iterative
procedure as described in Section~\ref{sec-pr} they might have been
able to find a rule that self-consistently had an accuracy of $R$ when
Participants were (truthfully) told that it had accuracy $R$, for some
$R$ significantly greater than 0.5.

\subsection{\hspace*{-8pt}What Participants should believe when predictions 
are made using categories}\label{sec-bel}\vspace{-6pt}

Section~\ref{sec-pr} outlined a concrete, realistic scheme for how
a Predictor can make predictions that satisfy condition~(i) of
Section~\ref{sec-uncont} --- that is, predictions that are correct with
some specified probability, $R$, for both Participants who take one
box and those who take two. In this section, I discuss when
condition~(k) will also hold --- when Participants will rationally
believe the Predictor has probability $R$ of being correct despite
knowing facts in addition to their chosen action, possibly including
facts that the Predictor did not know.


I will give a detailed subjective Bayesian account of how you as a
Participant should reason, given the information you have, and making
some reasonable assumptions about your prior beliefs.  Of course, most
people, even if generally rational, and provided with ample
computational resources, will not actually make their choice by applying
formal Bayesian methods --- they will likely use some informal, heuristic
method, which they hope will approximate the correct inference. But
the formal treatment seems appropriate for checking condition~(k). I
hope the detailed formal presentation will also clarify the causal and
inferential structure of the problem, and hence what is the correct
decision method.

As will be seen, the less knowledge Participants have, the easier it
is to arrange for a valid Newcomb problem. But the more knowledge
Participants have, the more concrete will be their view of the
problem, facilitating clearer intuitions. Shafir and Tversky (1992)
demonstrate experimentally that, in Prisoner's Dilemma situations,
uncertainty about the action of the other prisoner can lead to a
decision (cooperate) differing from what the same person would do in
\textit{all} the concrete situations that the uncertainty might
resolve to (that is, if they know that the other prisoner cooperated,
or know that they did not).  They speculate that the same phenomenon
may explain why some people take only one box in Newcomb's problem.
So constructing Newcomb problems that are as concrete as possible
seems of interest.

What Participants should believe depends on what they know of the
method the Predictor uses.  If prediction is done using categories,
there are several things that a Participant might know about the
prediction method:\vspace{-8pt}
\begin{itemize}
\item[1)] The categories, $G_1,\ldots,G_K$, that the Predictor uses to make
          predictions.\vspace{-2pt}
\item[2)] The fractions of the population,
          $w_1,\ldots,w_K$, in each of these categories.\vspace{-2pt}
\item[3)] Which category they themselves are in, $G_s$.\vspace{-2pt}
\item[4)] The Predictor's estimates, $q_1,\ldots,q_K$,
          for the fractions of potential Participants in each category who 
          would take one box.\vspace{-2pt}
\item[5)] The probabilities, $p_1,\ldots,p_K$,
          with which the Predictor predicts one-boxing for each 
          category.\vspace{-8pt}
\end{itemize}
As for any Newcomb problem, we also assume, in accord with condition~(j),
that Participants know $R$, the accuracy of the Predictor, at least to
within some error tolerance, $\epsilon$, as well as the general setup
of the problem. I will use $B_0$ to denote this background knowledge.
At the moment, I will assume that your full background knowledge, $B$,
contains nothing of relevance outside $B_0$ and some specified 
subset of~(1) to~(5) above, but I relax this assumption in 
Section~\ref{sec-additional}.

\noindent \textbf{It's not a Newcomb problem if you know too much.} \ \
If Participants know all of (1) through~(5) above, condition~(k) for
a valid Newcomb problem will be
violated.  Just knowing (1), (3), and (5) suffices to make your
probability that the opaque box contains \$1M be independent of
whether you end up taking one box or two. In particular, if\vspace{-4pt}
\beq
B & \!=\! & B_0 \ \&\ \mbox{The Predictor uses categories $G_1,\ldots,G_K$}
  \nonumber\\ & &
\ \ \ \ \, \&\ \mbox{You are in category $G_s$}
  \nonumber\\ & &
\ \ \ \ \, \&\ \mbox{The Predictor predicts one-boxing for $G_i$ with 
                     probability $p_i$}
\eeq
then, recalling that $S_1$ is the event that the Predictor predicts one-boxing,
you should believe that
\beq
 \prs(S_1\, |\, A_1\and B) \ =\ 
 \prs(S_1\, |\, A_2\and B) \ =\ p_s 
\eeq
and so your probabilities for the Predictor being correct, $C$, given
that you take one box or two, should be
\beq
 \prs(C\, |\, A_1\and B) \ =\ p_s\ \ \ \ \mbox{and}\ \ \ \
 \prs(C\, |\, A_2\and B) \ =\ 1-p_s 
\eeq
which violates condition~(k).\footnote{\fsp Unless $p_s=0.5$, and $R=0.5$,
which is not an interesting situation.}

\noindent \textbf{It is a Newcomb problem if you don't know your own 
category.}\ \
However, if you don't know your own category, condition~(k) will be
satisfied even with knowledge of the $w_i$, $q_i$, and $p_i$.
Ignorance of your category is implausible if categories are based on
attributes such as age, height, or place of birth, but is quite
plausible if categories are based on things like brain volume or 
presence of certain gene variants.

Suppose, then, that your background knowledge, $B$, is
\beq
B & \!=\! & B_0 \ \&\ \mbox{The Predictor uses categories $G_1,\ldots,G_K$}
  \nonumber\\ & &
\ \ \ \ \, \&\ \mbox{These categories have proportions $w_i$}
  \nonumber\\ & &
\ \ \ \ \, \&\ \mbox{The proportion in category $G_i$ who take one box is $q_i$}
  \nonumber\\ & &
\ \ \ \ \, \&\ \mbox{The Predictor predicts one-boxing for $G_i$ with 
                     probability $p_i$}
\eeq
The $p_i$ that you know
must of course be consistent with the $w_i$, $q_i$, and $R$ that you
know, satisfying the equations of~(\ref{acc-eq-R}).  Some or all of
the $p_i$ might be 0 or 1.  Ignorance of your category is taken
to mean that your subjective probability that you are in $G_i$ is
equal to its population frequency, $w_i$. (So you don't just not
know your category with certainty, but also lack any evidence
regarding it beyond the $w_i$.) We can then write\vspace{1pt}
\beq
 \prs(C\, |\, A_1\and B) 
 & \!=\! & 
 \prs(S_1\, |\, A_1\and B) 
 \ =\
 { \prs(S_1 \and  A_1\giv B) \over\rule{0pt}{10pt} \prs(A_1\giv B)}
 \nonumber\\[6pt]
 & \!=\! &
 { \displaystyle \sum_{i=1}^K \prs(S_1 \and  A_1\and G_i\giv B) \over
   \displaystyle \sum_{i=1}^K \prs(A_1\and G_i\giv B)}
 \ =\  
 { \displaystyle \sum_{i=1}^K \prs(S_1 \and  A_1\giv G_i\and B) 
   \,\prs(G_i\giv B)
   \over
   \displaystyle \sum_{i=1}^K \prs(A_1\giv G_i\and B)\,\prs(G_i\giv B)}
 \ =\ 
 { \displaystyle \sum_{i=1}^K p_i\,q_i\,w_i \over 
   \displaystyle \sum_{i=1}^K q_i\,w_i }\ \ \ \ \ \ \ \
 \label{eq-knowp1}
\eeq
where here $G_i$ denotes the event that you are in category $G_i$. 
In similar fashion,\vspace{-6pt}
\beq
 \prs(C\, |\, A_2\and B) 
 \ \ =\ \ 
 \prs(S_2\, |\, A_2\and B) 
 \ \ =\ \ 
 { \displaystyle \sum_{i=1}^K (1\!-\!p_i)\,(1\!-\!q_i)\,w_i \over 
   \displaystyle \sum_{i=1}^K (1\!-\!q_i)\,w_i }
 \label{eq-knowp2}
\eeq
Since the Predictor has accuracy $R$, applying (\ref{acc-eq-R}) to
the right sides of equations~(\ref{eq-knowp1}) and (\ref{eq-knowp2}) gives
\beq
 \prs(C\, |\, A_1\and B) \ =\ R\ \ \ \ \mbox{and}\ \ \ \
 \prs(C\, |\, A_2\and B) \ =\ R 
\label{eq-correct1}
\eeq
showing that condition~(k) is satisfied.

A crucial step in this derivation is the replacement of
$\prs(S_1\and A_1\giv G_i\and B)$ by $p_i\,q_i$ on the basis that
$S_1$ and $A_1$ are conditionally independent, with
$\prs(S_1\giv G_i\and B)=p_i$ and $\prs(A_1\giv G_i\and B)=q_i$.
The probabilistic independence here follows from causal
independence, with the Predictor's randomized prediction not being
causally influenced by anything else, and your choice not being
influenced by the prediction (assuming, as we do, that you do not
do something like peek inside the opaque box before choosing). That
\mbox{$\prs(A_1\giv G_i\and B)=q_i$} is a consequence of assuming
that $B$ contains no other information relevant to your choice, and
that the Predictor's $q_i$ estimates are good. The latter assumption
can be weakened to an assumption that you have no better information
on the true prevalence of one-boxing in each category than the
Predictor has. (Similar comments of course apply to the replacement of
$\prs(S_2\and A_2\giv G_i\and B)$ by $(1\!-\!p_i)\,(1\!-\!q_i)$.)

The assumption that your background knowledge contains nothing
relevant other than the $G_i$, $w_i$, $q_i$, and $p_i$ is
unrealistic.  As mentioned above, the effect of knowing other relevant
information is discussed below.

However, knowing \textit{less} than is
assumed above still results in a valid Newcomb problem.  Rather than
$B$ including knowledge of the $w_i$, $q_i$, and $p_i$, we can exclude
any or all of these from $B$ and still conclude that $\prs(C\, |\,
A_1\and B) \ =\ \prs(C\, |\, A_2\and B) \ =\ R$.  This is a consequence
of~(\ref{eq-correct1}) holding for \textit{any} values of the 
$w_i$, $q_i$, and $p_i$.  In general, if $X$ and $Y$ are any events, and
$Z_1,\ldots,Z_n$ are events of which exactly one must hold, then if
$\Pr(X\giv Y\and Z_{\ell})=a$ for all $\ell$,
then $\Pr(X\giv Y)=a$ also.\footnote{\fsp
From the definition of conditional probability, if
$\Pr(X\giv Y\and Z_{\ell})=a$
then $\Pr(X\and Y\and Z_{\ell})=a\Pr(Y\and Z_{\ell})$, from which\vspace{-2pt}
\beq
 \Pr(X\giv Y) \ =\ \ {\Pr(X\and Y) \over\rule{0pt}{8pt} \Pr(Y)} 
 \ =\ {\displaystyle \sum_{\ell=1}^n \Pr(X\and Y\and Z_{\ell}) 
       \over \displaystyle \sum_{\ell=1}^n \Pr(Y\and Z_{\ell}) }
 \ =\ {\displaystyle \sum_{\ell=1}^n a \Pr(Y\and Z_{\ell}) 
       \over \displaystyle \sum_{\ell=1}^n \Pr(Y\and Z_{\ell}) }
 \ =\ {\displaystyle a \sum_{\ell=1}^n \Pr(Y\and Z_{\ell}) 
       \over \displaystyle \sum_{\ell=1}^n \Pr(Y\and Z_{\ell}) }
 \ =\ a
\eeq\vspace{-20pt}
}
The same is true when the $Z_{\ell}$ are replaced by possible values of
continuous variables such as the $w_i$, $q_i$, and $p_i$, regardless of
what subjective beliefs a Participant may have regarding these quantities.

\noindent \textbf{Inference when you know your category and the $w_i$, but
not the $q_i$.}\ \
We saw above that if you know your own category, knowing
the $p_i$ as well would invalid the situation as a Newcomb problem.
It is also unlikely that a Newcomb problem can be valid if you know
your category and both the $q_i$ and the $w_i$, since these would tell 
you a lot about
the $p_i$, due to the requirement that the Predictor have accuracy $R$,
even though as seen in Section~\ref{sec-pr} they may not
completely determine the $p_i$ when $K>2$.

Supposing that a Participant knows the categories, $G_1,\ldots,G_K$,
used by the Predictor, a significant degree of ignorance of $q_1,\ldots,q_K$ 
seems plausible, but not guaranteed.  It is \textit{possible} that a
Participant might essentially duplicate the Predictor's efforts
at estimating the fractions of people in each category who would take
one box. But absent such efforts, a Participant
may well have little knowledge of how likely people in each category
are to take one box.
Nozick's (1969) observation that ``\ldots people
seem to divide almost evenly on the problem, with large numbers
thinking that the opposing half is just being silly'' is evidence of
this, since those thinking the others are silly must not have realized
that the contrary view is common. Stalnaker (2018, p.~190) recounts
his initial reaction to Newcomb's problem:\vspace{-7pt}
\begin{quotation}\noindent
I thought, "It is obvious that one should take both boxes, and there is
no puzzle about how a predictor could get it right 90\% of the time, since
surely almost everyone will make this choice, so the predictor can safely
predict that everyone will, and be about 90\% accurate.'' That was my
first reaction\ldots I was of course factually mistaken\ldots
\end{quotation}
So Stalnaker initially had incorrect beliefs about how likely people
were to take one box. After realizing his mistake, he presumably
became quite uncertain about such matters. It seems reasonable to
think that anyone who has not deliberately set out to gather relevant
data has little idea what the prevalence of one-boxing is in any
particular category of Participants. Also, a Predictor who sets out to
create a valid Newcomb problem could deliberately choose to use
categories for which Participants are ignorant of $q_i$. At a minimum,
one can easily imagine that Participants are ignorant of the $q_i$,
and ask how a Participant should reason given such ignorance, without
thinking that this question must be irrelevant to any real problem.

So, suppose that you know the categories used and their population
frequencies, $w_i$, as well as your own category, $G_s$ --- 
(1), (2), and (3) in the list above.  That is, your background 
knowledge is
\beq
B & \!=\! & B_0 \ \&\ \mbox{The Predictor uses categories $G_1,\ldots,G_K$}
  \nonumber\\ & &
\ \ \ \ \, \&\ \mbox{These categories have proportions $w_i$} 
  \nonumber\\ & &
\ \ \ \ \, \&\ \mbox{You are in category $G_s$}
\eeq
What should you as a Participant infer in this situation?

Since you don't know $q_1,\ldots,q_K$, in a
Bayesian approach to inference and decision you should formulate a
prior probability distribution for these quantities, whose density I
will write as $f(q_1,\ldots,q_K)$, that reflects your knowledge (or
ignorance) of how people in these different categories might choose in
a Newcomb problem.  After learning that the Predictor has accuracy
$R$, you should modify your prior to eliminate values for the $q_i$
that do not (in conjunction with the $w_i$) allow for an accuracy of
$R$, renormalizing the distribution on the remaining values to again
integrate to one.  I will use $f_{R,w}$ to denote the density for this
modified prior distribution.

As discussed above, I will assume that you have little knowledge of
people's choices.  One plausible prior distribution expressing this
ignorance is a uniform distribution over $[0,1]$ for each of the $q_i$,
independently.  More generally, one might consider prior distributions
in which the $q_i$ are independent, and all have the same
distribution, so that $f(q_1,\ldots,q_K)=g(q_1) \cdots g(q_K)$ for some
univariate density function $g$. Further specifying that
$g(q)=g(1\!-\!q)$ would express an ignorance of which of one-boxing
and two-boxing is more common.

One can determine whether some accuracy $R$ is possible with given
$w_i$ and $q_i$ by fixing these values in~(\ref{acc-eq-R}) and then
seeing whether these two equations have a solution in which the $p_i$
(the randomization probabilities used by the Predictor) are all in the
interval $[0,1]$. For $K=2$ categories, with several values of $w_1$
(and hence of $w_2=1\!-\!w_1$), the top row of plots in
Figure~\ref{fig-2cat1} show how the maximum possible $R$ varies with
$q_1$ and $q_2$.


\begin{figure}

\vspace{-10pt}

\begin{center}
\includegraphics[scale=0.8]{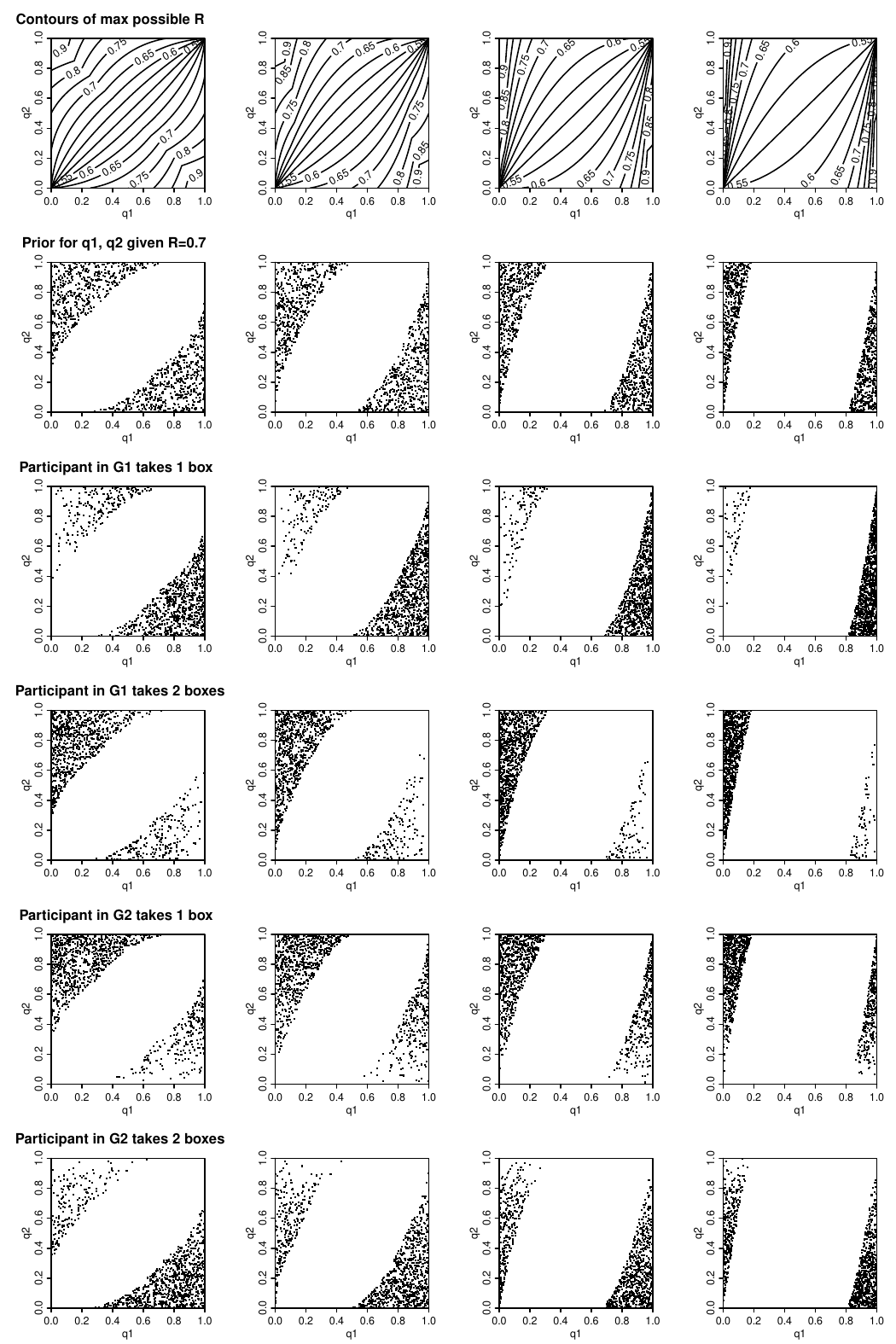}\vspace{-4pt}
\end{center}

\hspace{92pt}
$w_1=0.5$\hspace{57pt} $w_1=0.6$\hspace{57pt} $w_1=0.7$\hspace{57pt} $w_1=0.8$ 

\vspace{-4pt}

\caption{Two-category examples with $G_1$ proportions ($w_1$) of 0.5,
0.6, 0.7, and 0.8 (with $w_2=1\!-\!w_1$).  The top plots show the
maximum possible accuracy, $R$, a Predictor can achieve as
the fractions of one-boxers in $G_1$ and $G_2$ ($q_1$ and $q_2$) vary. Next
is a sample of 1000 points from the prior distribution for $q_1$ and $q_2$
uniform on the region for which $R=0.7$ is possible. Below that are
samples of 1000 points from posterior distributions that a Participant
with this prior in category $G_1$ or $G_2$ would have after choosing to
take one box or two.}\label{fig-2cat1}

\end{figure}

\begin{figure}

\begin{center}
\includegraphics[scale=0.8]{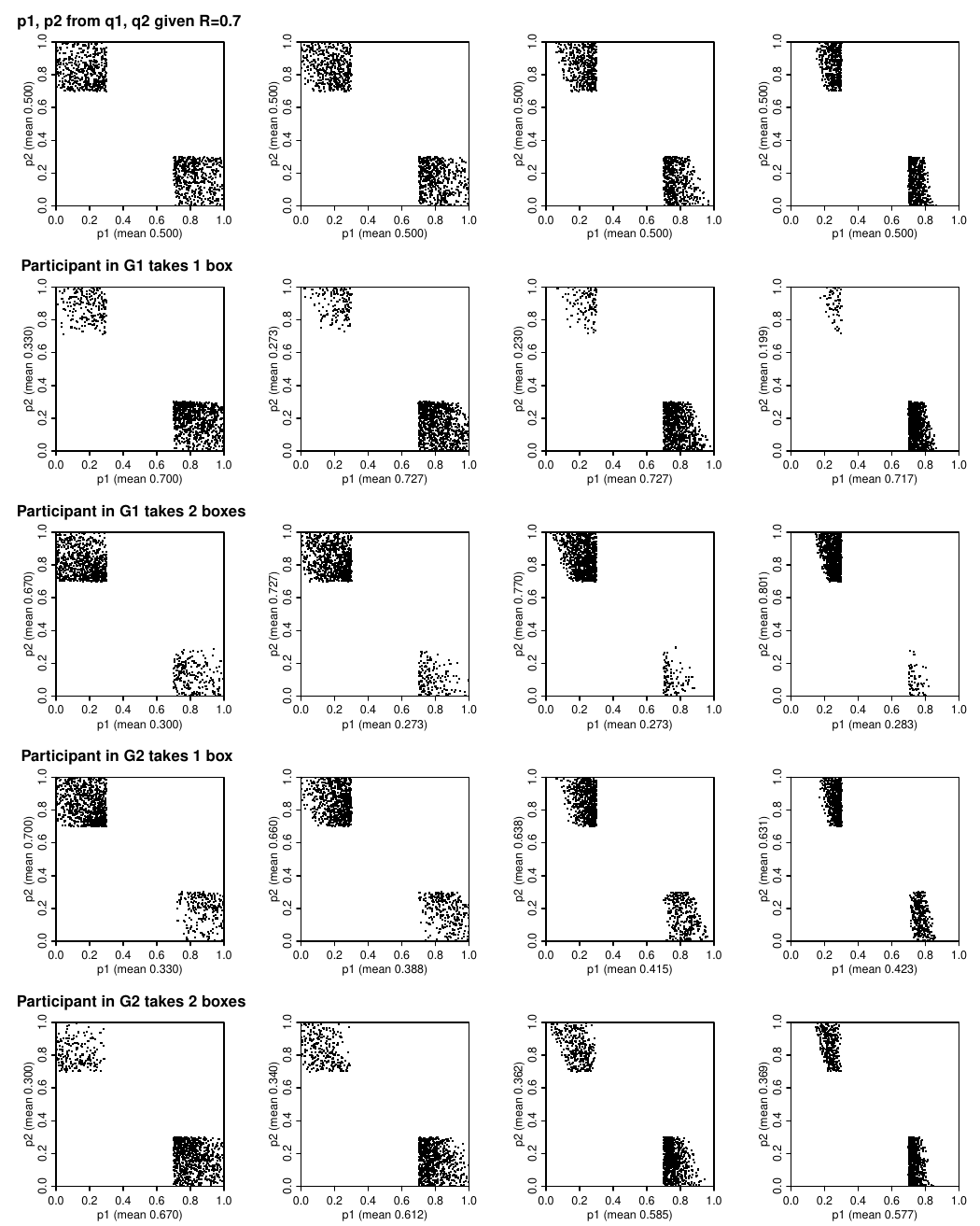}\vspace{-2pt}
\end{center}

\hspace{92pt}
$w_1=0.5$\hspace{57pt} $w_1=0.6$\hspace{57pt} $w_1=0.7$\hspace{57pt} $w_1=0.8$ 

\caption{Continuation of the two-category examples. The top plots show
1000 values for the probabilities, $p_1$ and $p_2$, with which the
Predictor predicts one-boxing, for Participants in categories $G_1$
and $G_2$, derived from values for $q_1$ and $q_2$ taken from the
uniform distribution conditional on accuracy of $R=0.7$ being possible
(see second row of Figure~\ref{fig-2cat1}). The plots in the four
lower rows show 1000 points from the posterior distribution for $p_1$
and $p_2$ that a Participant in category $G_1$ or $G_2$ would have
after choosing to take one box or two. In each plot, the means of
$p_1$ and $p_2$ are also shown, estimated using 1,000,000 sampled
points (with accuracy of about $\pm0.001$).}\label{fig-2cat2}

\end{figure}

The second row of Figure~\ref{fig-2cat1} shows a sample of 1000 random
points uniformly sampled from the region of $q_1$ and $q_2$ for
which $R=0.7$ is possible. This region corresponds to that for which
the maximum possible $R$ is at least 0.7, which can be seen in the
contour plots of the first row. These 1000 points are a representation
of the beliefs that you as a Participant should have if you begin with
a uniform distribution for $q_1$ and $q_2$, independently, and then
learn that the Predictor has an accuracy of $R=0.7$,
eliminating values for $q_1$ and $q_2$
for which this is impossible, leaving the uniform
distribution over the remaining values.\footnote{\fsp There are 
some implicit assumptions here, including that the Predictor will be content
with some accuracy $R$ even if it turns out that a much higher accuracy
is likely possible, and that the Predictor does not select the
categories used for prediction in a way that favours some $q_1$, $q_2$
values over others.}

When $K=2$, for a given $R$, the values of $q_1$ and $q_2$, together
with $w_1$ and $w_2$, determine what randomization probabilities,
$p_1$ and $p_2$, the Predictor should use for predicting that a
Participant in category $G_1$ or $G_2$ will take one box. (This is
illustrated in Figure~\ref{fig1} for a particular set of values for
the $w_i$ and $q_i$.)  It follows that the distributions for $q_1$ and
$q_2$ illustrated in the second row of Figure~\ref{fig-2cat1}
determine corresponding distributions for $p_1$ and $p_2$, which are
illustrated in the top row of Figure~\ref{fig-2cat2}. The values of
$p_1$ and $p_2$ shown there are all either at least $0.7$ or no more
than $0.3$, since when $K=2$ an accuracy of $R$ is not possible if
either $p_1$ or $p_2$ is less than $R$ and greater than
$1\!-\!R$.\footnote{\fsp Suppose that $p_1$ is in $(1\!-\!R,R)$.  Then
the probability that the prediction is correct must be less than $R$
for both one-boxers and two-boxers in $G_1$.  Whatever $p_2$ might be,
the probability that the prediction is correct must be less than $R$
for either one-boxers or two-boxers (or both) in $G_2$ (since either
$p_2<R$ or $1\!-\!p_2<R$, assuming $R>1/2$). So whatever $w_1$ and
$w_2$ are, the overall probability that the prediction is correct will be
less than $R$ for either one-boxers or two-boxers, contradicting
condition~(i) of Section~\ref{sec-uncont}.  The argument is similar if
it is $p_2$ that is in $(1\!-\!R,R)$. But when $K>2$, it is possible
for some $p_i$ to be in $(1\!-\!R,R)$.}

After you take one box or two boxes (that is, perform action $A_1$ or
$A_2$), but before you open the opaque box and see what's in it, your
beliefs should change to account for now knowing what action you took.
Taking one box or two boxes affects your beliefs through this action's
affect on your beliefs regarding the~$q_i$. If you know you are in
category $G_s$, your probability for a set of $q_i$ values should
increase in proportion to $q_s$ if you took one box, or in proportion
to $1\!-\!q_s$ if you took two boxes, since these are the
probabilities of these actions for someone in category $G_s$. Note
that although we began with a prior in which the $q_i$ were
independent, conditioning on the Predictor having accuracy $R$
introduces a dependence between the $q_i$ (as can be seen in the plots
in the second row of Figure~\ref{fig-2cat1}), and hence accounting for
having taken one box or two will change the probability distribution
for all the $q_i$, not just for $q_s$.

For the two-category examples of Figure~\ref{fig-2cat1}, the third and
lower rows show the posterior distributions for $q_1$ and $q_2$ after
taking one box or two boxes, for a Participant in $G_1$ or $G_2$.  These
distributions are again presented by showing 1000 points sampled from
them.\footnote{\fsp The posterior plots were produced by sampling
points from the posterior distribution without the constraint that the
Predictor have accuracy $R$ (which is computationally easy, as discussed
more below), and then
discarding points that don't satisfy this constraint. Changing the
order in which evidence is accounted for does not affect the final
result in Bayesian inference.\vspace{-10pt}}

The second and lower rows of Figure~\ref{fig-2cat2} show the posterior
distributions for $p_1$ and $p_2$ for a Participant in $G_1$ or $G_2$
who takes one box or two, that are derived from the posterior
distributions for $q_1$ and $q_2$. The means for $p_1$ and $p_2$ are
shown in the axis titles, from which one can obtain the Participant's
subjective probability that the predictor is correct.  For a
Participant in $G_1$, this is the mean of $p_1$ if the Participant
takes one box, and one minus the mean of $p_1$ if the Participant
takes two boxes.  For a Participant in $G_2$, the mean of $p_2$ or one
minus the mean of $p_2$ gives the probability of the Predictor being
correct.

\begin{figure}

\vspace{-10pt}

\begin{center}
\includegraphics[scale=0.8]{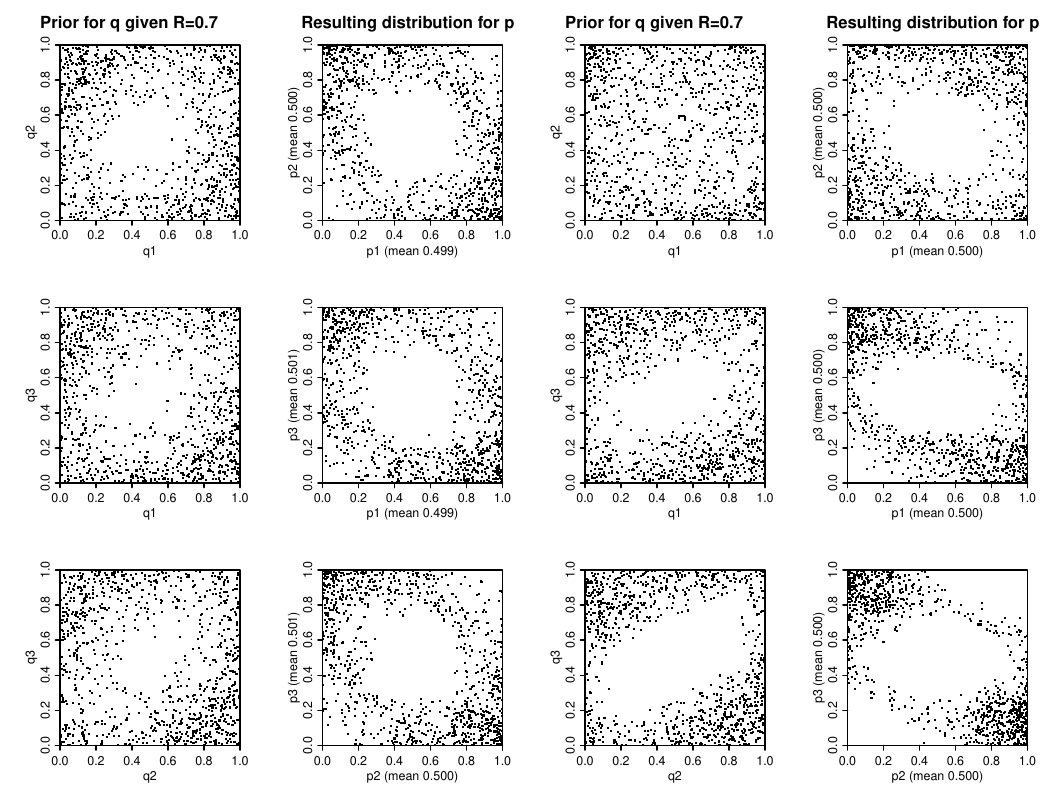}\vspace{4pt}

\includegraphics[scale=0.8]{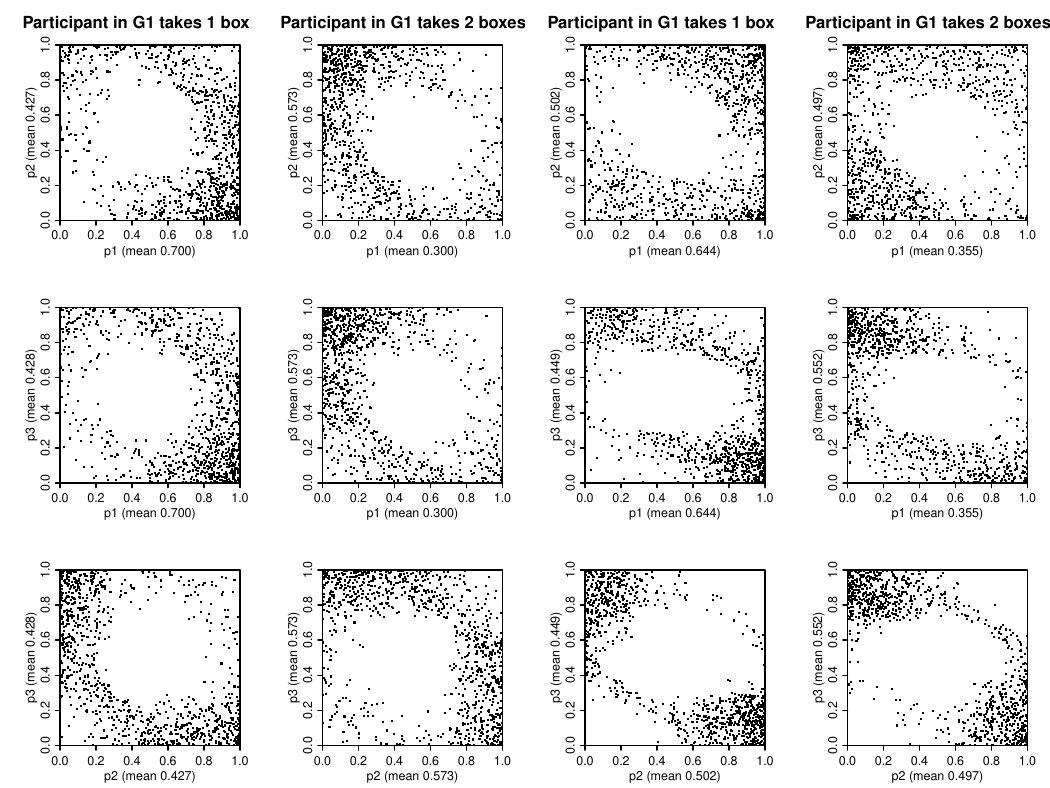}\vspace{-5pt}

\vspace{-8.35in}\hspace{9pt}\rule{0.5pt}{8.35in}\vspace{3pt}

\end{center}

\hspace{90pt}
$w_1=1/3,\ w_2=1/3,\ w_3=1/3$\hspace{49pt} $w_1=0.2,\ w_2=0.3,\ w_3=0.5$

\vspace{-4pt}

\caption{Three-category examples, using two sets of values for $w_1$, $w_2$, and
$w_3$, on the left and the right. Distributions for $q_1$, $q_2$, and $q_3$ and
for $p_1$, $p_2$, and $p_3$ are shown as three scatterplots of pairs of
variables, arranged vertically. On the top, 1000 points are shown from the
uniform prior distribution for $q_1$, $q_2$, and $q_3$ over the region where
predictive accuracy of $R=0.7$ is possible, along with 1000 values for $p_1$,
$p_2$, and $p_3$ \mbox{[ \textit{continued on next page} ]}\label{figK3}}

\end{figure}

\begin{figure}

\vspace{-10pt}

\begin{center}
\includegraphics[scale=0.8]{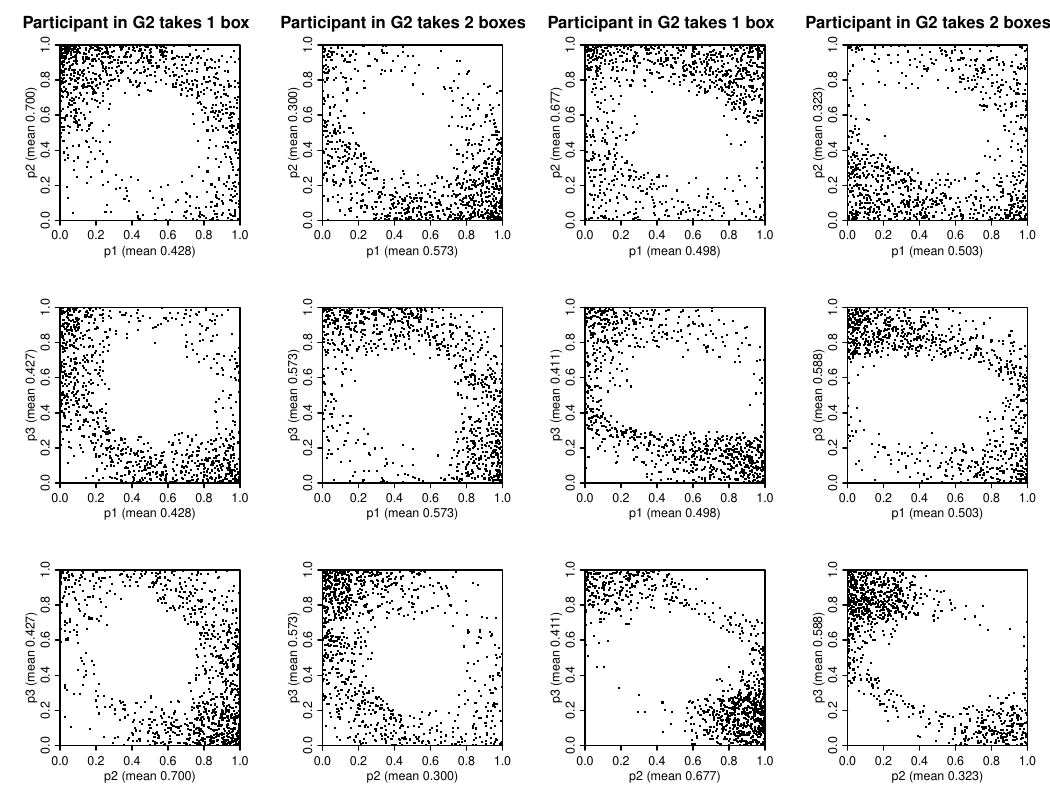}\vspace{4pt}

\includegraphics[scale=0.8]{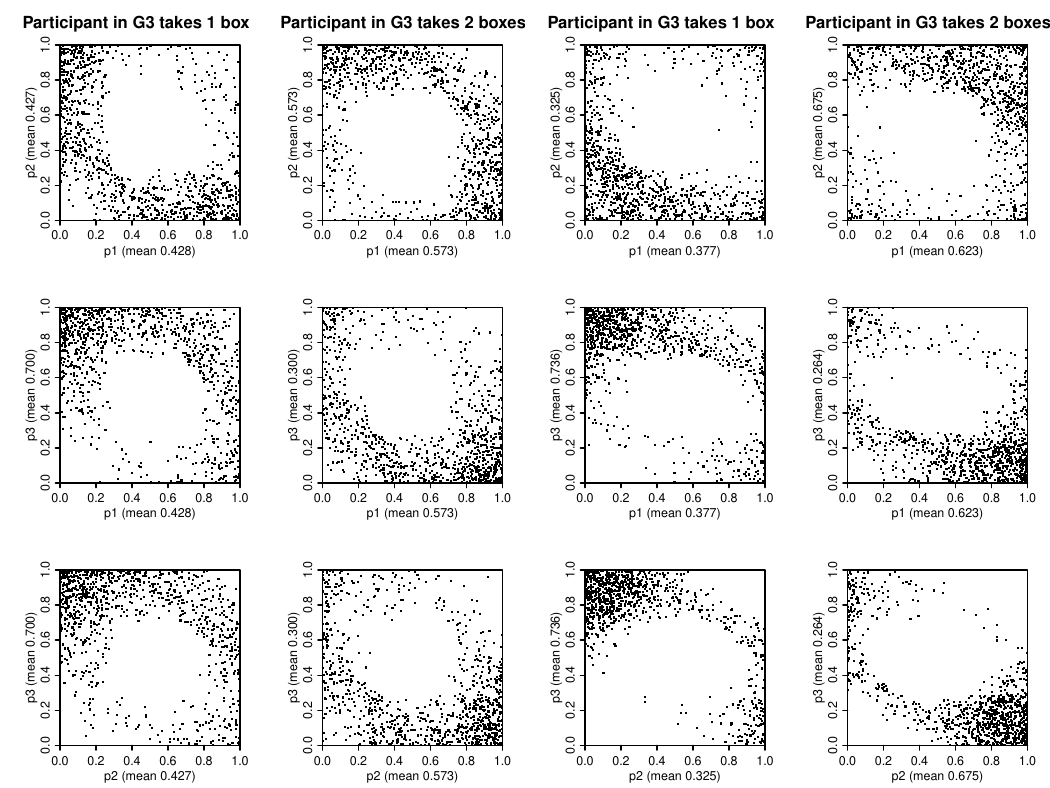}\vspace{-5pt}

\vspace{-8.35in}\hspace{9pt}\rule{0.5pt}{8.35in}\vspace{3pt}

\end{center}

\hspace{90pt}
$w_1=1/3,\ w_2=1/3,\ w_3=1/3$\hspace{49pt} $w_1=0.2,\ w_2=0.3,\ w_3=0.5$

\vspace{6pt}

Figure \ref{figK3}: [ \textit{continued from previous page} ] 
that give accuracy of $R=0.7$, sampled uniformly from
those that achieve this accuracy for each of the $q$ values.
Remaining plots show the posterior distribution of
$p_1$, $p_2$, and $p_3$ based on this prior for a 
Participant in each of $G_1$, $G_2$, and $G_3$, who has taken one box or two.
The means of
$p_1$, $p_2$, and $p_3$ are also shown, estimated using 1,000,000 sampled
points (giving roughly $\pm 0.001$ accuracy).

\end{figure}

Figure~\ref{figK3} illustrates prior and posterior distributions for
the $q_i$ and $p_i$ when $K=3$, both when all categories have the same
prevalence (all $w_i$ equal to 1/3), and in an example with unequal
prevalences ($w_1=0.2$, $w_2=0.3$, $w_3=0.5$).  The prior and
posterior distributions for the $q_i$ and $p_i$ in these figures are
displayed by showing the three pairwise scatterplots of the three
values for samples of 1000 points.

Here, as for the $K=2$ examples, prior distributions for the $q_i$ are
assumed to be uniform, constrained by the requirement that accuracy of
$R=0.7$ be possible.  When $K>2$, the $q_i$ do not uniquely determine the
$p_i$, so a Participant's subjective distributions for the $p_i$ depend on the
Participant's beliefs about how the Predictor will choose among the
$p_i$ that result in accuracy $R$.  I assume these beliefs
are captured by a uniform distribution over the region of $p_i$ having
accuracy $R$. (This does not mean that a
Participant believes that the Predictor actually chooses the $p_i$
randomly, only that their ignorance is expressed by this
distribution.)

\noindent \textbf{It's a Newcomb problem if you know the $w_i$, and
they're all equal.}\ \ For $K=2$, Figure~\ref{fig-C2} shows subjective
probabilities of the Predictor being correct found from the posterior
means of the $p_i$ in Figure~\ref{fig-2cat2}, which are based on a
uniform prior for $q_1$ and $q_2$. Probabilities are shown for a
Participant in $G_1$ or $G_2$ who has taken one box or two boxes, for
several values of $w_1$ and $w_2$.  Notice that these probabilities
exactly equal $R=0.7$ when $w_1=w_2=0.5$.

For $K=3$, the posterior probabilities that a Participant should have
for the Predictor being correct are shown in Figure~\ref{fig-C3}.
These are found from the posterior means of $p_1$, $p_2$, or $p_3$,
for Participants who know they are in $G_1$, $G_2$, or $G_3$, given
that they take one box or two, which are shown in Figure~\ref{figK3}.
As was the case for the $K=2$ examples, one can see in
Figure~\ref{fig-C3} that when the $w_i$ are all equal, a Participant's
posterior probability for the Predictor being correct equals $R$
exactly after taking either one box or two, regardless of which
category they know they are in.  Hence condition~(k) is satisfied in
this case.

When the $w_i$ are not all equal, we see in Figures~\ref{fig-C2}
and~\ref{fig-C3} that the posterior probabilities of the prediction
being correct are not equal to $R$ (which is 0.7 here), so
condition~(k) for a Newcomb problem is not satisfied exactly, though when
the $w_i$ do not differ greatly, the departure from $R$ is fairly
small.

One can also see in these Figures that the posterior probability that
the Predictor is correct is the same (within sampling error) after
taking one box as after taking two, though different for Participants
in different categories.  

As will be seen below, these properties hold more generally and can be
shown theoretically.

\begin{figure}[t]

\begin{center}
\begin{tabular}{|cc|c@{~~}cc|c@{~~}cc|}
\hline
\rule{0pt}{12pt}
 &  
 & \multicolumn{3}{c|}{~~Know you're in $G_1$~~}
 & \multicolumn{3}{c|}{~~Know you're in $G_2$~~} \\
\rule{0pt}{11pt}
   $w_1$ & $w_2$ &
 & Take 1 & Take 2 &
 & Take 1 & Take 2 \\
\hline
\rule{0pt}{12pt}
0.5 & 0.5 &
& 0.700 & 0.700 &
& 0.700 & 0.700
\\
\rule{0pt}{12pt}
0.6 & 0.4 &
& 0.727 & 0.727 &
& 0.660 & 0.660
\\
\rule{0pt}{12pt}
0.7 & 0.3 &
& 0.727 & 0.727 &
& 0.638 & 0.638
\\
\rule{0pt}{12pt}
0.8 & 0.2 &
& 0.717 & 0.717 &
& 0.631 & 0.631
\\
\rule{0pt}{12pt}
0.9 & 0.1 &
& 0.708 & 0.709 &
& 0.624 & 0.624
\\
\rule{0pt}{12pt}
0.95 & 0.05 &
& 0.704 & 0.704 &
& 0.620 & 0.620
\\
\hline
\end{tabular}
\end{center}

\caption{Posterior subjective probabilities of the Predictor being correct for
two-category examples. For each $w_1$, $w_2$,
the table shows the probability for the Predictor
being correct that a Participant who knows they are in $G_1$ or $G_2$
should have, once they have taken one box or two,
based on the uniform prior for $q_1$ and $q_2$ over the region where
the stated accuracy of $R=0.7$ is possible. Probabilities are
obtained from the mean values of $p_1$ and $p_2$ in
Figure~\ref{fig-2cat2}, together with additional runs in which
$w_1=0.9$ and $w_1=0.95$.}\label{fig-C2}

\end{figure}

\begin{figure}[b]

\begin{center}\vspace{5pt}

\begin{tabular}{|ccc|c@{~~}cc|c@{~~}cc|c@{~~}cc|}
\hline
\rule{0pt}{12pt}
 & & 
 & \multicolumn{3}{c|}{Know you're in $G_1$}
 & \multicolumn{3}{c|}{Know you're in $G_2$}
 & \multicolumn{3}{c|}{Know you're in $G_3$} \\
\rule{0pt}{10pt}
   $w_1$ & $w_2$ & $w_3$ &
 & \,Take 1$\!\!$ & $\!\!$Take 2 &
 & \,Take 1$\!\!$ & $\!\!$Take 2 &
 & \,Take 1$\!\!$ & $\!\!$Take 2 \\
\hline
\rule{0pt}{12pt}
1/3 & 1/3 & 1/3 &
& 0.700 & 0.700 &
& 0.700 & 0.700 &
& 0.700 & 0.700
\\
\rule{0pt}{12pt}
0.2 & 0.3 & 0.5 &
& 0.644 & 0.645 &
& 0.677 & 0.677 &
& 0.736 & 0.736
\\
\hline
\end{tabular}
\end{center}

\caption{Posterior subjective probabilities of the Predictor being correct for
Participants in the three-category examples. Obtained using the mean values
of $p_1$, $p_2$, and $p_3$ from Figure~\ref{figK3}, in the same way as for
Figure~\ref{fig-C2}.}\label{fig-C3}

\end{figure}

The effects of using a non-uniform prior distribution for the $q_i$
are shown in Figure~\ref{fig-C4}, when there are two categories, and
either $w_1=w_2=0.5$ or $w_1=0.8$ and $w_2=0.2$.

This table shows what a Participant should believe when using various
prior distributions in which the $q_i$ are independent (before
conditioning on accuracy $R$), with each having a distribution from
the Beta family. The Beta($a$,$b$) distribution for $q_i$ on $[0,1]$ has
density function proportional to $q_i^{a-1}(1\!-\!q_i)^{b-1}$, and has
mean $a/(a\!+\!b)$. The Beta(1,1) distribution is the uniform
distribution, which was assumed for the examples above. When $a$ and
$b$ are greater than 1, the distribution is more concentrated than
the uniform distribution. A value for $a$ less than one
produces a peak near zero; when $b$ is less than one, there is a peak
near one.  Beta($a$,$b$) distributions with $a=b$ are symmetrical
around $1/2$.

\begin{figure}[t]

\begin{center}\vspace{5pt}

\begin{tabular}{|l|cc|c@{~~}cc|c@{~~}cc|}
\hline
\rule{0pt}{12pt}
 &&  
 & \multicolumn{3}{c|}{\!\!Know you're in $G_1$\!\!\!}
 & \multicolumn{3}{c|}{\!\!Know you're in $G_2$\!\!\!} \\
\rule{0pt}{10pt}
 & $w_1$ & $w_2$ &
 & \,Take 1$$ & $$Take 2 &
 & \,Take 1$$ & $$Take 2
\\
\hline
\rule{0pt}{12pt} Same prior for $q_1$ and $q_2$: &
0.5 & 0.5 &
& 0.700 & 0.700 &
& 0.700 & 0.700
\\
\rule{0pt}{12pt} ~~~Beta(0.7,0.7) &
0.8 & 0.2 &
& 0.719 & 0.720 &
& 0.622 & 0.621
\\
\hline
\rule{0pt}{12pt} Same prior for $q_1$ and $q_2$: &
0.5 & 0.5 &
& 0.700 & 0.699 &
& 0.700 & 0.700
\\
\rule{0pt}{12pt} ~~~Beta(1.3,1.3) &
0.8 & 0.2 &
& 0.716 & 0.715 &
& 0.637 & 0.637
\\
\hline
\rule{0pt}{12pt} Same prior for $q_1$ and $q_2$: &
0.5 & 0.5 &
& 0.699 & 0.700 &
& 0.700 & 0.700
\\
\rule{0pt}{12pt} ~~~Beta(1.1,1.3) &
0.8 & 0.2 &
& 0.702 & 0.729 &
& 0.692 & 0.566
\\
\hline
\rule{0pt}{12pt} Same prior for $q_1$ and $q_2$: &
0.5 & 0.5 &
& 0.700 & 0.700 &
& 0.700 & 0.700
\\
\rule{0pt}{12pt} ~~~Beta(0.7,0.9) &
0.8 & 0.2 &
& 0.705 & 0.732 &
& 0.684 & 0.550
\\
\hline
\rule{0pt}{12pt} Different priors for $q_1$ and $q_2$: &
0.5 & 0.5 &
& 0.722 & 0.722 &
& 0.678 & 0.678
\\
\rule{0pt}{12pt} ~~~Beta(1,1) and Beta(1.3,1.3) &
0.8 & 0.2 &
& 0.723 & 0.723 &
& 0.610 & 0.610
\\
\hline
\rule{0pt}{12pt} Different priors for $q_1$ and $q_2$: &
0.5 & 0.5 &
& 0.750 & 0.750 &
& 0.650 & 0.650
\\
\rule{0pt}{12pt} ~~~Beta(0.7,0.7) and Beta(1.3,1.3) &
0.8 & 0.2 &
& 0.730 & 0.730 &
& 0.579 & 0.579
\\
\hline
\rule{0pt}{12pt} Different priors for $q_1$ and $q_2$: &
0.5 & 0.5 &
& 0.623 & 0.751 &
& 0.751 & 0.623
\\
\rule{0pt}{12pt} ~~~Beta(1.1,1.3) and Beta(1.3,1.1) &
0.8 & 0.2 &
& 0.686 & 0.729 &
& 0.731 & 0.507
\\
\hline
\rule{0pt}{12pt} Different priors for $q_1$ and $q_2$: &
0.5 & 0.5 &
& 0.619 & 0.748 &
& 0.748 & 0.620
\\
\rule{0pt}{12pt} ~~~Beta(0.7,0.9) and Beta(0.9,0.7) &
0.8 & 0.2 &
& 0.690 & 0.729 &
& 0.720 & 0.493
\\
\hline
\rule{0pt}{12pt} Different priors for $q_1$ and $q_2$: &
0.5 & 0.5 &
& 0.708 & 0.631 &
& 0.690 & 0.756
\\
\rule{0pt}{12pt} ~~~Beta(1,1.5) and Beta(0.7,1.3) &
0.8 & 0.2 &
& 0.684 & 0.738 &
& 0.750 & 0.515
\\
\hline
\rule{0pt}{12pt} Different priors for $q_1$ and $q_2$: &
0.5 & 0.5 &
& 0.655 & 0.804 &
& 0.722 & 0.538
\\
\rule{0pt}{12pt} ~~~Beta(0.5,1.3) and Beta(1,1.5) &
0.8 & 0.2 &
& 0.619 & 0.759 &
& 0.769 & 0.287
\\
\hline
\end{tabular}
\end{center}

\caption{Posterior subjective probabilities of the Predictor being correct in
two-category examples when Participants have non-uniform prior 
distributions for $q_1$ and $q_2$.  The initial priors are Beta distributions
with indicated parameters, which are then restricted to values
of $q_1$ and $q_2$ for which $R=0.7$ is possible. 
}\label{fig-C4}

\end{figure}

Conveniently, if you are in $G_s$ and your prior for $q_s$ is
Beta($a$,$b$), then your posterior distribution for $q_s$ after taking one
box is Beta($a\!+\!1$,\,$b$), and your posterior distribution after
taking two boxes is Beta($a$,\,$b\!+\!1$). These distributions ignore
that the Predictor has accuracy $R$. Accounting for this by
eliminating $q_i$ values for which an accuracy of $R$ is not possible
modifies the distribution of $q_s$ and introduces dependencies between
the $q_i$.\footnote{\fsp Beta distributions can be sampled efficiently,
after which values of the $q_i$ for which accuracy $R$ is not
possible can be discarded.  In this way, one can obtain samples from a Beta 
prior constrained by accuracy $R$, and from the resulting
posterior after taking one box or two, 
allowing reasonably fast computation
of the posterior probabilities shown for the examples here.\vspace{-6pt}}

One can see in Figure~\ref{fig-C4}, as well as in Figures~\ref{fig-C3}
and~\ref{fig-C2} above, that when the $w_i$ are all equal and the
prior distributions for the $q_i$ are all the same, the posterior
probability that the Predictor is correct, both after taking one box
and after taking two, exactly equals $R$ (within estimation error of
$\pm0.001$), regardless of which category the Participant knows they are in.

This makes intuitive sense. If the priors 
for the $q_i$ (before conditioning on $R$) are all the
same, and independent, and the $w_i$ are all the same, nothing
distinguishes one category from another.  (Recall that at this point
we are assuming there is no other relevant information.) When all
categories look the same, knowing which category you are in is useless
information.  So the situation is the same as if you did not know your
category, which as discussed above leads to the posterior probability
that the Predictor is correct being exactly equal to $R$, satisfying
condition~(k).

In fact a stronger result can be proved, in which the $q_i$ are not
necessarily independent in the prior before conditioning on $R$:\ \ If
all the $w_i$ are the same and the $q_i$ have a joint prior (before
conditioning on $R$) which is invariant to a circular shift of the
indexes, $i$, then the posterior probability that the Predictor is
correct will be exactly equal to $R$, whether you take one box or two
(assuming that you have no other information). This is proved as
Claim~A in the Appendix.

One can also see that when the priors for each of the $q_i$ are all
symmetrical (even if different), a Participant's posterior probability
that the Predictor was correct is the same after taking one box as
after taking two boxes, though it differs with the category the
Participant knows they are in, and may not equal $R$.  This is
intuitive since a symmetric prior treats one-boxing and two-boxing
the same. It is proved as Claim~B in the Appendix.

To summarize, when you know your category, and you know that the $w_i$ are
all equal, your posterior probability that the Predictor was correct
after choosing one box or two will equal $R$ as long as your priors
for the $q_i$ are independent and all the same, or more generally
satisfy a shift symmetry. When different $q_i$ have
different prior distributions, however, your probability that the
Predictor was correct can depart from $R$, sometimes by only a small
amount, but sometimes more substantially,
as illustrated by the last prior in Figure~\ref{fig-C4}.

\noindent \textbf{It's a Newcomb problem if you don't know the $w_i$,
and have a symmetric prior for them.}\ \ When categories are defined
by demographic attributes --- for example, sex and country of
residence --- it's plausible that many Participants will have a good
idea of the fraction, $w_i$, of the population in each category. But
they are unlikely to know the $w_i$ when categories are defined by
personal history and characteristics --- few will have any clear idea,
for example, of what fraction of people have high blood pressure, play
tennis, and read the Illiad in school.

In this situation, your background knowledge is
\beq
B & \!=\! & B_0 \ \&\ \mbox{The Predictor uses categories $G_1,\ldots,G_K$}
  \nonumber\\ & &
\ \ \ \ \, \&\ \mbox{You are in category $G_s$}
\eeq
Your belief regarding whether the Predictor will be
correct will depend on your prior distributions for the $w_i$ and for
for the $q_i$. If your prior distributions for both (before
knowing anything else) are
symmetrical with regard to the categories (shift symmetric), you 
know nothing that distinguishes one category from another, and hence
knowledge of your own category tells you nothing. So your beliefs
in this situation should be the same as when you do not know your category;
hence you should believe with probability
$R$ that the Predictor 
was correct. This is shown in detail as
Claim~C in the Appendix.\footnote{\fsp
Numerical demonstrations of how shift symmetric priors for $w$ and for $q$ 
lead to believing the Predictor is correct with probability $R$ can be derived
from results in Figures~\ref{fig-C2}, \ref{fig-C3}, and~\ref{fig-C4},
along with these results with categories permuted.
With known $w$ the probability of correctness
equals $R$ only when all $w_i=1/K$. For unequal $w_i$, suppose you know the set
of $w_i$ values, but not which value goes with which category. For example,
for the second row in Figure~\ref{fig-C2} your prior (before
knowing anything else) is that either
$w_1=0.6$ and $w_2=0.4$ or $w_1=0.4$ and $w_2=0.6$, both having probability
$1/2$. Note that this prior is shift symmetric. Knowing you are in category 
$G_s$ shifts your probabilities for these two possibilities by the
corresponding factors, $w_s$.
If your prior for $q$ is symmetric between
one-boxing and two-boxing, your decision to take one box
or two will not further shift these probabilities. So we can find the
probability the Predictor is correct as a weighted average of the probabilities
with and without swapping categories. For the second row of 
Figure~\ref{fig-C2}, we get that $\prs(C\giv A_1\and G_1\and B)\ =\
\prs(C\giv A_2\and G_1\and B)\ =\ 
\prs(C\giv A_1\and G_2\and B)\ =\ 
\prs(C\giv A_2\and G_2\and B)\ =\ 
0.6\times0.727\ +\ 0.4\times0.660\ =\ 0.7002$, which equals $R$ to
within round-off error. Similarly, for the second row of Figure~\ref{fig-C3}
we can compute the probability of correctness for any $A_i$ and $G_i$ as

\vspace{-7pt}

\parbox{7in}{
\beq
0.2\times0.644\ +\ 0.3\times0.677\ +\ 0.5\times0.736\ \ =\ \ 0.6999
\ \ \approx\ \ R\ \ =\ \ 0.7 \nonumber
\\[-20pt]\nonumber
\eeq
}
\label{foot-unk-w}}

In conclusion, when considering only your possible
knowledge of the categories used by the Predictor, their prevalences
(the $w_i$), the one-boxing probabilities for each category (the
$q_i$), and which category you yourself are in ($G_s$), there are many
scenarios in which you should rationally believe that the Predictor is
correct with the specified probability $R$, or close to it, even after
making a decision to one-box or two-box ($A_1$ or $A_2$). Cases where
this is not true include when you know that you are in a
low-prevalence category (see the sixth row in Figure~\ref{fig-C4},
with $w_2=0.2$, for example), or you know your category and have a
strong prior belief about what Participants in that category are
likely to do (for example, the last row in Figure~\ref{fig-C4}).

However, actual Participants will have much additional knowledge about
themselves, perhaps unknowable to the Predictor.  The next section
considers how this affects whether a situation is a valid Newcomb
problem.

\subsection{\hspace*{-8pt}What Participants should believe when they know
additional information}\vspace{-6pt}\label{sec-additional}

Condition (k) of Section~\ref{sec-uncont} is a crucial
characteristic of any Newcomb problem that is relevant to human
decision making. As a human, you know many, many things about yourself
that will not be known to any Predictor whose knowledge is less than
that needed for a detailed simulation of your thoughts (a situation
discussed in Section~\ref{sec-simul} below).  This additional information
could sometimes lead you to think that despite the Predictor having
accuracy $R$ overall (condition~(i) of Section~\ref{sec-uncont}),
they might be unable to predict well what a person with your
characteristics will do.  You then might not believe that the Predictor is
correct with probability $R$ regardless of your choice of one box or
two --- that is, you will not take your decision as establishing with
probability $R$ that the opaque box contains \$1M, if you take one box, or
is empty, if you take two. In this section, I will demonstrate that
condition~(k) can sometimes hold even when a Participant has such
additional information, as well as showing situations where
condition~(k) does not hold.

Actual Participants will have a vast amount of additional information
about their backgrounds, attitudes, and deliberations, which they will
certainly not be able to account for using formal inference.  I will,
unavoidably, look at a simplified situation in which the additional
information that you, a Participant, know is which of $L$ categories,
$H_1,\ldots,H_L$ you are in.  For example, with $L=4$, these
categories might be\vspace{-5pt}
\beq
  \ \ H_1 & = & \mbox{You have read Nozick's 1969 paper\ \&\ 
                  You have developed a headache while deliberating} 
  \nonumber\\[-1pt]
  \ \ H_2 & = & \mbox{You have read Nozick's 1969 paper\ \&\ 
                  You have not developed a headache while deliberating} 
  \nonumber\\[-1pt]
  \ \ H_3 & = & \mbox{You have not read Nozick's 1969 paper\ \&\ 
                  You have developed a headache while deliberating} 
  \nonumber\\[-1pt]
  \ \ H_4 & = & \mbox{You have not read Nozick's 1969 paper\ \&\ 
                  You have not developed a headache while deliberating} 
  \nonumber
\eeq

\vspace{-10pt}

\noindent These categories are separate from the $K$ categories,
$G_1,\ldots,G_K$, that the Predictor uses. You may also know of those
categories, and know which one you are in.  

Each combination of categories will have some prevalence in the population
from which Participants are drawn.  For simplicity, I will initially take 
membership in the Predictor's categories, $G_1,\ldots,G_K$ to be independent
of membership in $H_1,\ldots,H_L$, so the fraction of potential
Participants who are in both $G_i$ and $H_j$ is $w_iu_j$, where, as before $w_i$
is the prevalence of $G_i$, and $u_j$ is the prevalence of $H_j$. 
For the moment, suppose that you know the $w_i$ and $u_j$.

Your full background knowledge,
$B$, referenced in condition~(k), might then be\vspace{-6pt}
\beq
\ \ B 
 & \!=\! & B_0 \ \,\&\,\ \mbox{The Predictor uses categories $G_1,\ldots,G_K$}
    \ \,\&\,\ \mbox{You have considered categories $H_1,\ldots,H_L$}
  \nonumber\\[-1pt] & &
\ \ \ \ \, \,\&\,\ \mbox{The proportions of Participants in $G_1,\ldots,G_K$
                     are $w_1,\ldots,w_K$}
  \nonumber\\[-1pt] & &
\ \ \ \ \, \,\&\,\ \mbox{The proportions of Participants in $H_1,\ldots,H_L$
                     are $u_1,\ldots,u_L$}
  \nonumber\\[-1pt] & &
\ \ \ \ \, \,\&\,\ \mbox{You are in category $G_s$}
  \ \,\&\,\ \mbox{You are in category $H_t$}
\eeq

\vspace{-10pt}

\noindent Here, $B_0$ is knowledge about the setup, including the Predictor's
accuracy, $R$.

The background knowledge $B$ above does not include one-boxing
probabilities for Participants in different combinations of categories,
which Participants will not know in a genuine Newcomb problem, though
of course you will have some prior distribution for these unknown
quantities. Define
$r_{i,j}$ to be the probability that a Participant in $G_i$ and $H_j$ will
take one box. Then the one-boxing probabilities, $q_i$, for Participants
in $G_1,\ldots,G_K$, used by the Predictor will be\vspace{-10pt}
\beq
  q_i & = & \sum_{j=1}^L\, u_j r_{i,j}
\eeq
You know that these $q_i$ must be such that, together with the $w_i$,
they allow the Predictor to have accuracy $R$. This constrains the
values of the $r_{i,j}$.  For the example below, I assume that in
your prior distribution, before conditioning on $R$, the $r_{i,j}$
are independent and uniformly distributed. After accounting
for your knowledge of the Predictor's accuracy, the $r_{i,j}$ have
a more complex distribution, which will imply a distribution for the~$q_i$.

\begin{figure}

\vspace{-10pt}

\begin{center}
\includegraphics[scale=0.8]{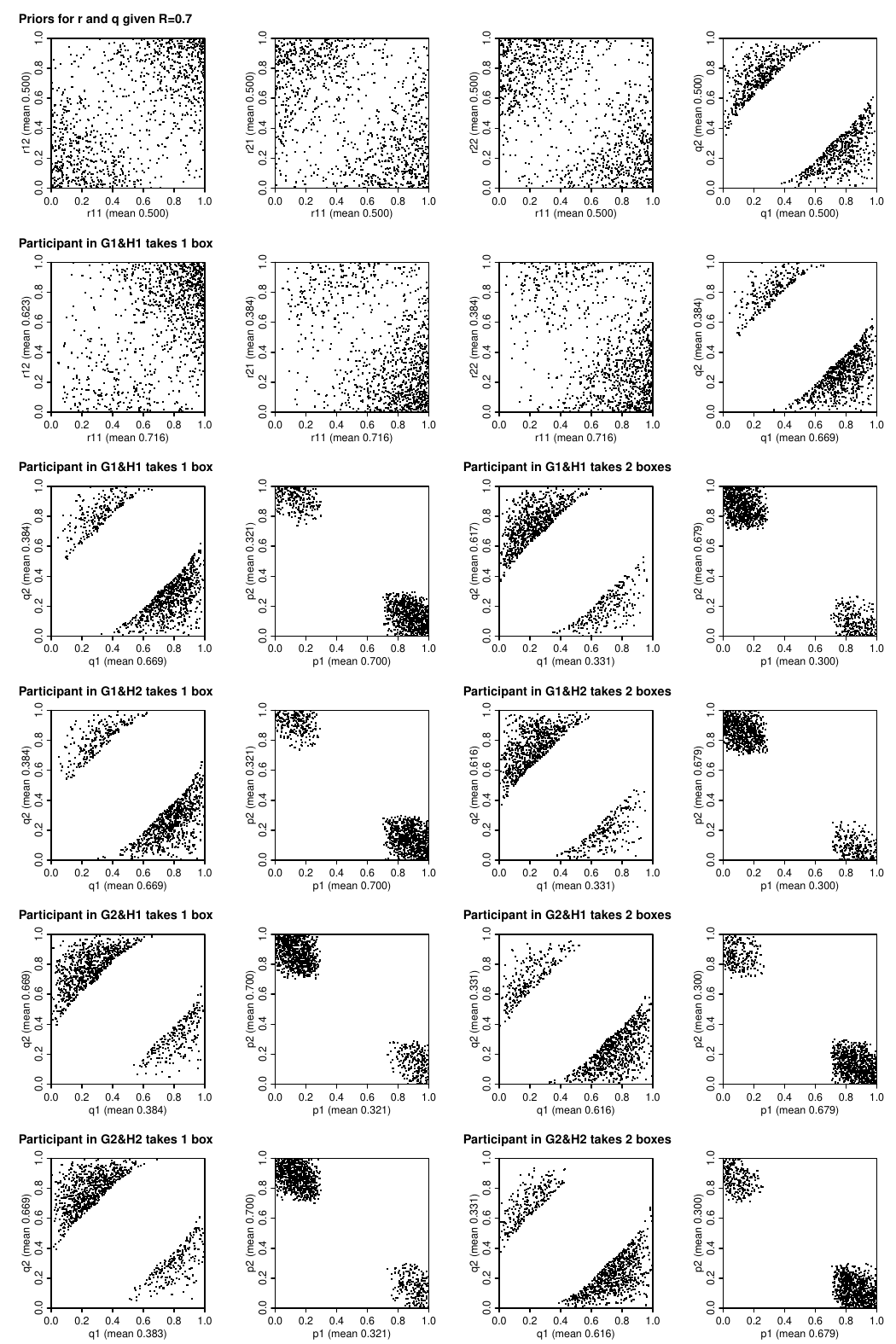}\vspace{4pt}

\end{center}

\caption{Scenario with $w_1=0.5$, $w_2=0.5$, $u_1=0.5$, $u_2=0.5$}\label{figX}

\end{figure}

\begin{figure}

\vspace{-10pt}

\begin{center}
\includegraphics[scale=0.8]{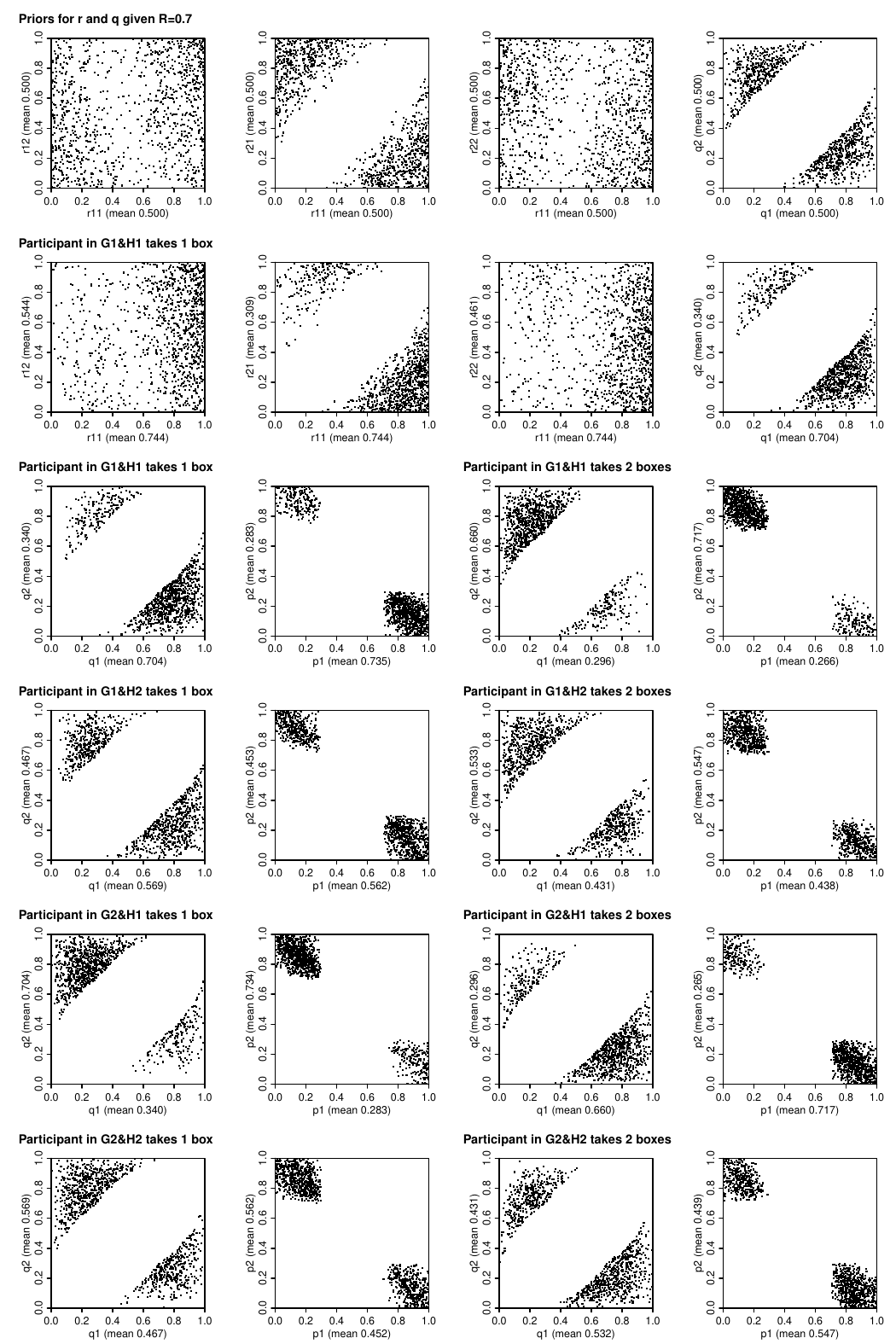}\vspace{4pt}

\end{center}

\caption{Scenario with $w_1=0.5$, $w_2=0.5$, $u_1=0.8$, $u_2=0.2$.}\label{fiXy}

\end{figure}

\begin{figure}

\vspace{-10pt}

\begin{center}
\includegraphics[scale=0.8]{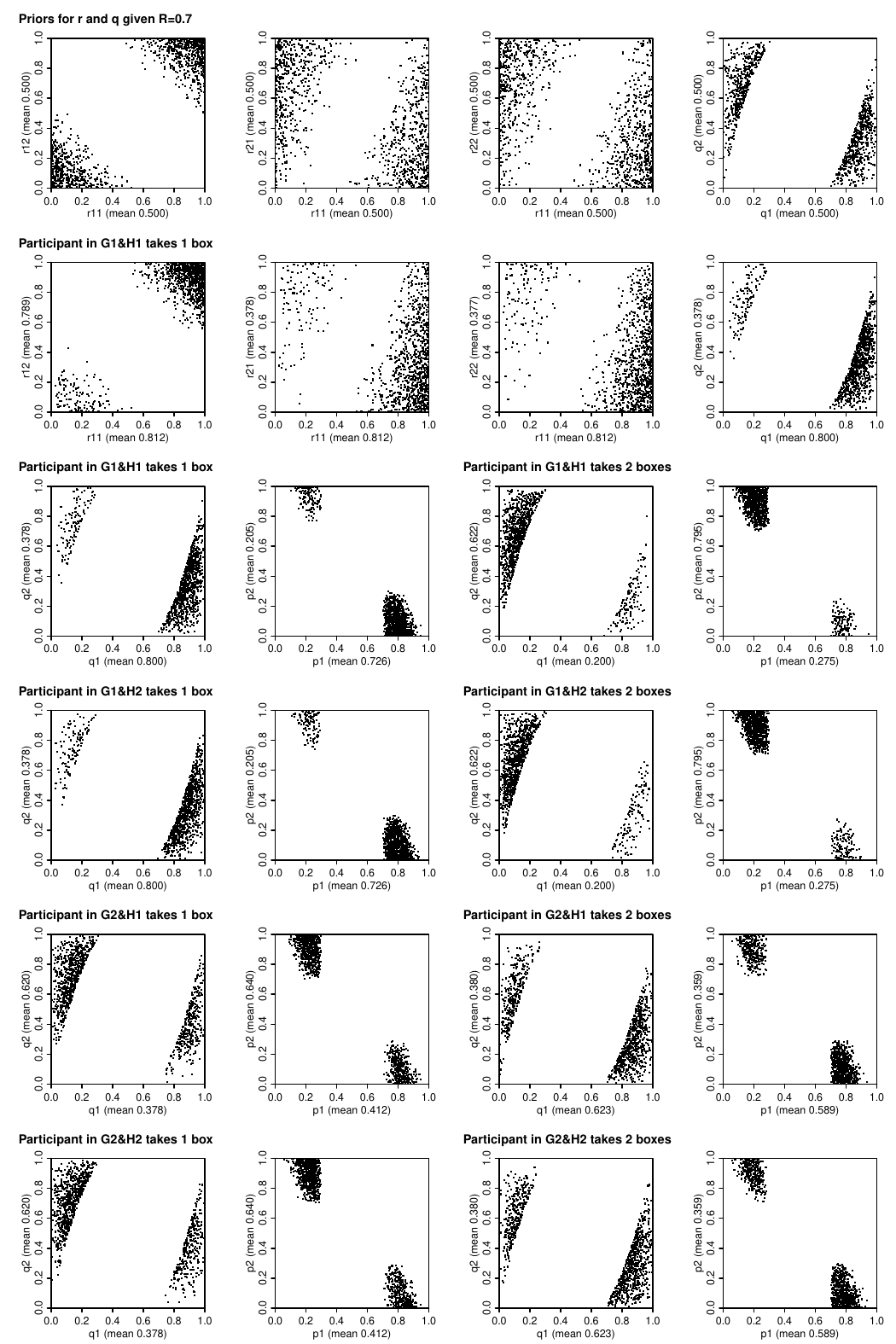}\vspace{4pt}

\end{center}

\caption{Scenario with $w_1=0.7$, $w_2=0.3$, $u_1=0.5$, $u_2=0.5$.}\label{figXz}

\end{figure}

Figures~\ref{figX} to~\ref{figXz} show prior and posterior
distributions for the $r_{ij}$, $q_i$, and $p_i$ when $K=2$ and $L=2$
with various values for the $w_i$ and $u_j$.  The resulting
probabilities for the Predictor being correct that you as a
Participant should hold after taking one box or two are shown in
Figure~\ref{figXtbl}, which are found from the mean values of $p_1$
and $p_2$ in the previous figures (along with results of a run with
$u_1=0.6$ that is not shown).\footnote{\fsp In the upper-right plots
in these figures, one can see that conditioning on $R=0.7$ changes the
uniform prior distribution of $r_{11}$ and $r_{12}$ to one that
favours $q_1=u_1r_{11}+u_2r_{12}$ being close to zero or close to one,
since this helps with prediction. The top-middle plots show that $r_{11}$
is negatively correlated with $r_{12}$ and $r_{22}$, reflecting that
differing values for $q_1$ and $q_2$ (which these contribute to) make
prediction easier. The lower plots show how distributions change
when a Participant in each combination of categories takes one or two
boxes.}

We can see that when $w_1=w_2=0.5$ and $u_1=u_2=0.5$, a Participant in any
combination of categories will believe with probability $R=0.7$ that
the Predictor was correct after taking one box, or after taking
two. That is, condition~(k) is satisfied.  However, when either $w_1
\ne w_2$ or $u_1 \ne u_2$, the posterior probabilities of the
Predictor being correct do not exactly equal $R$, with the difference
being greater when $u_1$ and $u_2$ differ more.  


\begin{figure}[t]

\begin{center}
\begin{tabular}{|cccc|cc|cc|cc|cc|}
\hline
\rule{0pt}{12pt}
 & & &
 & \multicolumn{2}{c|}{Know you're}
 & \multicolumn{2}{c|}{Know you're}
 & \multicolumn{2}{c|}{Know you're}
 & \multicolumn{2}{c|}{Know you're} \\
\rule{0pt}{12pt}
 & & &
 & \multicolumn{2}{c|}{~~in $G_1$ \& $H_1$}
 & \multicolumn{2}{c|}{~~in $G_1$ \& $H_2$}
 & \multicolumn{2}{c|}{~~in $G_2$ \& $H_1$}
 & \multicolumn{2}{c|}{~~in $G_2$ \& $H_2$} \\
\rule{0pt}{13pt}
   $w_1$ & $w_2$ & $u_1$ & $u_2$
 & Take 1 & Take 2
 & Take 1 & Take 2
 & Take 1 & Take 2
 & Take 1 & Take 2 \\
\hline
\rule{0pt}{12pt}
0.5 & 0.5 & 0.5 & 0.5 
& 0.700 & 0.700
& 0.700 & 0.700
& 0.700 & 0.700
& 0.700 & 0.700
\\
\rule{0pt}{12pt}
0.5 & 0.5 & 0.6 & 0.4 
& 0.725 & 0.725
& 0.662 & 0.662
& 0.725 & 0.725
& 0.662 & 0.662
\\
\rule{0pt}{12pt}
0.5 & 0.5 & 0.8 & 0.2 
& 0.735 & 0.734
& 0.562 & 0.562
& 0.734 & 0.735
& 0.562 & 0.561
\\
\rule{0pt}{12pt}
0.7 & 0.3 & 0.5 & 0.5 
& 0.726 & 0.725
& 0.726 & 0.725
& 0.640 & 0.641
& 0.640 & 0.641
\\
\hline
\end{tabular}
\end{center}

\caption{Participants' posterior subjective probabilities of the
Predictor being correct after taking one box or two, in scenarios
in which categories $G_1$ and $G_2$ are known
to the Predictor and Participant, categories $H_1$ and $H_2$ are known
to the Participant, and the Participant knows the $w_i$ and $u_j$.
A Participant's prior for the $r_{ij}$, before conditioning on
the Predictor's accuracy, $R=0.7$, is uniform over $[0,1]$.
Correctness probabilities are from the posterior distributions of $p_1$
and $p_2$ shown in Figures~\ref{figX} to~\ref{figXz}, plus an additional
run not shown.}\label{figXtbl}

\end{figure}

Claim D in the Appendix describes more general shift symmetry
conditions for prior distributions on the $r_{ij}$, the $w_i$, and
the $u_j$ that ensure your posterior probability for the Predictor
being correct equals $R$, regardless of which combination of
categories you know you are in, and after taking either one box or two
boxes (condition~(k)). As for earlier claims, an intuitive explanation
for this is that these symmetry conditions make knowledge of the
categories you are in of no value, since nothing distinguishes one
category from another, and we saw in Section~\ref{sec-bel} that
condition~(k) holds when you don't know your category.

I'll next consider scenarios, again with $K=2$ and $L=2$, in which the
$r_{ij}$ are not independent in the prior, and the $w_i$ and $u_j$ are
not known.  These will confirm that when the priors satisfy the symmetry
conditions of Claim~D, Participants will believe that the
Predictor is correct with probability $R$.\footnote{\fsp One can also verify 
that Claim
D is correct for a modification of the example in the second row of
Figure~\ref{figXtbl}, in which we assume the $u_j$ are not known, but have prior
distributions that give probability $1/2$ to $u_1=0.6$ (hence $u_2=0.4$)
and to $u_1=0.4$ (hence $u_2=0.6$). Since the uniform prior for the $r_{ij}$
does not favour one-boxing or two-boxing, the posterior distribution
for the two possible values of $u_1$ given that you know your
categories are $G_1$ and $H_1$
will give probability 0.6 to $u_1=0.6$ and probability 0.4 to
$u_1=0.4$.  The posterior probability of the Predictor being correct
given that you take one box will then be the weighted average of the posterior
probability from Figure~\ref{figXtbl} and the probability from that
figure with $H_1$ and $H_2$ swapped:
\beq
   0.6\times0.725\ +\ 0.4\times0.662\ \ =\ \ 0.6998
\eeq
which equals $R=0.7$ to within roundoff error. One can similarly confirm
that Claim D holds for the other possible categories and actions, and
for similarly-modified versions of the other examples in Figure~\ref{figXtbl}.}

\begin{figure}

\vspace{-10pt}

\begin{center}
\includegraphics[scale=0.8]{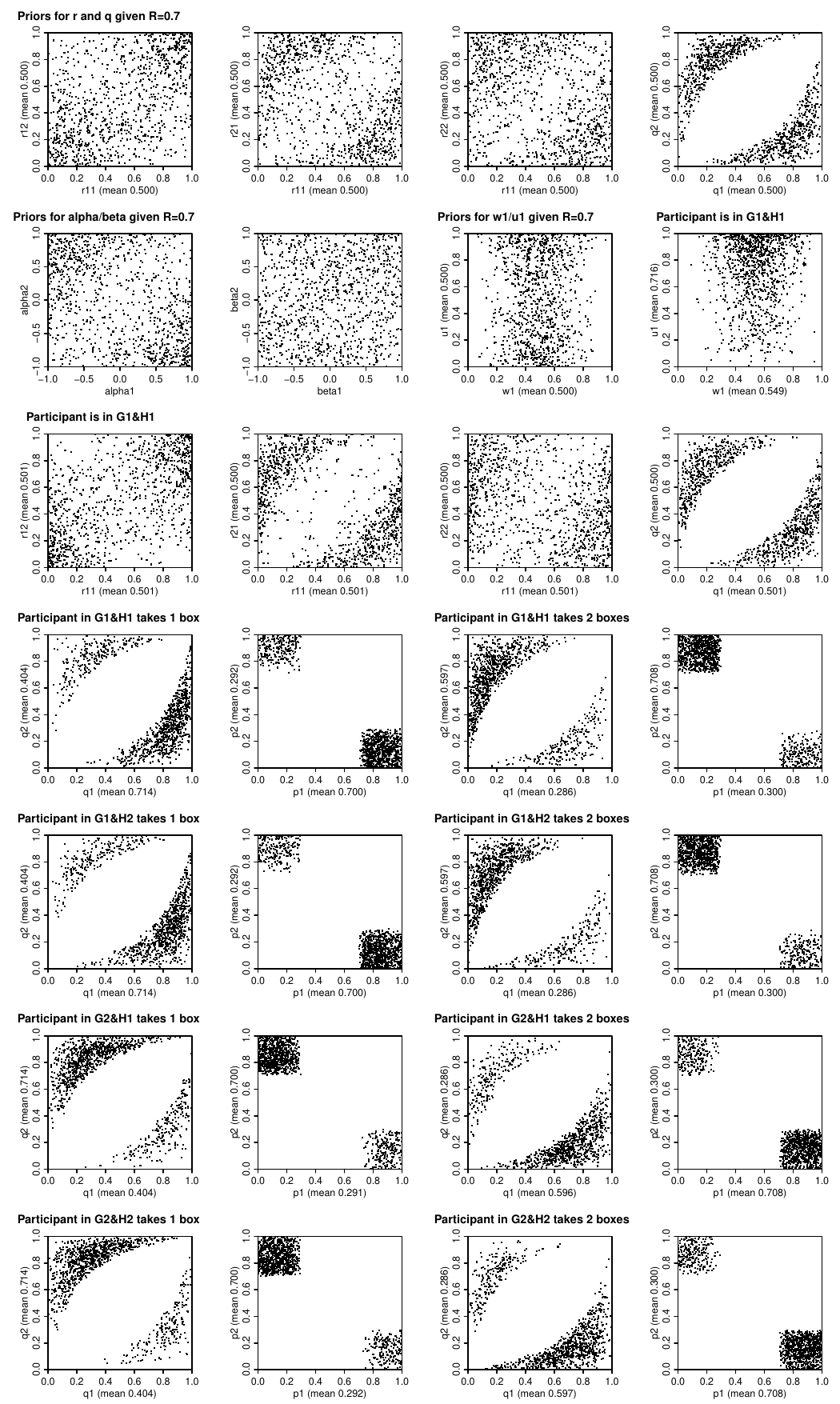}\vspace{4pt}

\end{center}

\vspace{-14pt}

\caption{Priors for $w_1$ and $u_1$ are uniform; prior for $r_{ij}$ is 
Beta$(2.5\!+\!\alpha_i\!+\!\beta_j,2.5\!-\!\alpha_i\!-\!\beta_j)$.}\label{figY}

\end{figure}

Figure~\ref{figY} shows prior and posterior distributions when the
priors for $w_1$ and $u_1$ are independent and uniform
over $[0,1]$, and the prior distribution for each $r_{ij}$ is
$\mbox{Beta}\,(2.5+\alpha_i+\beta_j,\,2.5-\alpha_i-\beta_j)$, where
the $\alpha_i$ and $\beta_j$ are regarded as unknown, with prior
distributions uniform on $[-1,+1]$.  These priors satisfy the
symmetry condition for Claim~D.  Due to use of common $\alpha_i$ and
$\beta_j$ values, there is prior dependence between $r_{i1}$ and
$r_{i2}$, and similarly between $r_{1j}$ and $r_{2j}$ --- a
typical situation in which the categories $G_1,\ldots,G_K$ and
$H_1,\ldots,H_L$ are both separately related to one-boxing
probabilities, rather than only their interaction mattering.

Looking at the mean values of $p_1$ and $1\!-\!p_1$ for Participants
in $G_1$ taking one or two boxes, and of $p_2$ and $1\!-\!p_2$ for
Participants in $G_2$, we can see that a Participant will believe
that the Predictor is accurate with probability $R=0.7$ in this
scenario.

\begin{figure}

\vspace{-10pt}

\begin{center}
\includegraphics[scale=0.8]{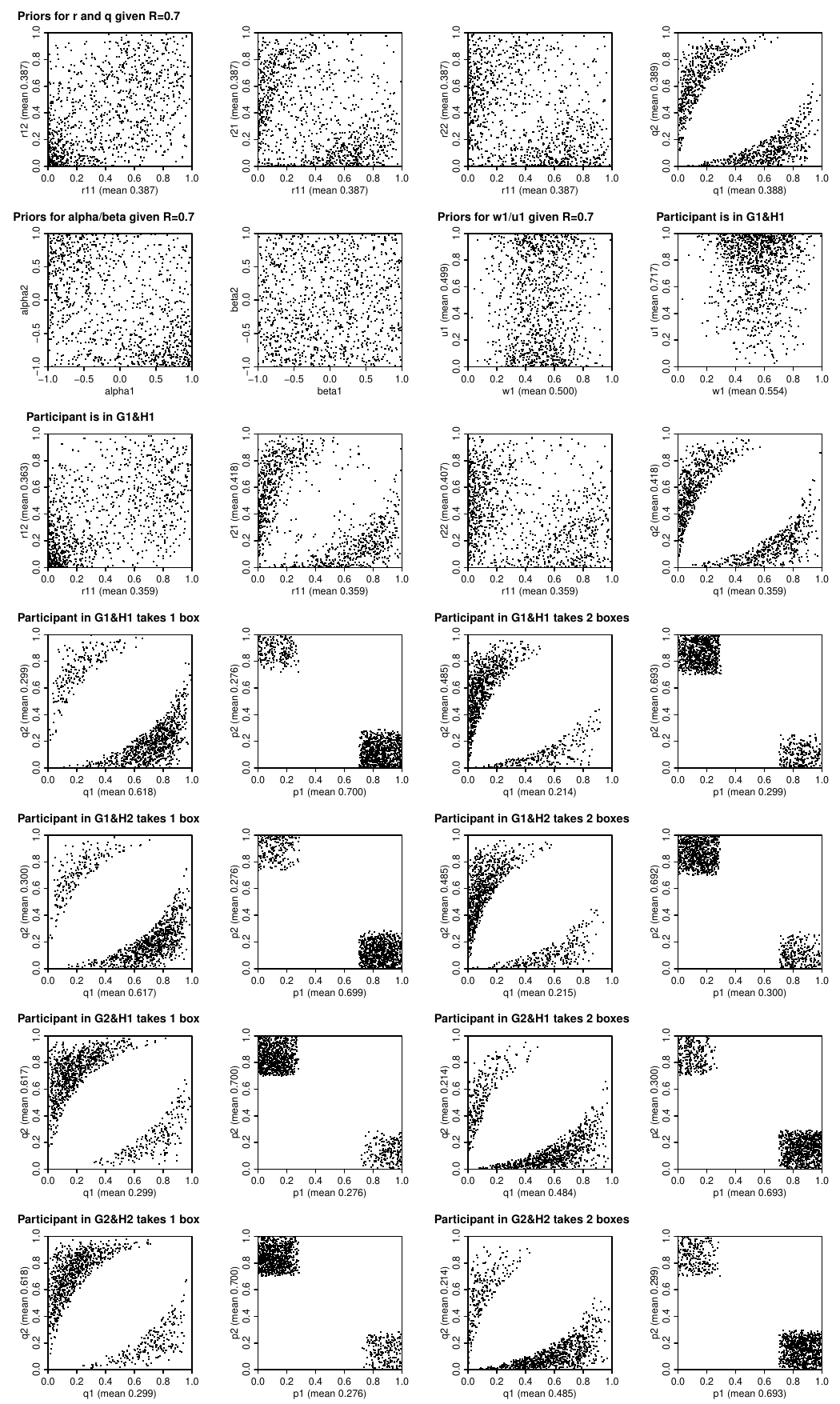}\vspace{4pt}

\end{center}

\vspace{-14pt}

\caption{Priors for $w_1$ and $u_1$ are uniform; prior for $r_{ij}$ is 
Beta$(2.2\!+\!\alpha_i\!+\!\beta_j,3.2\!-\!\alpha_i\!-\!\beta_j)$.}\label{figYx}

\end{figure}

In Figure~\ref{figYx}, this scenario is modified so the prior
distribution of each $r_{ij}$ is
$\mbox{Beta}\,(2.2+\alpha_i+\beta_j,\,3.2-\alpha_i-\beta_j)$, with the
$\alpha_i$ and $\beta_j$ again having uniform priors on $[-1,+1]$.  A
Participant with this prior believes that two-boxing is
generally more likely than one-boxing, though this needn't be true for
any particular combination of categories.  Since all $r_{ij}$ have
the same prior, given the common $\alpha_i$ and $\beta_j$, this prior
still satisfies the symmetry conditions of Claim~D. And as can be seen,
the Participant's posterior probability for the Predictor being correct
is indeed equal to $R=0.7$.

\begin{figure}

\vspace{-10pt}

\begin{center}
\includegraphics[scale=0.8]{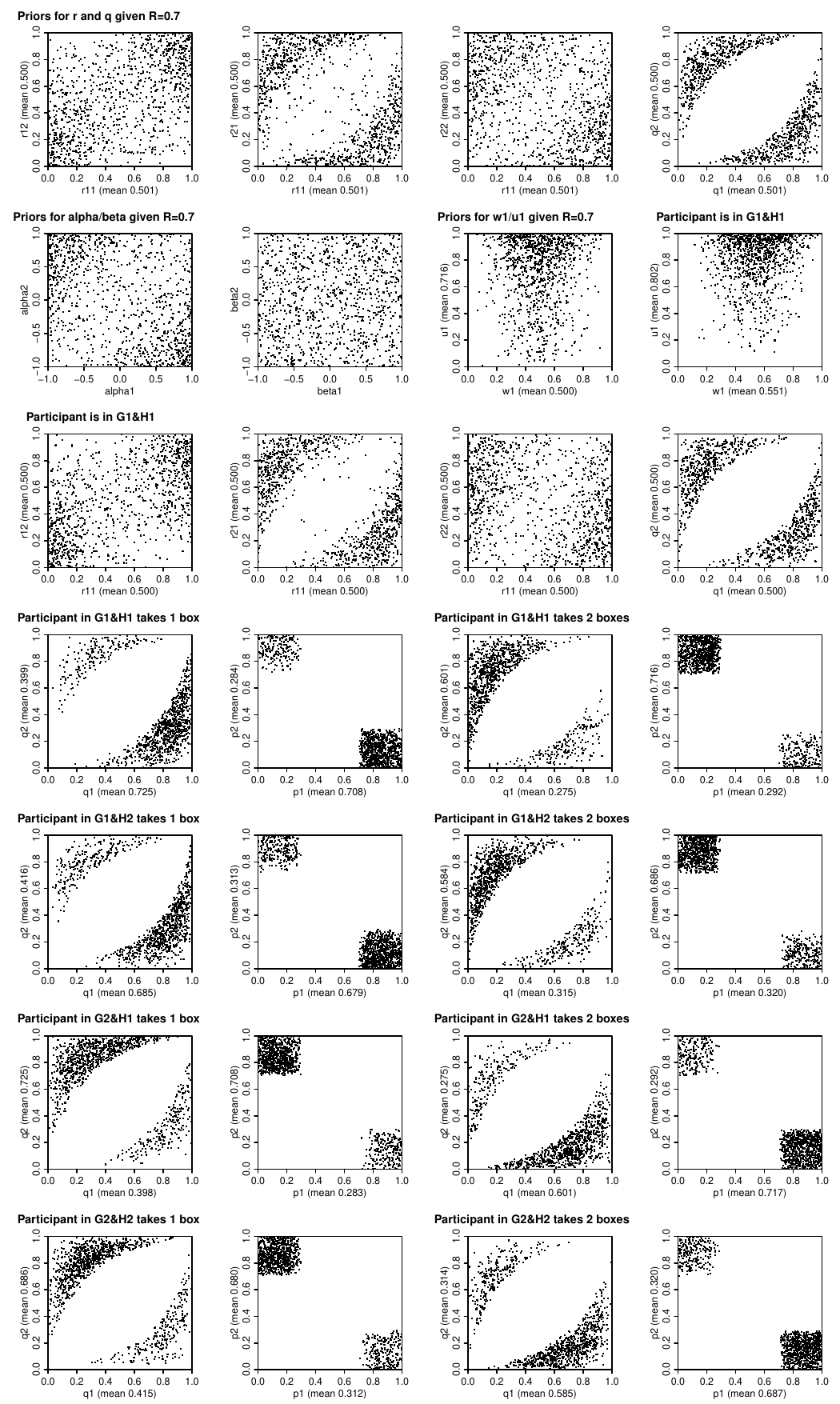}\vspace{4pt}

\end{center}

\vspace{-14pt}

\caption{Prior for $w_1$ is uniform; prior for $u_1$ is Beta(2,1); 
prior for $r_{ij}$ is
Beta$(2.5\!+\!\alpha_i\!+\!\beta_j,2.5\!-\!\alpha_i\!-\!\beta_j)$.}\label{figYy}

\end{figure}

\begin{figure}

\vspace{-10pt}

\begin{center}
\includegraphics[scale=0.8]{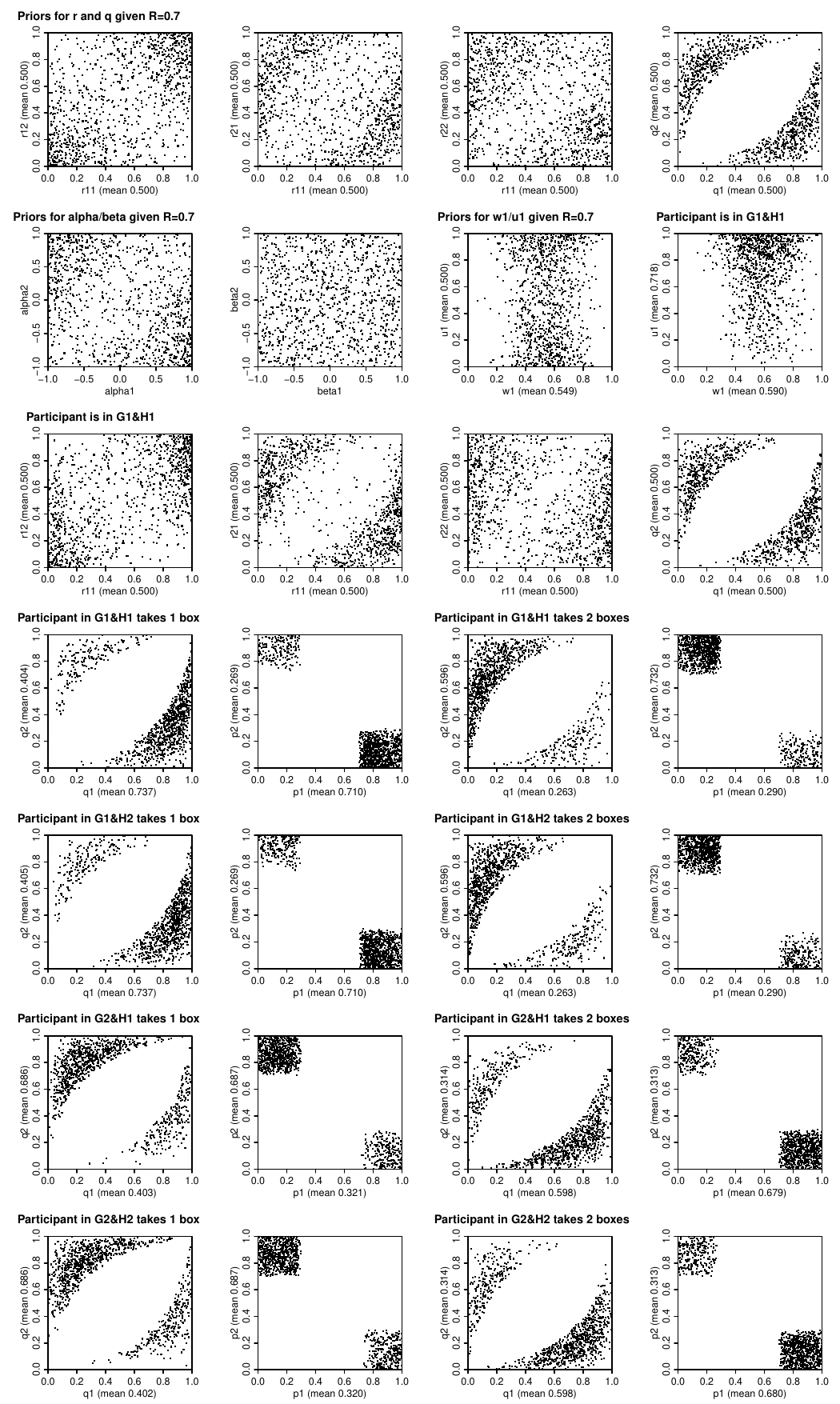}\vspace{4pt}

\end{center}

\vspace{-14pt}

\caption{Prior for $w_1$ is Beta(2,1); prior for $u_1$ is uniform;
prior for $r_{ij}$ is
Beta$(2.5\!+\!\alpha_i\!+\!\beta_j,2.5\!-\!\alpha_i\!-\!\beta_j)$.}\label{figYz}

\end{figure}

The results for these two scenarios are summarized in the top two rows
of Figure~\ref{figYtbl}, followed by results for scenarios in which
the conditions for Claim~D to apply do \textit{not} hold --- because
the prior lacks symmetry between $u_1$ and $u_2$ (Figure~\ref{figYy}),
or lacks symmetry between $w_1$ and $w_2$ (Figure~\ref{figYz}), or
both (in combination with two priors for $r_{ij}$, both not
shown). When Claim~D does not apply, we can see that the Participant
will not believe the Predictor is correct with probability
$R=0.7$. However, the departure from $R$ is fairly small, at least for
the fairly small departures from symmetry tested here.

\begin{figure}[t]

\begin{center}
\begin{tabular}{|cll|cc|cc|cc|cc|}
\hline
\rule{0pt}{12pt}
 & & 
 & \multicolumn{2}{c|}{Know you're}
 & \multicolumn{2}{c|}{Know you're}
 & \multicolumn{2}{c|}{Know you're}
 & \multicolumn{2}{c|}{Know you're} \\
\rule{0pt}{12pt}
 &  &  
 & \multicolumn{2}{c|}{~~in $G_1$ \& $H_1$}
 & \multicolumn{2}{c|}{~~in $G_1$ \& $H_2$}
 & \multicolumn{2}{c|}{~~in $G_2$ \& $H_1$}
 & \multicolumn{2}{c|}{~~in $G_2$ \& $H_2$} \\
\rule{0pt}{13pt}
   $\!r_{ij}$ prior & $\ w_i$ prior & $\ u_j$ prior
 & Take 1 & Take 2
 & Take 1 & Take 2
 & Take 1 & Take 2
 & Take 1 & Take 2 \\
\hline
\rule{0pt}{12pt}
2.5\ \ 2.5 & Uniform & Uniform
& 0.700 & 0.700
& 0.700 & 0.700
& 0.700 & 0.700
& 0.700 & 0.700
\\
\rule{0pt}{12pt}
2.2\ \ 3.2 & Uniform & Uniform
& 0.700 & 0.701
& 0.699 & 0.700
& 0.700 & 0.700
& 0.700 & 0.701
\\
\rule{0pt}{12pt}
2.5\ \ 2.5 & Uniform & Beta(2,1)
& 0.708 & 0.708
& 0.679 & 0.680
& 0.708 & 0.708
& 0.680 & 0.680
\\
\rule{0pt}{12pt}
2.5\ \ 2.5 & Beta(2,1) & Uniform
& 0.710 & 0.710
& 0.710 & 0.710
& 0.687 & 0.687
& 0.687 & 0.687
\\
\rule{0pt}{12pt}
2.5\ \ 2.5 & Beta(2,1) & Beta(2,1)
& 0.718 & 0.719
& 0.689 & 0.689
& 0.695 & 0.694
& 0.669 & 0.669
\\
\rule{0pt}{12pt}
2.2\ \ 3.2 & Beta(2,1) & Beta(2,1)
& 0.698 & 0.736
& 0.652 & 0.719
& 0.724 & 0.664
& 0.695 & 0.641
\\
\hline
\end{tabular}
\end{center}

\caption{Participants' posterior subjective probabilities of the Predictor 
being correct after taking one box or two, in scenarios
in which categories $G_1$ and $G_2$ are known
to the Predictor and Participant, categories $H_1$ and $H_2$ are known
to the Participant, and the Participant does not know 
the $w_i$ and $u_j$. A Participant's prior for the one-boxing 
probabilities, $r_{ij}$, is either
Beta$(2.5+\alpha_i+\beta_j,\,2.5-\alpha_i-\beta_j)$ or
Beta$(2.2+\alpha_i+\beta_j,\,3.2-\alpha_i-\beta_j)$, as indicated,
with the $\alpha_i$ and $\beta_j$ having uniform priors over $[-1,+1]$.
Correctness probabilities are from the posterior distributions of $p_1$
and $p_2$ shown in Figures~\ref{figY} to~\ref{figYz}, along with two additional
runs not shown.}\label{figYtbl}

\end{figure}

What can we conclude from these scenarios regarding whether
Newcomb's problem can be realized in the real world of today?

It is not just plausible, but pretty much certain, that using the
method of Section~\ref{sec-pr}, a Predictor who has ordinary
biographical information on Participants and crude measurements of
their brains of the sort that are currently feasible will be able to
achieve predictions satisfying condition~(i), with an overall
accuracy, $R$, substantially greater than the value of 0.5005 needed
for the recommendations of EDT and CDT to differ. All that is required
for this is that the frequency of one-boxing or two-boxing vary
among different categories of people to the degree that numerous other
behavioural traits clearly vary.

Ensuring that condition~(k) holds is more challenging.  We have seen
that ignorance by Participants can lead to them believing that the
Predictor has accuracy $R$.  This can take the form either of them not
knowing what category they are in, or not knowing what characterizes
each category.  A Predictor who is motivated to produce a valid
Newcomb problem that satisfies condition~(k) could arrange for
Participants to be ignorant --- either by not revealing the
categories used, or by choosing categories Participants have
scant knowledge~of.

The biggest threat to condition~(k) seems to be Participants' private
knowledge, including knowledge of the state of their deliberations.
As seen above, such private knowledge does not necessarily lead to a
violation of condition~(k). But since real Participants will have a
vast amount of private knowledge, it may seem hard to guarantee
this. You might, for example, have been exposed as a child to fringe
religious beliefs, which you think have led to your thought processes
differing in important ways from most people.  Or, during your
deliberations, you might happen to think of what seems to you to be a
particularly clever and unusual argument for taking two boxes, and on
that basis doubt that the Predictor, who can only go by what category
you are in, will have predicted you will take two boxes and hence put
nothing in the opaque box. However, the Predictor can mitigate the
risk that a Participant will think they are not typical of their
category by not revealing what categories they use, and/or by using
narrow categories, in which some ``unusual'' characteristics might
actually be common.

What is essential to my argument here is not that a valid Newcomb
problem with prediction done using categories can actually be realized
--- this is ultimately an empirical question, which is sensitive to
context, and is anyway unlikely to be
resolved due to practical issues --- but that it is not hard to
\textit{imagine} a Newcomb problem with no fantastic elements.
This is sufficient to motivate the question of what is the rational
choice to make when faced with such a problem.

\section{Why you should take both boxes when prediction is based on
categories}\vspace{-10pt}


I will argue here that when the Predictor operates by using statistics
on how often people in different categories take one box versus two,
it is rational for you as a Participant to take both boxes, as
recommended by CDT.

My argument must of course be addressed to those who presently would
take one box, perhaps because they think that is what EDT recommends.
Noting Nozick's (1969) admonition --- ``Nor will it do to just repeat
one of the arguments, loudly and slowly'' --- I will try to form
bridges from the ``smoking lesion'' problem, where the CDT result is
generally accepted as correct, to an impersonal version of Newcomb's
problem, and finally to Newcomb's problem itself, when predictions
are based on statistics for categories, aiming to show how the causal
intuition generally accepted for the smoking lesion problem
extends to Newcomb's problem as well.

In addition to the EDT argument for one-boxing, and the CDT argument
for two-boxing, I will consider two other well-known arguments.

The ``Why ain'cha rich?''\footnote{\fsp Note for non-native English
speakers: ``Why ain'cha rich?'' means ``Why aren't you rich?'' in a
non-standard English dialect.} argument (Gibbard and Harper 1976;
Ahmed 2014, Section~7.3.1) is addressed to a two-boxer who claim theirs
is the rational choice. To this, a one-boxer may reply ``If
you're so smart, why ain'cha rich?'', while gripping their \$1M that the
two-boxer didn't get.  Most two-boxers do indeed leave with only
\$1K, while most one-boxers leave with \$1M, and you have no reason to
think you are exceptional in this regard.  So if you decide to take
both boxes, it seems you're also deciding to be poor.

The ``You could have done better'' argument (Ahmed 2014,
Section~7.4.3) is addressed to one-boxers, and claims that they all
could have done better (by \$1K) if they had taken two
boxes. Furthermore, they all ought to have known this when they made
their decision. One can dramatize this by imagining a friend of yours
is on the other side of the boxes, who can see into the open back of
the opaque box (Nozick 1969, pp.~116-117; Ahmed 2014,
Section~7.4.2). Your friend is not allowed to speak to you, but you
know what advice they would give you if they could, regardless of what
they see --- take both boxes!  And even though the friend is imaginary,
shouldn't you nevertheless follow this advice?

I will consider how the analogues of each of these arguments fare for
the succession of problems that lead up to the standard Newcomb problem,

\subsection{\hspace*{-8pt}The smoking lesion 
                          problem}\label{sec-smoke}\vspace{-6pt}

Several Newcomb-like problems involving medical conditions have been
proposed, starting with Nozick's (1969, p.~125) example of someone who
may or may not have inherited a gene that causes both a horrible
disease in middle age and a preference for an academic career, and who
is deciding whether or not to go to graduate school. These problems
share a structure in which there is a common cause of both some
undesirable condition (in the future) and of a tendency to make some
decision, but in which the decision has no causal effect on whether
the undesirable condition develops.

The ``smoking lesion'' problem has become the most-discussed example
of this sort (perhaps because some have thought that some people
once believed its premise).  Ahmed (2014, p.~90) describes
it as follows:\vspace{-7pt}
\begin{quotation}\noindent
Susan is debating [ whether to smoke or not ]. She believes that
smoking is strongly correlated with lung cancer, but only because
there is a common cause --- a lesion that tends to cause both smoking
and cancer. Once we fix the [ absence or presence ] of this lesion,
there is no additional correlation between smoking and cancer. Susan
prefers smoking without cancer to not smoking without cancer, and she
prefers smoking with cancer to not smoking with cancer. Should Susan
smoke? It seems clear that she should.\vspace{-7pt} \end{quotation}
Note that although the correlation is strong, we needn't assume that
it is perfect --- there is some small change that a smoker will not
get cancer, or that a non-smoker will.

As far as I am aware, Ahmed's judgement that Susan should smoke, given
her beliefs, is universally accepted.  It is the conclusion that CDT
leads to, with ``smoking'' being a dominant action, better regardless
of whether Susan has the lesion or not.

The analogue of ``Why ain'cha rich?'' for the smoking lesion problem
is a non-smoker telling a smoker with cancer ``If you're so smart, why are
ya dying?'', to which the smoker may convincingly reply ``Because I
have the lesion, through no fault of my own''.  And the smoker can
rightly tell the non-smoker ``You could have done better'' since
the non-smoker has foregone a pleasure for no benefit. 
These judgements seem uncontroversial.

However, an outside observer who sees that Susan has decided to smoke
will rationally conclude that Susan is likely to get cancer.  This
presents a challenge for EDT, since if we assume that Susan greatly
prefers not smoking without cancer to smoking with cancer, naive
application of EDT will lead Susan to refrain from smoking, on the
basis that smoking would be bad news regarding her prospects of
avoiding cancer. 

A common response of EDT advocates to the smoking lesion problem is
the ``tickle defense'' --- that Susan will smoke only if she has a
desire (a ``tickle'') for cigarettes, and that when she recognizes
that she has this tickle, that in itself will convey to her the bad
news that she will likely get cancer. Whether she then actually smokes
or not will not change this news, so since she enjoys smoking, she
should do so.

This strikes me as being perhaps too clever. Is this really why nearly
everyone thinks you should smoke in this scenario?  Could it instead
be that people apply CDT in an intuitive way, and the tickle defense
is an after-the-fact attempt to remove causality from their reasoning?

Regardless, this more sophisticated application of EDT could be seen
as showing that EDT and CDT do not differ in their recommendations for
the smoking lesion problem, but as Ahmed (2014, p.~99) remarks, it
seems more useful to consider that if you can sense whether or not you
have a ``tickle'', you aren't in a real smoking lesion situation ---
one that would shed light on whether EDT or CDT (or some other
decision theory) is correct.

In general, as discussed by Ahmed (2014, Sections~4.2 and~4.3), Eells
(1981; 1982, Chapter~6), and Berm\'udez (2018), EDT will recommend the
intuitively correct action in a problem with the structure of
smoking lesion as long as the common cause of your choice and of
the undesirable result operates on your choice by influencing your
beliefs or preferences, and you are (sufficiently) aware of these
beliefs and preferences. Your actual choice then tells you nothing
more about the common cause, and hence nothing more about whether the
undesirable result will happen.

If this were the whole story, the smoking lesion problem could be
dismissed as having no relevance to real Newcomb problems. However,
there is a path other than through beliefs and preferences by which
the lesion could influence whether you smoke --- via what decision
theory you employ. Nozick (1969, pp.~127--130) discusses a scenario of
this nature, and appears to consider it as plausible as any of this
sort. Eells and Ahmed, however, both mention this possibility, but
then dismiss it as impossible.  Here is Ahmed (2014,
p.~93):\vspace{-8pt}
\begin{quotation}
Note finally that if the lesion does cause smoking [~via effects on
the smoker's desires and beliefs~] then it does not cause smoking only
amongst people who \textit{know} themselves to be facing
\textit{Smoking Lesion}. People that have the lesion will, because of
its presence, smoke more than those that don't have it, whether or not
they have any idea that this is what causes them to smoke and
whether or not they have any idea of what else it causes. \ldots

This is what makes it absurd to think that the lesion causes smoking
in the only other way that it could cause smoking: that is, by 
promoting adherence to a decision rule. \ldots if the case
is at all realistic, the lesion must promote smoking in cases where
CDT and EDT are not in dispute: thus as I said, the lesion equally
promotes smoking amongst people who have no idea that smoking is evidence
of its presence. It could only do this if it promotes smoking through
some route that does not run via the agent's decision rule. \ldots\vspace{-8pt}
\end{quotation}
This reasoning seems correct for the version of the smoking lesion
problem that Ahmed is envisioning, in which some people are unaware
of the correlation between smoking and cancer, or are unaware that
this correlation is non-causal. But one can equally well imagine that 
everyone knows these things, so everyone's initial beliefs are the same.
One can also easily imagine that everyone has a desire to smoke. 
The observed correlation could then only be due to people having different
decision rules.\footnote{\fsp Oddly, Ahmed himself has argued that the
tickle defense can be undermined by a common cause affecting how
one deliberates (e.g., what decision rule one uses) in the context
of Prisoner's Dilemma problems, as I discuss in Section~\ref{sec-pd}.}

Here's a story one could tell. At some time in the past, everyone
mistakenly believed that there was a correlation between smoking and
cancer, and that this (actually non-existent) correlation was not
causal.  Everyone also desired to smoke, though much less strongly
than they desired to avoid cancer.  These beliefs and desires led some
people to smoke, and others to refrain from smoking.  Later research
firmly established a correlation (at that time, and up to the
present) between smoking and cancer, and found that this correlation
is not causal (a trial with enforced smoking behaviour that was
randomly assigned showed no difference in cancer rates). Interviews of
smokers and non-smokers find no difference in their beliefs and
desires regarding smoking. These facts are now well known to everyone.

The only possible explanation for the presently-observed correlation
in this story seems to be that some common cause (such as presence or
absence of a ``lesion'') affects both whether someone (consciously or
unconsciously) uses a decision rule that leads them to smoke, and
whether or not they develop cancer.  Those with the lesion tend to use
a rule (such as CDT) that recommends smoking, and they also tend to
develop cancer. While this story is of course rather implausible, it
seems not much more implausible than the usual smoking lesion story.
I'll call a story like this a ``smoking lesion affecting decision rule''
problem, in contrast to the ``smoking lesion affecting desire'' version
that Ahmed has in mind.

Eells (1982, Chapter 6) has a different objection to the idea that
the common cause in problems like smoking lesion could produce the 
observed correlation
via an effect on the decision rule used. He says (pp.~146--147):\vspace{-8pt}
\begin{quotation}\noindent
\ldots Given that an agent is rational and given that he has
some particular pair of probability and desirability assignments,
whether a common cause is present or absent is irrelevant to which act
is (rationally) determined by him to be the correct one. \ldots
the assumption that an agent is rational is enough to ensure that the
presence or absence of a common cause will not affect which decision rule
is used. \ldots

So I think it is reasonable to assume that the agent, knowing 
some decision theory, believes that the way in which the common
cause causes rational people to decide on the symptomatic act is by
affecting their probabilities and desirabilities in such a way that their
decision rules (which the common cause plays no role in their choosing)
yield the recommendation to perform the symptomatic act. He
believes that the common cause can affect a rational person's probabilities
and desirabilities but not the choice of decision rule.\vspace{-8pt}
\end{quotation}

This reasoning depends crucially on the assumption that all people
involved behave rationally. In particular, they must all have
resolved the issue of whether EDT, CDT, or some other
decision theory is appropriate for problems of this sort. It's not
enough for them to have a generally rational disposition, and 
to be \textit{trying} to behave rationally.
It must be that none of them have made a mistake. But one thing the debate on 
Newcomb's problem has made quite clear, regardless of one's view of it,
is that many people 
make mistakes regarding decision theory, and persist in these mistakes
for decades, despite careful thought.

In any case, if assuming perfect rationality eliminates any difference
in what EDT and CDT recommend, we should simply consider that version
of the smoking lesion problem to be uninteresting.  Our aim should be
so see whether there is some version that is at least somewhat
plausible and for which the recommendations of EDT and CDT differ.

In the smoking lesion affecting decision rule scenario I presented above, we
may assume that all people involved \textit{believe} that they are
rational, and in particular, you believe this of yourself when
deliberating.  You also know that the presence of the cancer-causing
lesion makes it more likely that someone will use a decision rule
(perhaps CDT) that recommends they smoke (we suppose that everyone has
a desire to smoke), and \textit{also} leads them to think that their
decision rule is rational. This is quite compatible with the absence
of the lesion inclining people to use a decision rule that recommends
abstaining, while they too think that they are behaving rationally.

In this situation, deciding to smoke should in fact lead you to think
that you will develop cancer --- there is no informative ``tickle''
preceding your actual decision --- and hence it seems that EDT will
recommend that you not smoke, whereas CDT will recommend that you do.
Should the widely-accepted common sense judgement that you should
smoke in the usual smoking lesion problem also apply to this smoking
lesion affecting decision rule scenario?

It may be useful to consider a similar but more ordinary situation.
You notice that, in your view, some people behave rationally, others
irrationally, as evidenced by decisions they make that are, in your
judgement, good or bad (in relation to their expressed or evident beliefs
and desires). You also notice that the rational people tend to get
cancer at a higher rate than the irrational people.  Though surprising,
there is nothing all that strange or inexplicable about this.  The
rational people with cancer and the healthy irrational people might 
differ in some gene that affects both thought processes and cancer risk,
or perhaps incipient cancer affects the brain in a
way that (in your view) promotes rationality. In this situation,
would you believe that you can avoid cancer by acting in a way that
you regard as irrational? One may wonder if that is even coherent or
possible (Silverberg 1965).

In a smoking lesion affecting decision rule situation, your initial 
impulse will presumably be to smoke, since you enjoy it. If this is
also your final decision, you will believe that you are likely to
develop cancer, while the irrational (in your view) non-smokers will
not. But as in the situation described above, and as in the usual smoking
lesion problem where the lesion affects desire, I think this is not a
good reason to change your decision, due to your knowledge of the causal
structure of the problem.

If you do decide to smoke, and later get cancer, can you respond to
the ``If you're so smart, why are ya dying?'' objection in the same
way as before --- ``Because I have the lesion, through no fault of my
own''?  This seems just as reasonable to me as it was when the lesion
affected your desire, rather than your decision rule. However, it's
less clear that you can say that a non-smoker ``could have done
better''.  Non-smokers without the lesion lack the (in your mind)
clear-headed ability to reason about choices that is possessed by
those with the lesion, so they ``could'' have done better only through
an unusual effort at rationality (which by hypothesis is rare).

These are intuitive judgements. Perhaps an advocate of EDT would
instead think that you should not smoke.  But how would they defuse
the intuition that one should smoke in the smoking lesion affecting
desire problem, in which I suppose they \textit{would} smoke?
Alternatively, there might be arguments for why EDT does not actually
recommend abstaining in the smoking lesion affecting decision rule
problem, but I will not speculate about what such arguments there
might be.

The smoking lesion affecting decision rule scenario brings us closer to
Newcomb's problem, in which beliefs and preferences are fixed by the
problem specification (see conditions~(j) and~(l) of
Section~\ref{sec-cond}). But Newcomb's problem involves a prediction
of what \textit{you} will do, an aspect that perhaps invites
confusion.  Before addressing that, I will look at a simpler
Newcomb problem without such individualized prediction.

\subsection{\hspace*{-8pt}An impersonal Newcomb 
problem}\label{sec-impn}\vspace{-6pt}

Consider an impersonal Newcomb scenario in which the Predictor makes
the same prediction for all Participants, based simply on the
proportion, $q$, of potential Participants who would take one box
rather than two.  The Predictor has a good estimate of $q$, obtained
as described below.  Participants are given a value, $R$, between
$0.5$ and $1$, and told that $q$ is either $R$ or $1\!-\!R$, a claim
that they believe for what they think are good reasons.  (Perhaps a
trusted friend has examined evidence regarding $q$, and has assured
them of this.) The Predictor always puts \$1M in the opaque box if
$q>0.5$, and always leaves it empty if $q < 0.5$.  (I assume 
$q$ does not equal $0.5$ exactly, which could happen only by a
highly unlikely coincidence.)

The Predictor can find a suitable $q$ for the impersonal Newcomb
problem iteratively, starting with some initial guess for
$q$, with the corresponding $R$ equal to the maximum of $q$ and
$1\!-\!q$. The Predictor then samples potential Participants,
persuades them that they are in an impersonal Newcomb situation with
this $R$, and observes their decisions. If the observed proportion of
one-boxers is within some small tolerance of $q$, this $q$ and $R$ can
be used for the impersonal Newcomb problem. Otherwise, some different
value for $q$ can be tried, repeating until a suitable $q$ is 
found.\footnote{\fsp This procedure will find some suitable $q$ not
equal to $0.5$ (with the corresponding $R$ greater than $0.5$) as long 
as the proportion of one-boxers ($q$) varies continuously with the 
value of $R$ that Participants are told holds, and the proportion $q$ when 
they are told that $R=0.5$ is not equal to $0.5$ (it should in
reality be close to zero). Note that there could be more than one value of $q$
which is consistent with the behaviour of
Participants when told the corresponding value of $R$.
The continuity assumption could be false
if Participants attach psychological significance to values for $R$
that are round numbers, but one can easily imagine that they don't.
It is logically possible that the $R$ found will not be greater
than the threshold of $0.5005$ needed for EDT and CDT to differ in their
recommendations, but that seems quite unlikely in practice.\vspace{-10pt}}

Similarly to the standard Newcomb problem (as discussed in
Section~\ref{sec-pr}), the Predictor will need to deceive the
potential Participants who are sampled (or their friends), presenting
fraudulent evidence that the true proportion of one-boxers is either
$R$ or $1\!-\!R$, when initially this is not actually known. But
once a suitable $q$ is found, the Participants' belief that $q$
is either $R$ or $1\!-\!R$ will be true.

Since the Predictor will have 100\% accuracy for one of the two
actions, and 0\% accuracy for the other, this impersonal Newcomb
scenario does not satisfy condition~(i) required for the standard Newcomb
problem. However, condition~(k), regarding subjective probabilities of
the Predictor being correct, may nevertheless be satisfied, and it is
this condition that directly affects reasoning about whether to take
one box or two.

To see how condition~(k) may hold, suppose that you are a
Participant, and that you know nothing other than the description of the 
setup above and the value of $R$ (so the extra information $B$ is just $B_0$).
Then condition~(k) is that
\beq
  \prs(C\giv A_1\and B)\ \,=\,\ \prs(S_1\giv A_1\and B)\ \,=\,\ R
  \ \ \ \ \ \mbox{and}\ \ \ \ \
  \prs(C\giv A_2\and B)\ \,=\,\ \prs(S_2\giv A_2\and B)\ \,=\,\ R
\eeq
Given how the Predictor operates, the event $S_1$ is the same as
$q>0.5$, which in turn is the same as $q=R$. 
Using this and the definition of conditional probability, you
should reason that
\beq
  \prs(C\giv A_1\and B)
   &\!\!=\!\!& \prs(q=R\giv A_1\and B)
   \\[9pt]
   &\!\!=\!\!& {\prs(q=R\and A_1\giv B) \over \rule{0pt}{11pt} \prs(A_1\giv B)}
   \\[9pt]
   &\!\!=\!\!& {\prs(A_1\giv q=R\and B)\,\prs(q=R\giv B) \over 
              \rule{0pt}{11pt} \prs(A_1\giv B)}
   \\[9pt]
   &\!\!=\!\!& {\prs(A_1\giv q=R\and B)\,\prs(q=R\giv B) \over \rule{0pt}{11pt}
   \prs(A_1\giv q=R\and B)\,\prs(q=R\giv B)
     \ +\ \prs(A_1\giv q=1\!-\!R\and B)\,\prs(q=1\!-\!R\giv B)\,}\ \ \ \
\label{eq-simp1}
\eeq
\rule{0pt}{12pt}When $B$ has no additional information, it is 
reasonable for your prior distribution on $q$ to be
$\prs(q=R\giv B)\,=\,\prs(q=1-R\giv B)\,=\,0.5$. If you 
also see yourself as typical (as you will if $B=B_0$), you should have a 
subjective probability that you will take one box that matches the proportion
of one-boxers in the population, $q$, so that
$\prs(A_1\giv q=x\and B)=x$ and $\prs(A_2\giv q=x\and B)=1\!-\!x$. 
Equation~(\ref{eq-simp1}) then simplifies to
\beq
  \prs(C\giv A_1\and B)
   & = & {0.5\,R \over \rule{0pt}{11pt} 0.5\,R\ +\ 0.5\,(1\!-\!R)}
   \ \,=\,\ R
\eeq
Similarly, the event $S_2$ is the same as $q<0.5$, which is the same as 
$q=1\!-\!R$, so
\beq
  \prs(C\giv A_2\and B)
   &\!\!=\!\!& \prs(q=1\!-\!R\giv A_2\and B)
   \\[9pt]
   &\!\!=\!\!& {\prs(q=1\!-\!R\and A_2\giv B) \over 
                \rule{0pt}{11pt} \prs(A_2\giv B)}
   \\[9pt]
   &\!\!=\!\!& {\prs(A_2\giv q=1\!-\!R\and B)\,\prs(q=1\!-\!R\giv B) 
             \over \rule{0pt}{11pt} \prs(A_2\giv B)}\ \ \ \ \
   \\[9pt]
   &\!\!=\!\!& {\prs(A_2\giv q=1\!-\!R\and B)\,\prs(q=1\!-\!R\giv B) 
    \over \rule{0pt}{11pt}
     \prs(A_2\giv q=1\!-\!R\and B)\,\prs(q=1\!-\!R\giv B)\ +\ 
    \prs(A_2\giv q=R\and B)\,\prs(q=R\giv B)\,}\ \ \ \
   \\[9pt]
   &\!=\!& {0.5\,(1-(1\!-\!R)) \over \rule{0pt}{11pt} 
          0.5\,(1-(1\!-\!R))\ +\ 0.5\,(1\!-\!R)}
   \ \,=\,\ R
\eeq
So, with no additional knowledge beyond $B_0$, 
condition~(k) is satisfied for the impersonal Newcomb problem.

Ignorance of whether $q$ is $R$ or $1\!-\!R$ is plausible, and could
be made more likely by choosing a suitable population of potential
Participants, about whom Participants know little.\footnote{\fsp One
should not, for example, use as the population of potential
Participants a class of students who all know each other. Using
the population of all philosophy students would also not be ideal.
Participants are more likely to have little knowledge of $q$ if
they are from a population such as ``Canadians who have purchased
a can of tomato soup in the last week''.} Whether
Participants should see themselves as typical involves the same issues
as discussed for the standard Newcomb problem in
Section~\ref{sec-additional}, but unlike for the standard Newcomb problem,
Participants cannot be ignorant of the category they are in --- there
is only one category, of all Participants in the population that is
sampled.  This might make a realization satisfying condition~(k) difficult.

However, the Predictor could try to reduce the impact of the
additional information that a Participant might have by a suitable choice
of the population from which Participants are sampled.  The Predictor
might, for example, exclude unusual people who might not see
themselves as typical.  They can also refrain from telling Participants exactly
what population was sampled.  So I think it is not too hard to imagine a
concrete realization of the impersonal Newcomb scenario satisfying 
condition~(k), even if in practice this condition might often be violated.

If you are a Participant in an impersonal Newcomb scenario, CDT
recommends that you take both boxes, for the same reason as in the
usual Newcomb scenario --- your decision has no causal effect on
whether or not the opaque box contains \$1M, so taking both gets you
\$1K more than taking only one.  In contrast, when condition~(k) is
satisfied, EDT would seem to recommend taking one box, again for the
usual Newcomb reason that if you reach for just one box, you will
believe (with probability $R$) that you will obtain \$1M, whereas if
you reach for both boxes, you will believe that you will obtain
only \$1K.

The ``Why ain'cha rich'' argument against two-boxing does not work in
the impersonal Newcomb problem, since if anyone gets \$1M from the
opaque box, everyone does. And whether or not Participants get \$1M
from the opaque box, the two-boxers leave with \$1K more than the
one-boxers, who clearly ``could have done better''. So the two-boxers
are always richer than the one-boxers, even though two-boxers
\textit{anticipate} being poor, and one-boxers \textit{anticipate}
being rich. The apparent contradiction comes from one-boxers being
rich when they are the majority of potential Participants, two-boxers being
poor when they are the majority, and both groups taking their own
choice as indicative of who forms the majority.

\subsection{\hspace*{-8pt}What to do in Newcomb's problem when
 prediction is based on categories}\vspace{-6pt}

I now turn to Newcomb's problem in its standard form, with prediction
done using categories, hoping that my examination of the smoking
lesion and impersonal Newcomb problems above will help guide intuitions
in this apparently more puzzling scenario.

I assume here that the Predictor operates as described in
Section~\ref{sec-pr}, and that all the conditions of Newcomb's
problem in Section~\ref{sec-uncont} are satisfied, particularly
including condition~(k), as discussed in Section~\ref{sec-bel}.

I emphasize that in the scenario I discuss here, prediction of a
Participant's choice does not involve demons, or mysterious alien
beings, or an Omega with apparently godlike powers, or even devices
employing future technology that we might think are possible. The
Predictor operates as described in Section~\ref{sec-pr}, using
categories defined via ordinary observations. Accordingly, there is
no excuse for failure to determine the correct choice. We should not
resort to saying that decision theory can't be expected to give an
answer in a scenario this bizarre. The scenario is fundamentally mundane,
even if difficult and costly to realize in practice.

I will argue that for Newcomb's problem with prediction based on
categories, CDT gives the right answer --- take both boxes --- for the
right reason.  The intuitive expression of this reason, which also
justifies smoking in smoking lesion problems and taking both boxes in
an impersonal Newcomb problem, is that the probabilistic dependence
between the right choice and an undesirable circumstance (cancer,
empty box) is due to uncertainty regarding a common cause of both, and
hence, the fact that this choice leads you to expect the undesirable
result should not be a reason to avoid it, when the choice is
otherwise desirable (pleasure of smoking, extra \$1K).

Note that although one can perhaps view causal relationships as
objective features of the world, for a common causal relationship to
explain your (subjective) probabilistic dependence between choice and
circumstance it is necessary that before your choice you have
subjective uncertainty regarding the nature or presence of the common
cause.  This underlies the tickle defense --- once you recognize the
tickle, you no longer have uncertainty about whether the lesion is
present, though it's causal effects remain. 

Causal reasoning (or at least its result) is well-accepted for the
smoking lesion problem in which the lesion affects desire for
smoking. There, the common cause is explicitly mentioned, and seems at
least somewhat plausible.  It is commonplace that a physical thing
like a ``lesion'' can have adverse medical consequences. It is a bit
far-fetched to think that this lesion will also produce a desire to
smoke, but that some physical circumstances affect our desires is also
commonplace, such as when we are more thirsty after being out in the
hot sun, so it is not too hard to imagine that the lesion can do this.
The apparently unanimous agreement that one should smoke in this
scenario seems to me to be due to this clear causal story, even though
the tickle defense can be used to justify it using EDT without reference 
to causation.

In the smoking lesion affecting decision rule problem, and in the standard and
impersonal Newcomb problems, the desires of Participants are fixed by
the problem specification, as are their beliefs concerning the scenario. 
Different choices by Participants must therefore be driven by their (conscious
or unconscious) decision rules, or by how well they apply these rules.
Here, there would seem to be some conflict of intuitions regarding
predictability. We like to think that our reasoning processes are
free, but we do concede that they are affected by factors such
as our education. We also rely on a high degree of predictability of
other people, without which social life would be impossible.

For you to regard the lesion as a common cause of smoking and cancer in the
smoking lesion affecting decision rule problem, you must believe that a physical
circumstance can profoundly affect your thinking.  This is not a
commonplace experience, but is not unknown, as illustrated by the
effects of brain injuries or strokes. So I think most people will still
have a causal intuition that leads to seeing smoking as the correct action,
despite this apparently being contrary to EDT.

In the standard and impersonal Newcomb problems, causal reasoning may
be impeded not just by confused intuitions about predictability of
thoughts, but also because the common causes involved may be obscure
or speculative.

In the impersonal Newcomb problem, the proximate cause
of the Predictor putting \$1M in the opaque box or not is whether
their estimate of $q$, the proportion of one-boxers in the population,
equals $R$ or $1\!-\!R$. This estimate was based on the actions of a
sample of people from the population. These people's actions, and the
resulting estimate of $q$, have a causal influence on your choice only via $R$,
whose value is compatible with both $q=R$ and $q=1\!-\!R$.  Whether
$q$ equals $R$ or $1\!-\!R$ may be influenced by common factors 
that influence both your choice and the choices of the sample of people 
used in estimating $q$, or which influence whether you and they are part 
of the sampled population.  In reality, there will be many, many such 
common factors, most of which you will be unaware of.

Consider an impersonal Newcomb problem with the population being all
adults in Toronto. Suppose you are selected to participate, and are
convinced (by suitable evidence) that with $R=0.8$ the proportion of
one-boxers in this population ($q$) is either $R$ or $1\!-\!R$.  If
you see yourself as a typical Torontonian, you should believe that
your choice will match the majority choice with probability
$R=0.8$. But what will you think is the explanation for this, in terms
of common causes?

Perhaps Toronto's temperate climate has an influence on how people
think?  If so, it would affect both you and the people on whom the
estimate of $q$ was based. Of course, you don't know if this is the
case, and if so which choice the Toronto climate nudges people toward.
Genes could be a causal factor. If you are from a long-time Toronto
family, many other Torontonians may be related and share with you
crucial genes relating to cognition. If you're from elsewhere, it may
be that some genetic factors influence both your decision to come to
Toronto, and your choice (one way or another) in an impersonal Newcomb
problem, with these genes having such effects in other people as well.

The possibilities seem endless. Perhaps you have been influenced by a
book you bought at a Toronto bookstore? Other Torontonians may have
purchased the same book. Clearly, you cannot possibly consider all
possible common factors. A great benefit of causal reasoning is that
you don't have to --- \textit{whatever} causal explanation underlies
the dependence between your choice and the majority choice in the
subjective probability distribution you have prior to making your
decision (which is driven also by your knowledge that $q$ is either
$R$ or $1\!-\!R$), CDT recommends that you take both boxes. The CDT
recommendation would be the same even if you were aware of some
dominant influence on your choice that is unique to you, so that the
probabilistic dependence disappeared.  As with the tickle defense,
this awareness would merely bring the recommendation of EDT into line
with that of CDT (and invalidate the situation as a genuine impersonal
Newcomb problem).

I suspect, however, that there are some people who would take only one
box in an impersonal Newcomb scenario, and some of these would nevertheless
smoke in a smoking lesion scenario, both when the lesion affects
desire and when it affects the decision rule used. To the extent that
this reflects intuitive judgement, rather than formal application
of EDT or some other theory, I would explain this (to me) irrational
choice as resulting from ``quasi-magical thinking'' (Shafir and
Tversky 1992) in situations of uncertainty. In the smoking lesion
scenarios, the causal relationship is known --- only the presence or
absence of the lesion is uncertain. There is vastly greater
uncertainty regarding causal influences in the impersonal Newcomb
problem, which Shafir and Tversky's results show could explain how a
person might decide differently than they would in \textit{any}
situation where this uncertainty had been resolved, with the common
causes being as clearly established as in the smoking lesion problem.

There is even more uncertainty about causal structure in the standard
Newcomb problem with prediction done using categories.  If you
know the categories the predictor uses, and which category you are in,
all the variety of possible causal structures mentioned above for the
impersonal Newcomb problem apply to your inference regarding $q_s$ for
your category, and in addition, you will be uncertain about $q_i$ for
the other categories, which also affect how the Predictor
predicts your choice. If you are ignorant of your category, and perhaps
also what categories the predictor uses, you will have an even less clear
idea of how common causes produce an association between your choice
and the prediction of your choice. When combined with conflicting
intuitions about the predictability of actions, one can see how some
people will adopt quasi-magical reasoning, that leads them to take
only one box.

Nevertheless, if hypothetically the exact causal connections that link
your choice to the Predictor's prediction could be determined, it
seems clear to me that you should take both boxes, just as it is clear
that you should smoke in the smoking lesion problem. And since this is
true for any specific set of causal connections, I think it
is true too when there is uncertainty regarding these connections.

This is also the view of Shafir and Tversky (1992, p.~461), whose
``first interpretation'' below describes prediction using 
categories.:\vspace{-8pt}
\begin{quotation}\noindent
Excluding trickery, there are two interpretations of the Predictor's
unusual powers. According to the first interpretation, the Predictor
is simply an excellent judge of human character. Using some database
(including, e.g., gender, background, and appearance), a predictor
might be able to predict the decision maker's response with
remarkable success. If this interpretation is correct, then you have
no reason to take just one box: however insightful the Predictor's
forecast, you will do better if you take both boxes rather than one
box only. The second interpretation is that the Predictor has
truly supernatural powers\dots\vspace{-8pt}
\end{quotation}
Shafir and Tversky see taking two boxes as justified by
``consequential'' logic, which seems to correspond to CDT, though they
also emphasize (p.~450) the ``sure-thing'' principle (also discussed
by Pearl (2016), for example) that ``if we prefer $x$ to $y$ given any
possible state of the world, then we should prefer $x$ to $y$ even
when the exact state of the world is not known'', whose denial leads
to the quasi-magical thinking they observe among some decision makers
who are uncertain of the situation.

Pollock (2010, p.~70) also describes prediction using 
categories:\vspace{-8pt}
\begin{quotation}\noindent
We are talking about a ``quasi-real-world problem'' in the sense that
the decision maker can assume that what is going on must be
consistent with at least the broad contours of the way we think the
world works. Accordingly, we can assume that the predictor is not a
magician, but [is] relying upon objective cues to make his
prediction. He must know about some objective property of
decision-makers that strongly correlates with how they choose in the
Newcomb Problem. This might be a complex of personality traits, or a
gene, or early childhood experience, or whatever.\vspace{-8pt}
\end{quotation}
Pollock then tells a story where you find out that the common cause
allowing accurate prediction is a gene variant.  By taking a genetic
test, you then find out whether you have the ``one box'' or ``two
box'' variant of the gene (which correspond to the Predictor having
put \$1M or nothing in the opaque box). In one scenario, the genetic
test is free, in another it costs \$500. Pollock claims that after
learning that you have either variant, the rational choice would be to
take two boxes, and that this is uncontroversial.\footnote{\fsp I
suspect that, as for Newcomb's problem with both boxes transparent
(see footnote~\ref{foot-yud}), this is not entirely true. But in this
discussion I will suppose that taking two boxes when you know which
gene variant you have is widely accepted to be correct.} 
Applying a
similar logic to that of Shafir and Tversky, and that of myself above, he
concludes that you should also take two boxes if you do not know which
variant of the gene you have, and hence also that you should not pay
\$500 to be tested.

This reasoning is disputed by Ahmed (2014, Section~7.4.2):\vspace{-8pt}
\begin{quotation}\noindent
Before one gets the information and after one gets it one is facing
\textit{different problems}. Before getting the information the
agent takes her acts to be symptomatic of what is in the box;
\textit{after} getting it she does not. So there is nothing wrong
with taking different acts to be rational \textit{ex~post} and 
\textit{ex~ante}. At least that is so unless changes in the
symptomatic bearing of one's acts on the relevant events are
practically irrelevant if unaccompanied by changes in what one
knows about their causal bearing. But to assume that at the outset
would be to beg the question against EDT.\vspace{-8pt}
\end{quotation}
Ahmed supports this position with two examples that are rather
dissimilar to Newcomb's problem. The issue seems to arise 
in more relevant form in the smoking lesion affecting decision rule
problem. (Recall, though, that Ahmed thinks this problem is
impossible, as discussed in Section~\ref{sec-smoke} above.)

Suppose that before deciding whether to smoke in a smoking lesion affecting
decision rule scenario, you could pay a small sum in order to find out
whether or not you have the lesion. Should you do so?\footnote{\fsp As
for all smoking lesion problems, we assume (unrealistically) that
knowledge of whether or not you have the lesion, and hence are likely
to get cancer, is of no value or disvalue in itself. Also note that,
unlike when the lesion affects your actions via your desires, there is
no ``tickle'' of desire whose presence or absence has already
informed you of whether you have the lesion.}
CDT says that paying to find out would be a waste of money, since CDT
would recommend smoking whatever the result, and hence also recommends
smoking when you do not know whether you have the lesion. But it seems
that EDT would recommending paying to find out whether you have the
lesion, and then recommend smoking regardless of the result. This
seems clearly irrational to me.\footnote{\fsp Of course, by hypothesis, both 
those with the lesion using CDT and those without it using EDT believe that
they are rational.}

Does this intuition extend to Newcomb problems (standard or
impersonal)?  It's first necessary to clarify that the possibility of
testing for the ``gene variant'', or other causal factor, must be
rare. If many Participants have this option, the data on which
predictions are based would be compromised. So assume that it is only
by some strange bit of luck (occurring after the prediction is made)
that you have the option to pay \$500 for this information. Should
you pay?

When presented in this fashion, the logic that recommended not paying
for information in the smoking lesion affecting decision rule problem seems
persuasive here as well. You'll end up taking two boxes regardless of
what gene variant you learn that you have, so why not just take two
boxes now, and save your money?

This reasoning has not persuaded Ahmed, and many other one-boxers.  I
speculate that this is because the causal structure of an actual
non-fantastic Newcomb problem would be very complex, as discussed
above. It will not be a matter of whether or not you have some gene
variant, but rather which of numerous possible common influences have
affected both you and the relevant group of people from whom the
Predictor gathered data. Failure to appreciate this leads either
to a focus on unrealistic ``single gene'' scenarios, which people
rightly suspect, or to abstract scenarios with no concrete picture of
how the Predictor operates, in which discussion wanders untethered by
any reliable intuition.

In Pollock's ``single gene'' story where the Predictor uses your
variant of a gene to predict your choice, he at first assumes that the
gene only has an influence on your intuitions:\ \ ``The gene is not
making either group of decision-makers irrational --- just inclining
them towards different behavior when they have not completely solved
the problem'' (Pollock 2010, p.~70). He then concludes that you should
take both boxes, regardless of whether you know which gene variant you
have. He argues that the ``Why ain'cha rich?'' argument does not
apply, and says that instead ``\ldots if we suppose the two versions
of the gene are equally distributed among decision-makers, the
rationally informed decision-maker will on average receive
\$501,000''.

But if rational Participants who take two boxes do not believe that
they will likely get only \$1K, the scenario is not a Newcomb 
problem.\footnote{\fsp Eells (1981) also argues that rational
Participants who know their beliefs and desires will have a subjective
probability distribution in which their actions are independent of
the Predictor's prediction (see his equations~(15) and~(16)).  But
this holds only if all Participants are perfectly rational, and as I discuss 
in Section~\ref{sec-uncont}, this cannot be the case in a valid
Newcomb problem.}
Pollock later acknowledges this issue (p.~75), but maintains that if
the Predictor uses some more complex property of Participants to
predict your choice even when you rationally deliberate, then
``You[r] choices are entirely determined by whichever property you
have, and that may force you to take one box rather than two''. 
(He does maintain that taking two boxes \textit{would}
be the right choice, if you actually had a choice.)

The argument seems strange. He says that ``Although it may seem to you
that you are making voluntary choices, in fact you are not''. But if
so, surely your choices are involuntary regardless of whether or not
the Predictor makes use of the property that predicts them. This
diversion into the long-running debate on determinism and free will
seems irrelevant. But it does suggest that there is something wrong
with how the causal structure of the problem is being visualized.

Here is a better story.  The population from which Participants are
drawn is a town settled by Norwegians and Italians.  Most of the
Italians eat spaghetti; most of the Norwegians do not.  The Norwegians
are mostly Protestant; the Italians are mostly Catholic. The
Norwegians have all read \textit{Peer Gynt}; the Italians have all
read \textit{The Divine Comedy}. The Norwegians are more closely
related to other Norwegians than to Italians, and the Italians are
more closely related to other Italians than to Norwegians.  There are
many other differences, some of which perhaps go along with different
initial inclinations regarding what to do in a Newcomb problem, some
with different examples of deliberation they are familiar with, some
with different degrees of motivation to think carefully about a
decision, some with skill at logical reasoning --- and some of which
are of no relevance.

You have only vague knowledge of all this. You also have no idea what
information the Predictor uses to make predictions. Perhaps they have
access to church membership lists. Or maybe they know who has taken
which books out of the town library. Or they may know everyone's
scores at bowling, a popular pastime of everyone in the town. Or they
may run several restaurants and know who eats what. But somehow, they
manage to make predictions with accuracy $R=0.8$.

As you reach for both boxes, you (by assumption) think that the opaque
box is likely (with probability 0.8) to be empty (partly because you
know that it has been empty with that frequency when previous
Participants took both boxes).  But why should this be?  There are
possible causal explanations for the Predictor's accuracy. You may
guess that the Norwegian/Italian split might be involved, though there
are other possibilities, such as choice in Newcomb's problem being
related to bowling skill. Supposing that whether you are Norwegian or
Italian is relevant, you might wonder how exactly. There are many
possibilities. Perhaps reading \textit{The Divine Comedy} is
influential --- but if so, which way? Or might the books discussing
predestination you've talked about with your coreligionists have
influenced both your and their thoughts?

If you somehow knew both how the Predictor predicts and also all the
details, whatever they may be, of the causal relationships that link
your choice with other people's past choices that informed how the
Predictor predicts, I think it would be clear that taking two boxes is
the right choice, just as smoking is the right choice in the smoking
lesion problem. But in a smoking lesion problem, there are only two
possibilities --- you have the lesion or not --- which lets you
clearly visualize why you should smoke by imagining both
alternatives. In contrast, in any realistic Newcomb problem there will
be an enormous number of possible causal connections, so it takes a
greater feat of imagination to see that since taking both boxes is
best for \textit{all} such possibilities, taking both boxes is also
best when you don't know what the causal connections are.

\section{Prediction by simulating a Participant's 
thoughts}\label{sec-simul}\vspace{-10pt}

In Section~\ref{sec-prcat}, I discussed practically-realizable scenarios
in which reasonably good predictions of whether a Participant will
take one box or two can be made based on which of a fairly small set
of categories the Participant is in.  I expect that the accuracy, $R$,
of predictions made in this way, can be pushed higher by using a
larger number of categories, defined by combinations of an increasing
number of attributes.  As discussed at the end of
Section~\ref{sec-pr}, estimating $q_i$ for each of many categories
$G_i$ can still be feasible if the way attributes affect the $q_i$ can
be approximately modelled in a fairly simple way.

However, as the number of attributes becomes very large --- for
example, including many details of a Participant's brain anatomy,
potentially down to the strengths of individual neural synapses ---
fully utilizing the information in these attributes will be
feasible only with strong knowledge of how the human brain
operates. Only in this way does it seem possible to make very accurate
predictions using such detailed information about a unique human
individual --- information which will differ greatly from any previous
human that the Predictor may have data on, so that making good use of
it by modelling general patterns will not be possible even with
sophisticated statistical techniques.

To achieve very high predictive accuracy (say, $R=0.9999$), it seems
necessary to use neural data on a Participant to accurately simulate their
thoughts in detail, which will also require simulating their
environment.  Without such a detailed simulation, some Participants
with highly variable thoughts concerning what action to take will not be
predictable. For example, a Participant's choice of one box or two
might be affected by whether or not some past memory happens to
surface during their deliberations, or by whether or not their pencil
lead breaks while doing some calculation on paper.

In this paper, I will assume a physicalist and functionalist view of
consciousness.\footnote{\fsp Functionalism is discussed by, for
example, Levin (2023).  Alternatively, one might think that detailed
simulation of thought is not possible, so Newcomb's problem can only
be implemented by methods similar to those of
Section~\ref{sec-prcat}. Or one might think detailed simulation is
possible, but gives rise to a ``zombie'' (Kirk, 2023), and since you
know you are not a zombie, you know you are not such a simulation. In
that case, I think one should view prediction by such a zombie
simulation in the same way as prediction using categories, but I will
not argue for that position here.}  On this view, a detailed
simulation of a Participant will itself be a conscious being.  Since
this being has (we assume) thoughts that cannot be distinguished (by
themselves) from those of a real person, if you participate in such
a Newcomb scenario, you will not know whether you are this simulated
being, or the real human Participant.

If you as a Participant know that the Predictor uses thought
simulation to make the prediction, your uncertainty about whether you
are the real Participant or a simulation could be seen as a
violation of condition~(j) of Section~\ref{sec-uncont}, since you
will not be sure that you are a Participant in a standard Newcomb
setup --- in particular, you will not be confident that the decision
you make will precede the Predictor filling the (real) boxes.  If
this view of the problem specification is adopted, I would maintain
that a Newcomb problem with very high predictive accuracy is not
possible.  But if we do consider scenarios with thought simulation to
be valid Newcomb problems, I will argue that the correct decision may
be to take only the opaque box, though if you have more detailed
knowledge of the prediction process, randomizing your decision may
sometimes be best.\footnote{\fsp I will not consider here the possibility
that you suspect, but are not sure, that the Predictor uses thought
simulation. I assume you have reliable knowledge of this, or have
inferred it from the high accuracy of the Predictor being otherwise
inexplicable. I will also ignore a concern that would in reality loom
large if you know that the Predictor can accurately simulate your
thoughts --- the possibility that the entire Newcomb scenario does not
exist in reality, with there being no actual Participant, and no
actual monetary reward in the end.}

Newcomb scenarios with conscious simulated beings have been considered
before, by myself (Neal 2006, Section~2.5), by Aaronson (2005; 2013,
Chapter~19), and by others.  My discussion here shares the same basic
intuition of these past discussions --- if you are the simulated
Participant, taking one box gains the real Participant \$1M, which
outweighs the \$1K that is lost by not taking both boxes if you are
the real Participant --- but here I will analyse the problem in substantially
more detail than these past treatments.\footnote{\fsp
Strangely, the idea that the Predictor performs a detailed simulation
of a Participant's thoughts has appeared a number of times in the literature
on Newcomb's problem without being accompanied by any awareness that this
could mean that you don't know whether or not you are the simulated
entity. Lewis (1979), for example, says ``The potentially predictive 
process \textit{par excellence} is \textit{simulation}. To
predict whether I will take my thousand, make a replica of me,
put my replica in a replica of my predicament, and see whether my
replica takes \textit{his} thousand. \ldots A flesh-and-blood duplicate
made by copying me atom for atom would be one good sort of replica.''
}

\subsection{\hspace*{-8pt}Thought simulation and its limits}\vspace{-6pt}

Prediction by thought simulation would proceed as follows. While the
Participant is in the examination room, the Predictor uses some
advanced technology to make very detailed measurements of their brain
and body. The Predictor then quickly (faster than real life) runs a
simulation of the Participant's future thoughts during deliberation on
some powerful computer,\footnote{\fsp Alternatively, the Predictor
runs the simulation at whatever speed is possible, but slows down time
for the real Participant as they are moving to the deliberation
room, so that the prediction is made and boxes filled before they start
deliberating. This is possible according to currently-accepted
physical theory, for example by placing the Participant in a strong
gravitational field, though such methods are utterly impractical.}
and sets the contents of the two boxes according to the
results of this simulation before the Participant enters the
deliberation room.

For thought simulation to be possible, the Predictor must not only
have detailed knowledge of the Participant's body and brain, but also
of the contents of the examination and deliberation rooms (which must
be isolated from the outside world), since a Participant's environment
will obviously affect their thoughts.

Is such a simulation possible? Current technology is very, very far
from being adequate for either the measurements or the computation
required for such a simulation.  The simulation scenario is clearly
fantastic in this sense --- unlike prediction using a fairly small
number of categories, which merely poses some surmountable practical
difficulties. But we may still wonder whether such simulation is
possible in principle.

In a world operating according to classical physics, it would be.  The
required measurements could in principle be made arbitrarily precise
(down to measuring the position and velocity of every molecule),
without any disturbance to the Participant, and the future state of
the Participant and environment could then in principle be calculated
according to deterministic laws.

However, we do not live in the world of classical physics. According
to quantum physics, the entity that evolves through time
deterministically is the wave function, and according to the ``no
cloning'' theorem (Aaronson 2013, p.~125), this cannot be duplicated
without disturbance.  So perfect simulation is not possible even in
principle, according to currently-accepted physics.\footnote{\fsp Perfect
simulation may also be impossible if perfect isolation of the
deliberation room from the outside world is impossible.}

There remains the question of whether the physical processes of
thought might involve only coarse aspects of the brain and its
surroundings, which could be measured to adequate precision without
undue disturbance to the Participant, and used in a simulation that
has a high probability of duplicating the choice of the real
Participant.  This is an empirical question, to which no one currently
knows the answer.

My own opinion (somewhat, but not highly, informed) is that it is
likely possible (in principle) to measure and simulate with sufficient
accuracy to create a conscious entity quite similar to the
Participant, but that for some Participants this simulated person will
have a significant probability of making a different choice than the
real Participant.  The thoughts leading to a Participant's decision
could sometimes depend sensitively on details of their memories of the
past, or on the exact strength or timing of their reactions to
sensory input or to their internal thoughts or feelings.\footnote{\fsp
Aaronson (2013, Chapter 19), echoing the jargon of computational
complexity theory, refers to this as Newcomb's problem being
``you-complete''.} For such a Participant, even a slight inaccuracy in
the simulation could result in the simulated Participant deciding
differently from the real Participant.

Note that for a Newcomb problem in which the Participants are not
humans, but instead are artificial intelligences running on standard
digital computers (equivalent to Turing machines), a Predictor could
easily duplicate the state of a Participant, since digital computers
are designed to allow this despite the quantum aspects of their
hardware. If the environment in which the digital Participants
deliberate was also artificial (without, for example, video input from
the world outside the computer), it would be easy to run the duplicate
in the same artificial environment, on a faster computer than that
running the real Participant, and hence obtain a perfect prediction
of the Participant's choice before they ``enter'' the (artificial)
deliberation room.

This is an interesting problem, but it is \textit{not} the standard
Newcomb problem, which is about how \textit{humans} should make
decisions.  It is not clear that the correct decision for an AI in a
Newcomb problem is the same as the correct decision for a human.  Even
considering this question may require resolving issues that do not arise
with human Participants, such as whether or not running an AI
Participant and a duplicate of it on two computers, with identical
artificial environments, results in two conscious beings, or only
one.\footnote{\fsp If you are inclined to say two, would running a single
instance on a computer that redundantly performs all operations in
duplicate in order to correct hardware errors also produce two
conscious beings?  If you are inclined to say only one, would running
the duplicate in a very slightly different environment (say with
barely perceptible differences in some colours), producing a tiny
chance of different behaviour, mean that there were now two conscious
beings? Does the answer matter for anything?}

Despite the almost-certainly insurmountable difficulties of doing so,
I will assume below that the Predictor can actually simulate a
Participant's thoughts closely, though not absolutely perfectly, and
hence attain a very high predictive accuracy.  The question then is
what you should do in such a situation.

\subsection{\hspace*{-8pt}Take one box when prediction uses a thought 
simulation}\label{sec-simdecision}\vspace{-6pt}

Suppose that you are deliberating in a Newcomb scenario in which you
know that the Predictor predicts your decision by running a simulation
of your thoughts. While deliberating, you do not know whether
you are the real, flesh-and-blood ``you'', or the simulated ``you''.  You
know that the Predictor has an accuracy, $R$, that is only slightly
less than one.\footnote{\fsp I take an accuracy close to one to be typical
of thought simulation scenarios, since it is hard to imagine that
the simulation is good enough that you can't tell whether you are
the simulated or real Participant, but only allows for low accuracy
such as $R=0.7$. It would be possible, though, that the simulation itself
is highly accurate, but that random errors are introduced when a
careless assistant to the Predictor fills the boxes, reducing the
$R$ that is achieved.\vspace{-5pt}}
Suppose also, as usual for a Newcomb problem, that you have
no reason to think you are less predictable than other
Participants. Should you decide to ``take'' one box or two boxes?

If you are the real Participant, your decision has the causal effect
of delivering the contents of the boxes you decide to take to the
real Participant. If you are the simulated Participant, your
decision instead has the causal effect of determining whether the
Predictor puts \$1M in the (real) opaque box, which will
subsequently be delivered to the real Participant. (The Predictor
puts \$1M in the opaque box if you ``take'' one box, but puts in
nothing if you ``take'' two.)

If you are the real Participant, you will be able to enjoy whatever
payoff is contained in the boxes you take.  If you are the simulated
Participant, your existence will presumably be terminated soon after
you make your decision.  What are your desires if you turn out to be
the simulated Participant?

We can consider the possibility that conditional on being the
simulated Participant (which you are never sure of) you care nothing
about what happens to the real Participant, and so are indifferent
to what is in the boxes taken by the real Participant.  The outcome
for you as a simulation is termination regardless.  If this is your
view, you should make your decision on the assumption that you are the
real Participant, since if not, your decision will have no effect on
your fate. EDT then recommends taking one box, and CDT recommends
taking two, for the usual reasons.  

However, I think it is unrealistic to think that a simulated
Participant will have no concern for the real Participant.  Even if
you are a simulation, your thoughts are nearly identical to those of
the real Participant. Furthermore, you are never sure that you
\textit{aren't} the real Participant. So not having sympathy for the
real Participant would seem bizarre.\footnote{\fsp I doubt that it is
even coherent for a simulated Participant to have no concern for the
real Participant. This would require an ethics based on an indexical
concern for ``me'' separate from a concern for people who resemble me
in various respects (perhaps very strongly resemble). But who I am
changes minute by minute, so this would imply a lack of concern for my
future self. Since the purpose of ethics is to motivate decisions
whose effects are all in the future, this stance seems to undermine the
egoism on which it is apparently based.\vspace{-12pt}}
I will assume, then, that regardless of whether you are the simulation
or the real Participant, you are motivated to help the real Participant.

Figure~\ref{fig-simpay} shows on the left the payoffs for taking one
box or taking two boxes in this scenario, for the four possible states
of the world, in which you are the real or simulated Participant and in
which the other ``you'' takes one or two boxes.

\begin{figure}
\begin{center}

\begin{tabular}{l|cc|cc}
 & \multicolumn{2}{c|}{You are real Participant} &
   \multicolumn{2}{c}{$\!\!\!\!$You are simulated Participant$\!\!\!\!\!\!$} 
\\[6pt]
 & $\!$Simulated you$\!\!$& $\!\!$Simulated you$\!\!$&~~Real you & Real you \\
 &  takes 1 box  & takes 2 boxes &  ~~takes 1 box & takes 2 boxes \\
 & $(W\and S_1)$ & $(W\and S_2)$ & ~~$(\barW\and A_1)$ & $(\barW\and A_2)$ 
\\[6pt]
\hline & & & & \\[-5pt]
$\!\!\!$Take 1 $(D_1)\!\!$
 &  \$1M       &    0   &  \$1M   & \$1M$\,+\,$\$1K\\[6pt]
$\!\!\!$Take 2 $(D_2)\!\!$
 & \$1M$\,+\,$\$1K &   \$1K &    0    & \$1K 
\\[4pt]
\hline & & & & \\[-5pt]
$\!\!$Difference~~& $\!\!\!\!-$\$1K &  $\!\!\!\!-$\$1K &   \$1M    & \$1M 
\end{tabular}~~\rule[-62pt]{2pt}{131pt}~~%
\begin{tabular}{cc}
 \multicolumn{2}{c}{$\!\!\!\!$Average of real and simulated$\!\!\!\!$} \\[6pt]
 Other you & Other you \\
  takes 1 box  & takes 2 boxes \\
  ($O_1$) & ($O_2$) \\[6pt]
\hline  & \\[-5pt]
 \$1M       & \!\!(\$1M+\$1K)/2\!\!\!\!\!\!  \\[6pt]
 \!\!\!(\$1M+\$1K)/2\!\! &   \$1K \\[4pt]
\hline  & \\[-5pt]
 \!\!\!(\$1M$-$\$1K)/2\!\! & \!\!(\$1M$-$\$1K)/2\!\!\!\!\!\!
\end{tabular}~~

\end{center}
\vspace{3pt}
\caption{Payoffs to the real ``you'' when a real or simulated
``you'' takes one box or two boxes (left), and average of the 
payoffs for actions of the real and simulated you (right).}\label{fig-simpay}

\end{figure}

We can analyse what Evidential Decision Theory (EDT) recommends you do
in this situation by adapting the methods of
Section~\ref{sec-recommend}.  As in that section, let $A_1$ and $A_2$
denote the events that the real Participant takes one box or two,
and let $S_1$ and $S_2$ denote the events that the Predictor predicted
that the real Participant will take one box or two. In this thought
simulation scenario, $S_1$ and $S_2$ are also the events that the
simulated Participant decided to take one box or two.

Let $W$ be the event that you are the real Participant, with $\barW$
being the event that you are the simulated Participant.  We can write
the four states of the world that are possible when you are deliberating
as $W\and S_1$, $W\and S_2$, $\barW\and A_1$, and $\barW\and A_2$. 
The events, $D_1$ and $D_2$, that you (whoever you are) take one box or 
take two can be written as\vspace{-6pt}
\beq
   D_1 \ =\ W\and A_1\ \mbox{or}\ \barW\and S_1\ \ \ \ \ \mbox{and}\ \ \ \ \
   D_2 \ =\ W\and A_2\ \mbox{or}\ \barW\and S_2
\eeq
The events of the other you taking one or two boxes are\vspace{-2pt}
\beq
   O_1 \ =\ W\and S_1\ \mbox{or}\ \barW\and A_1\ \ \ \ \ \mbox{and}\ \ \ \ \
   O_2 \ =\ W\and S_2\ \mbox{or}\ \barW\and A_2
\eeq
Note that $D_1\and W=A_1\and W$, $D_1\and \barW=S_1\and \barW$, 
$O_1\and W=S_1\and W$, $O_1\and \barW=A_1\and \barW$, $D_2\and W=A_2\and W$,
and so forth.

Since you do not know whether you are the real or simulated
Participant, you cannot decide by comparing $V(A_1)$ to $V(A_2)$, as
in Section~\ref{sec-recommend}, but should rather compare $V(D_1)$ to
$V(D_2)$.  Your decision might be affected by additional information,
$B$, that you know, but a requirement for a Newcomb scenario (condition (k)
of Section~\ref{sec-uncont})
is that this information does not change your subjective probability
that the Predictor is correct. In this thought simulation scenario,
I will also assume that $B$ is not informative about whether you are the
real Participant or not, so\vspace{-4pt}
\beq
  \prs(W\giv B)\ =\ \prs(\barW\giv B)\ =\ 1/2
\eeq
Your decision should also not be informative about whether you are
the real Participant, hence
\beq
  \prs(W\giv D_1\and B) & =& \prs(\barW\giv D_1\and B)\ \ =\ \ 1/2
\\[4pt]
  \prs(W\giv D_2\and B) & =& \prs(\barW\giv D_2\and B)\ \ =\ \ 1/2\ \
\eeq
Believing that the Predictor has accuracy $R$ means that you should have
conditional probabilities
\beq
  \prs(S_1\giv A_1\and W\and B) & =& \prs(S_1\giv D_1\and W\and B)\ \ =\ \ 
  \prs(O_1\giv D_1\and W\and B)\ \ =\ \ R
\\[4pt]
  \prs(S_2\giv A_2\and W\and B) & =& \prs(S_2\giv D_2\and W\and B)\ \ =\ \ 
  \prs(O_2\giv D_2\and W\and B)\ \ =\ \ R
\eeq
If the simulation is good, you should rationally also believe that, 
symmetrically,
\beq
  \prs(A_1\giv S_1\and \barW\and B) & =& \prs(A_1\giv D_1\and \barW\and B)\ =\
  \prs(O_1\giv D_1\and \barW\and B)\ \ =\ \ R
\\[4pt]
  \prs(A_2\giv S_2\and \barW\and B) & =& \prs(A_2\giv D_2\and \barW\and B)\ =\
  \prs(O_2\giv D_2\and \barW\and B)\ \ =\ \ R
\eeq
One can note that the above imply that
$\prs(O_1\giv D_1\and B) \ = \ \prs(O_2\giv D_2\and B) \ =\ R$.

We can now compute your subjective probabilities for the four possible
states of the world when deliberating, conditional on your taking one
box (decision $D_1$):
\beq
  \prs(W\and S_1\giv D_1\and B) & = &
    \prs(W\giv D_1\and B)\cdot\prs(S_1\giv D_1\and W\and B) \ \ =\ \ R/2 
\\[7pt]
  \prs(W\and S_2\giv D_1\and B) & = &
    \prs(W\giv D_1\and B)\cdot\prs(S_2\giv D_1\and W\and B)\ \ =\ \ (1\!-\!R)/2 
\\[7pt]
  \prs(\barW\and A_1\giv D_1\and B) & = &
    \prs(\barW\giv D_1\and B)\cdot\prs(A_1\giv D_1\and \barW\and B)\ \ =\ \ R/2 
\\[7pt]
  \prs(\barW\and A_2\giv D_1\and B) & = &
    \prs(\barW\giv D_1\and B)\cdot\prs(A_2\giv D_1\and \barW\and B)\ \ =\ \ 
     (1\!-\!R)/2 
\eeq
and similarly, conditional on taking two boxes (decision $D_2$), the possible 
states have these probabilities:
\beq
  \prs(W\and S_1\giv D_2\and B) & = &
    \prs(W\giv D_2\and B)\cdot\prs(S_1\giv D_2\and W\and B) \ \ =\ \ (1\!-\!R)/2
\\[7pt]
  \prs(W\and S_2\giv D_2\and B) & = &
    \prs(W\giv D_2\and B)\cdot\prs(S_2\giv D_2\and W\and B)\ \ =\ \  R/2 
\\[7pt]
  \prs(\barW\and A_1\giv D_2\and B) & = &
    \prs(\barW\giv D_2\and B)\cdot\prs(A_1\giv D_2\and \barW\and B)\ \ =\ \ 
  (1\!-\!R)/2 
\\[7pt]
  \prs(\barW\and A_2\giv D_2\and B) & = &
    \prs(\barW\giv D_2\and B)\cdot\prs(A_2\giv D_2\and \barW\and B)\ \ =\ \ R/2
\eeq

We can now compute the expected value of $D_1$ and $D_2$ according to EDT:
\beq
  V(D_1) & = & 1000000\cdot\prs(W\and S_1\giv D_1\and B) \nonumber\\
  &&         \ +\ 0 \cdot \prs(W\and S_2\giv D_1\and B) \nonumber\\
  &&         \ +\ 1000000 \cdot \prs(\barW\and A_1\giv D_1\and B) \nonumber\\
  &&         \ +\ 1001000 \cdot \prs(\barW\and A_2\giv D_1\and B) \\
  & = & 1000000\cdot R/2\ +\ 1000000\cdot R/2\ +\ 1001000\cdot (1\!-\!R)/2 \\
  & = & 1000000\cdot R\ +\ 1001000\cdot (1\!-\!R)/2
\\[6pt]
  V(D_2) & = & 1001000\cdot\prs(W\and S_1\giv D_2\and B) \nonumber\\
  &&         \ +\ 1000 \cdot \prs(W\and S_2\giv D_2\and B) \nonumber\\
  &&         \ +\ 0 \cdot \prs(\barW\and A_1\giv D_2\and B) \nonumber\\
  &&         \ +\ 1000 \cdot \prs(\barW\and A_2\giv D_2\and B) \\
  & = & 1001000\cdot (1\!-\!R)/2\ +\ 1000\cdot R/2\ +\ 1000\cdot R/2 \\
  & = & 1001000\cdot (1\!-\!R)/2\ +\ 1000\cdot R
\eeq
The difference is\vspace{-6pt}
\beq
   V(D_1)\ -\ V(D_2) & = & 999000\cdot R
\eeq
Since this is positive for any $R>0$, EDT recommends taking one box in this
thought simulation scenario.

In Section~\ref{sec-recommend}, Causal Decision Theory (CDT) was easy
to apply to Newcomb's problem without thought simulation, because
taking two boxes dominates taking one box --- it is better in every
state of the world.  However, when Prediction is by thought
simulation, and you do not know whether you are the real or
simulated Participant, this is not true, as can be seen in
Figure~\ref{fig-simpay}.  Taking two boxes dominates if you are the
real Participant, but taking one box dominates if you are the
simulated Participant, so there is no overall dominant action when you
do not know which you are.

In general, when no action is dominant, it is necessary to compute the
expected causal effect of each action, and choose the action with
higher expected utility.  When your decision is informative about
the state of the world, this can be confusing.  Here, however, there
is a simple solution.

Rather than compute both $U(D_1)$ and $U(D_2)$ and then see whether or
not $U(D_1)>U(D_2)$, we can instead compute $U(D_1)-U(D_2)$ and see
whether or not it is positive.  The last row in
Figure~\ref{fig-simpay} shows the difference between the payoff from
action $D_1$ and that from action $D_2$ in each of the four states of
the world.  $U(D_1)-U(D_2)$ can be found by averaging these
differences with respect to the subjective probabilities of these four
states. Note, however, that the difference in payoffs is the same
($-$\$1K) for $W\and S_1$ as for $W\and S_2$, and that it is also the
same (\$1M) for $\barW\and A_1$ as for $\barW\and A_2$.  We can therefore
find $U(D_1)-U(D_2)$ as
\beq
  U(D_1)-U(D_2) & = & (-\mbox{\$1K})\,\prs(W\giv B)\ +\ 
                      (\mbox{\$1M})\,\prs(\barW\giv B) \\
& &
                   (-\mbox{\$1K})\,(1/2)\ +\ 
                      (\mbox{\$1M})\,(1/2) \ =\ \ 499500
\eeq
Since this is positive, CDT recommends taking one box, for any value of $R$.

When $R$ is nearly one, $V(D_1)-V(D_2)$ is nearly 999000, whereas
$U(D_1)-U(D_2)$ is half that, which might seem strange.  But note that
$U(D_1)-U(D_2)$ is the benefit from taking one box rather than two
whenever ``you'' are in this Newcomb situation.  There are two of
``you'' who are in this situation (the real you and the simulated you),
so if they both decide to take one box, the total benefit (versus them
both taking two boxes) is $2\times499500=999000$.

Another way of looking at this problem is to note that you are equally
likely to be the real or simulated you, and that (assuming a good
simulation) all probabilities should be the same conditional on being
the real or being the simulated you. Because of this, one can simply
average the payoffs for the real and simulated you (on the left 
in Figure~\ref{fig-simpay}) and solve the decision problem with these
average payoffs, which are shown on the right in Figure~\ref{fig-simpay}.

Using this approach with EDT gives the same values for $V(D_1)$ and
$V(D_2)$ as above, with $V(D_1)>V(D_2)$. With the average payoffs,
$D_1$ dominates $D_2$, so CDT will also recommend taking one box.

According to EDT, the benefit of taking one box rather than two is
smaller when $R$ is smaller, whereas it is constant according to CDT.
This is because CDT is looking at the causal effect of ``you'' taking
one box, which turns out not to depend on whether the other ``you''
also takes one box, whereas EDT is looking at the expected result after
taking one box, which includes an expected benefit of the other
``you'' taking one box as well, which you have less reason to think
will happen if $R$ is small.

Since EDT and CDT both recommend taking one box when the Predictor
runs a simulation of your thoughts, this version of Newcomb's problem
is not directly helpful in judging which of these theories of decision
is correct.  However, if, as I maintain, Newcomb scenarios with very
high predictive accuracy are conceivable only with thought simulation,
the argument for EDT based on the intuitive appeal of one-boxing when
$R$ is close to one is undermined.  And as I argue above, when $R$ is
not so close to one, and prediction is done using categories, properly
considered intuition should favour two-boxing, justified by CDT.

\subsection{\hspace*{-8pt}Randomization may be best when prediction uses 
multiple thought simulations}\label{sec-multisim}\vspace{-6pt}

A Predictor who cannot simulate a Participant perfectly might simulate
the Participant several times, and predict based on what a majority of
the simulated Participants do. Of course, this makes no sense if the
simulations are deterministic, and hence all do the same thing.  But
when the Predictor cannot measure the Participant's brain perfectly,
nor perfectly simulate it, introducing randomness in the simulation
would be natural, as a way of representing this lack of knowledge.
Multiple simulations with different random choices will then be a
better guide to what the real Participant will do than a single
simulation.

Variation of decisions in multiple simulations might arise from
simulated Participants making different fundamental choices ---
affected, perhaps, by whether they are distracted at crucial points
when contemplating decision theory. I won't attempt to analyse what
you should (if not distracted!)\ do in such a situation. Instead, I
will consider only variation that results from Participants
deliberately randomizing, by physical or mental methods (as discussed
in Section~\ref{sec-rand}), with this randomization being reflected
in the results of the simulations, as well as the decision of the
real Participant.  It will turn out that randomization is
helpful, which refutes the idea that one should consider real and
simulated Participants to be the same person, who necessarily make the
same choice, which seems to be an intuition underlying some arguments
for taking one box in Newcomb's problem (even when prediction is not
done by thought simulation).

For simplicity, I will consider only a situation where the Predictor
runs three simulations of a Participant, and predicts based on what
a majority of these simulated Participants do.  I assume also that
Participants know that this is how the Predictor operates.

Suppose the real Participant and all simulated Participants
randomize their decision, taking one box with probability $p$.  A
majority (two or three) of the simulated Participants will take one
box with probability $q\,=\,3p^2(1\!-\!p)\,+\,p^3$ (there are three
ways for two to take one box, each way having probability
$p^2(1\!-\!p)$, and one way for all three to take one box, having
probability $p^3$). The resulting predictive accuracy is
$R\,=\,pq\,+\,(1\!-\!p)(1\!-\!q)$ (either both the prediction and the
actual decision are to take one box, or both are to take two). The
expected payoff to the real Participant is
$u\,=\,1000000q\,+\,1000(1\!-\!p)$, combining the expected payoff of
$1000000q$ from the opaque box when a majority of simulated
Participants take one box with the expected payoff of $1000(1\!-\!p)$
from the transparent box when the real Participant takes both boxes.

Looking at this situation from the perspective of an outside observer,
we might ask what value of $p$ maximizes the expected payoff to the
real Participant. The maximum will occur where $du/dp$ is zero:
\beq
  0 \ \ =\ \ {du \over dp} & = & 1000000\,{dq \over dp}\ -\ 1000 \\
  & = & 1000000\,(6p(1\!-\!p)\ -\ 3p^2\ +\ 3p^2)\ -\ 1000
        \ \ =\ \ 6000000\,p(1\!-\!p)\ -\ 1000
\eeq
The solution to this equation corresponding to the maximum for $u$ is
\beq
  p^* & = & {1\over2} \ +\ {\sqrt{1-4/6000} \over 2}
    \ \ =\ \ 0.999833306 \label{eq-opt-rand}
\eeq
This optimal $p^*$ leads to corresponding values
$q^*=0.999999917$, $R^*=0.999833222$, and $u^*=1000000.08$.

The expected gain of eight cents over the \$1M obtained when always taking one
box may seem underwhelming, but is sufficient to show that randomization
can be beneficial.

But will randomization appear attractive to Participants themselves,
when they are making their decisions?

For it to be rational for a Participant to decide randomly whether to
take one box or two, with some probability other than 0 or 1, it must
be that they will not regret the result of randomizing, whether it be
to take one box, or to take two.  The correct randomization
probability can be found from this condition.  I will assume that when
you, a simulated or real Participant, randomizes with probability
$p$, you will expect that other instances of you will also randomize
with that probability.\footnote{\fsp This is a simplification, of course,
but if your uncertainty regarding what randomization probability other
versions of you will use is centred on the probability you use, and is
symmetrical, the effect on your beliefs will be the same as if they
randomized with the same probability as you.}

Since I assume that it is your (known) randomization probability,
not the outcome of this randomization, that affects your belief about
what decisions other instances of you will make, EDT and CDT will here
produce the same evaluation of the expected payoff when randomization
results in taking one box or two. If you assume that other
simulated or real versions of you take one box with probability $p$,
the difference in expected payoff if your randomization results in 
taking one box versus two boxes will be
as follows:
\beq
  U(D_1)-U(D_2)\ \ =\ \ V(D_1)-V(D_2)\ \ =\ \
  -(1/4)\,\mbox{\$1K} \ +\  (3/4)\,2p(1\!-\!p)\,\mbox{\$1M}
\label{eq-int-rand}\eeq
The first term on the right reflects that taking one box loses \$1K if you are 
the real Participant, which has probability $1/4$, since there
are three simulated instances of you and only one real you.
The second term is the probability (3/4) that you are a simulated
Participant, times the probability that your choice makes a difference
to the Predictor's prediction, times the gain to the real Participant
of your ``taking'' one box in this situation.  Note that your choice
makes a difference only if the other two simulations are split,
one taking one box and the other taking two. The split can happen two
ways, each with probability $p(1\!-\!p)$.

Equating (\ref{eq-int-rand}) to zero and solving for $p$ produces the
same $p^*$ as in equation~(\ref{eq-opt-rand}). Moreover, this choice
of randomization probability is stable --- if you were to consider
using a slightly higher randomization probability,
equation~(\ref{eq-int-rand}) would become negative, so you would wish
to correct by lowering $p$ back to $p^*$, and you would similarly
correct back to $p^*$ if you considered using a $p$ less than $p^*$.

\section{Two Newcomb-like problems}\vspace{-10pt}

Several real-world problems have been put forward as being analogous
to Newcomb's problem (see Ahmed (2014, Chapter 4) and Berm\'udez
(2018) for examples).  I will discuss two --- the well-known
Prisoner's Dilemma problem, and the problem of whether to vote in a
large election. I will argue that both are actually more similar to
the impersonal Newcomb problem of Section~\ref{sec-impn} than they
are to the standard Newcomb problem, and that for both, the analogue
of taking two boxes, recommended by CDT, is the correct decision.

\subsection{Prisoner's Dilemma\label{sec-pd}}\vspace{-8pt}

Alice and Sue\footnote{\fsp The agents are often referred to as Alice
and Bob (for example, by Ahmed (2014, Section~4.6)), but since
similarity of these agents will turn out to be a central aspect, it
seems better for them to not differ in sex. Also, symbols referring to
Alice and Sue can use letters $A$ and $S$, matching the letters
used in analogous symbols for Newcomb's problem.\vspace{-10pt}}
are embezzling money from their employer.  By exploiting
a computer security flaw, they have added themselves to the list of
people entrusted to disburse funds. This highly illegal action will be
discovered in the regular security review at the end of the month, so
they both plan to leave the country for (separate) jurisdictions
without extradition treaties, just after they have received the
embezzled funds (and moved them to untraceable cryptocurrency
accounts).  Even though they are on the trusted list, financial
policies prevent them from authorizing more than one payment a month,
from authorizing a payment over \$1M, or from authorizing a payment to
themselves.  So they have each authorized a payment of \$1M to the
other. But on the morning of the day of this payment, before it has
been made, they both receive notification of a change of policy
allowing trusted employees to authorize payments of up to \$1K to
themselves.  They both have time to modify the payment they authorized
so that it is for \$1K rather than \$1M, and is made to themselves
rather than to the other, before the payments are made.  Now, they
must each decide (separately) whether to do this, thereby gaining an
extra \$1K for themselves, while eliminating the \$1M payment to the
other, or whether to leave the payment of \$1M to the other
unmodified.

This is an example of a ``Prisoner's Dilemma'' problem, with payoffs as
shown in Figure~\ref{fig-pd}. Here, as in all Prisoner's Dilemma
problems, Alice and Sue act separately, without communicating with
each other.  The payoffs are such that Alice does better to ``defect''
(action $A_2$, modifying the payment to be \$1K for herself) than to
``cooperate'' (action $A_1$, leaving the payment as \$1M for Sue),
regardless of what Sue does.  And the same is true for Sue --- action
$S_2$ is better than $S_1$ whatever Alice does.  

I have chosen the numerical payoffs in this example to facilitate an
analogy with Newcomb's problem.

\begin{figure}[t]

\begin{center}
\begin{tabular}{r|@{~~~}rl@{~~~~~~}rl}
& \multicolumn{2}{c}{Sue cooperates ($S_1$)} 
& \multicolumn{2}{c}{Sue defects ($S_2$)} \\
\hline
\rule{0pt}{15pt}%
Alice cooperates ($A_1$) &   \$1M,& \$1M        &   0,& \$1M+\$1K \\[4pt]
Alice defects ($A_2$)    &   ~~\$1M+\$1K,& 0    & ~~~\$1K,& \$1K           
\end{tabular}
\end{center}

\caption{Payoffs for a Prisoner's Dilemma problem. In each cell,
Alice's payoff is followed by Sue's payoff.}\label{fig-pd}

\end{figure}

As is traditional in Prisoner's
Dilemma problems, we assume that\vspace{-7pt}
\begin{enumerate}
\item[1)] Alice and Sue are motivated solely by monetary gain. When relevant,
      we will assume that their utility is linear in the money 
      gained.\vspace{-2pt}
\item[2)]
      Since Alice is focused exclusively on obtaining money, she has no concern 
      whatever for the welfare of Sue.  Similarly, Sue has no concern
      for the welfare of Alice.\vspace{-2pt}
\item[3)] Alice and Sue also have no concern for the welfare of whoever is
      the source of the money they may get.\vspace{-2pt}
\item[4)] Neither Alice nor Sue know much about the other's
          decision process (e.g., adherence to CDT or EDT).\vspace{-2pt}
\item[5)] Alice and Sue are both certain that they will not be in this situation
      again, and so have no motivation to try to influence the other's 
      decision in any future repetition.\vspace{-2pt}
\item[6)] Alice and Sue are both certain that the other will never be able
      to retaliate against them for anything they decide to do.\vspace{-2pt}
\item[7)] Alice and Sue are either certain that their actions will not become
      known to other people, or have no concern whatever with how their
      reputations will be affected by other people knowing of 
      their actions.\vspace{-2pt}
\item[8)] Both Alice and Sue know all the above, know that the other knows
      all the above, know that the other knows they know, etc.\vspace{-7pt}
\end{enumerate}
Note that the traditional names of ``cooperate'' and ``defect'' for
the possible actions of Alice and Sue, which I have used here, carry
moral connotations that are foreign to the assumed psychopathic nature
of these agents. Such moral considerations should not bias one's
judgement of what it is rational for these agents to do.\footnote{\fsp
One might wonder whether the motivations of Alice and Sue need be so
absolutely psychopathic. And indeed, it may be more realistic to suppose
that Alice has at least some preference for Sue to get \$1M, as long 
as this preference is less strong than her preference to get an
extra \$1K for herself. The CDT argument below for defecting would still apply,
and the only effect on the EDT argument below would be to recommend
cooperating for an even smaller shift in the probability that the other
cooperates.}

CDT recommends defecting to both Alice and Sue, since they each have
no control over what the other does, and regardless of what the other
does, they get \$1K more by defecting than by
cooperating. Unfortunately for them, if they both follow this advice,
they will both flee the country with only \$1K, when they would have
each had \$1M if they had both cooperated. Defecting does, however,
avoid the possibility of fleeing the country with nothing, as would be
the fate of one who cooperated while the other defected.

EDT may recommend that Alice and Sue cooperate,
if each takes their own action as indicative of what the other 
will do. Suppose that early in Alice's deliberations, when she thinks
she is equally likely to cooperate or defect, she thinks the probability 
that Sue will cooperate is $p$.  If the actions of Alice and Sue are 
positively associated, her subjective probabilities for
Sue cooperating conditional on her own possible actions should 
be\vspace{-2pt}
\beq
  \prs(S_1\giv A_1) \ =\ p+\delta\ \ \ \ \mbox{and}\ \ \ \
  \prs(S_1\giv A_2) \ =\ p-\delta
\label{eq-pd-delta}\eeq
for some $\delta>0$.  Alice's expected utilities for her two possible
actions are then\vspace{-2pt}
\beq
  V(A_1)\ =\ (p+\delta)\,\$1M\ \ \ \ \mbox{and}\ \ \ \
  V(A_2)\ =\ (p-\delta)\,\$1M\ +\ \$1K
\eeq
from which we get that\vspace{-5pt}
\beq
  V(A_1)-V(A_2) & = & 2\delta\,\$1M\ -\ \$1K
\eeq
which is greater than zero if $\delta > 0.0005$.  So,
if conditioning on Alice cooperating results in even a small shift
in Alice's subjective probability of Sue cooperating, EDT will recommend 
that Alice cooperate.

Lewis (1979) argues that the contrast in the recommendations of CDT
and EDT for Prisoner's Dilemma does not just \textit{resemble} how CDT
and EDT conflict in Newcomb's problem, but that for each agent,
Prisoner's Dilemma \textit{is} a Newcomb problem.  He regards each
agent as a ``simulation'' of the other --- perhaps not a very accurate
simulation, but good enough that cooperating is a sufficient
indication that the other will do likewise to justify cooperation according
to EDT.

Ahmed (2014, pp.~111--112) considers and rejects
an objection to viewing Prisoner's Dilemma
as a Newcomb problem
that is akin to the ``tickle defense'' for the smoking lesion problem (see
Section~\ref{sec-smoke}):\vspace{-8pt}
\begin{quotation}\noindent
\ldots suppose that Alice thinks that what causes
her and Bob [~Sue in the example I give~] to decide to take the 1,000 dollars
is some type of chancy mechanism that they have in common \ldots
won't the Tickle Defence apply here too? If, as you might think plausible, 
the mechanism influences Alice by operating on her \textit{inclination}
to take the 1,000 dollars, and if this is something for which she already has
pretty robust evidence, shouldn't her actual decision to take the 1,000
dollars be evidentially irrelevant to what Bob
does? Yes it should, \textit{if} that is how Alice takes the mechanism to 
operate. But what if, instead, she thinks that what she and Bob share
is a chancy mechanism of \textit{deliberation}? In that case, neither
knowledge of her own beliefs and desires at the outset
of reasoning, nor knowledge of her inclinations at any intermediate stage
of it, suffices to screen off her action from Bob's. For that, she
would also need to know in which option these beliefs and desires
eventuate. And she can only find out about that \textit{by} deciding what to do.
Prior to the decision, she must take either decision to be diagnostic of
her partner's, because indicative of a common reasoning process that 
concludes with it.\vspace{-8pt}
\end{quotation}
Accordingly, Ahmed considers that Prisoner's Dilemma problems can be 
analogous to Newcomb problems, and like them, exhibit a genuine 
conflict between the recommendations of CDT and EDT.\footnote{\fsp
Berm\'udez (2018) argues against Prisoner's Dilemma being equivalent
to a pair of Newcomb problems on the grounds that Prisoner's Dilemma is
a strategic choice problem in which by definition a player has no knowledge
of the other player's choice, contradicting the probabilistic dependence
between action and prediction in a Newcomb problem.  But this is merely
a terminological dispute, which says nothing about what Alice and Sue
should do in the scenario that this section opens with, or about whether
the recommendations of CDT and EDT conflict in such scenarios.}

Lewis (1979) analyses Prisoner's Dilemma scenarios in terms of the
average reliability of Sue's decision as a prediction of Alice's
decision (and the reverse, if we assume the situation is symmetrical).
If $R_1=\prs(S_1\giv A_1)$ and $R_2=\prs(S_2\giv A_2)$, then the
average reliability is $(R_1+R_2)\,/\,2$.  As Lewis notes, EDT will
recommend cooperating if this average reliability is greater than
$0.5005$, which is the same threshold as was found in
Section~\ref{sec-recommend} for Newcomb's problem, where I assumed
that $R_1=R_2=R$ (which a Predictor can arrange by suitable use of
randomization). The same threshold can be found from
equation~(\ref{eq-pd-delta}) above, since it gives $R_1=p\!+\!\delta$
and $R_2=1-(p\!-\!\delta)$, so $(R_1+R_2)\,/\,2=(1/2)+\delta$, which
is greater than 0.5005 when $\delta$ is greater than the threshold
of 0.0005 found above.

In Prisoner's Dilemma, $R_1=\prs(S_1\giv A_1)$ and $R_2=\prs(S_2\giv A_2)$ are
whatever they are, may not be equal, and cannot be changed,
so allowing $R_1$ and $R_2$ to differ is generally necessary.\footnote{\fsp
Not always, though. If $p$ happens to equal $1/2$ in the presentation 
above, then $\prs(S_1\giv A_1)=1/2+\delta$ and also 
$\prs(S_2\giv A_2)=1-\prs(S_1\giv A_2) = 1-(1/2-\delta) = 1/2+\delta$.}
This could be allowed in Newcomb's problem as well, but if $R_1$ and $R_2$
differ greatly, some intuitive judgements of whether one should take
one box or two might not hold. In particular, in such a Newcomb
problem, one of $R_1$ or $R_2$ might be less than $1/2$, while their
average is above $1/2$.  One might wonder whether supposed
adherents of EDT would really take only one box if $R_1=0.2$ and
$R_2=0.9$, with the average being $0.55$, well above the threshold of
$0.5005$ --- they would need to leave behind the \$1K in the
transparent box even though they believe there is an 80\% chance that
they will get nothing from the opaque box.

I suppose some EDT adherents \textit{would} take only one box in this
situation --- reasoning that it doubles their chance of getting \$1M
--- but somehow it seems to me that deciding to cooperate in an
analogous Prisoner's Dilemma situation is more likely. So even if
Prisoner's Dilemma is equivalent to Newcomb's problem for
\textit{rational} agents, it may not be for all agents.

More fundamentally, Prisoner's Dilemma differs from Newcomb's problem
in lacking \textit{individual} predictions.  Perhaps Sue can be seen
as a somewhat-good simulation of Alice, but if so, Sue is just as good
a simulation of any other embezzler of her type, such as Mary.  So the
``prediction'' of what Alice will do based on Sue's choice is the same
as would be predicted for Mary.

Due to this lack of individualized prediction, I consider Prisoner's
Dilemma to be more analogous to the impersonal Newcomb problem of
Section~\ref{sec-impn} than to the standard Newcomb problem.  I agree
with Ahmed (2014) that the recommendations of EDT and CDT conflict for
Prisoner's Dilemma problems, as they do for impersonal Newcomb
problems.  Unlike Ahmed, I consider the CDT recommendation to defect
to be the rational choice.

Suppose that Alice and Sue are members of some population of
embezzlers (or perhaps, of some hypothetical population of potential
embezzlers).  Suppose also that Alice sees herself as being typical of
this population, and that she knows nothing relevant about Sue's
deliberation process beyond her also being from this population. Then
if Alice cooperates, her subjective probability that Sue cooperates
will increase if she thinks that there are potentially some
incompletely-known common characteristics of the population they are
both members of (education level, for example) that may have
influenced her decision, and that if present would tend to lead Sue to
the same decision.\footnote{\fsp Note that it is essential for the
common factors to be incompletely known. If Alice knows
that each member of the population independently chooses whether to
cooperate, with a probability of cooperation known to her, then her
own decision will not affect her belief regarding whether Sue will
cooperate, regardless of what common physical factors may causally underlie
their decisions.}

The total effect of such common factors leads to some fraction, $q$,
of the population cooperating if they are ever in a Prisoner's Dilemma
situation.  Alice does not know $q$, since she does not have complete
knowledge of the presence or absence of possible common factors, nor
of their effects, but she will have some prior distribution for $q$,
before accounting for her own decision. If she cooperates, her
distribution for $q$ will shift towards higher values, since these
assign a higher probability to her decision to cooperate, which in
turn shifts higher her probability that Sue will cooperate (which is
the mean of her posterior distribution for $q$). The
reverse of course happens if Alice defects.\footnote{\fsp Suppose that
Alice's prior distribution for $q$ (before she makes a decision) is
Beta($a$,$a$). This prior can be roughly characterized as the
knowledge about $q$ that one would have after having seen $2a$
embezzlers in this situation, equally split between $a$ who cooperated
and $a$ who defected. The mean of the Beta($a$,$a$) distribution is
$a/(a\!+\!a)\,=\,1/2$, so $p$ in equation~(\ref{eq-pd-delta}) will be
$1/2$.  Alice's posterior distribution for $q$ after she cooperates
would be Beta($a\!+\!1$,$a$), with mean $(a\!+\!1)/(a\!+\!a\!+\!1)$,
which equals $p+\delta$ with $\delta=1/(4a\!+\!2)$. Her posterior
distribution after defecting would be Beta($a$,$a\!+\!1$), with mean
$p-\delta$.  One can compute that $\delta$ will be greater than
$0.0005$ if $a$ is less than $500$.  So EDT recommends cooperating as
long as Alice's prior knowledge of how people like Sue and herself
behave is equivalent to having seen less than 1000 instances of such
behaviour, equally split between cooperating and defecting.}

This is very similar to the causal situation in the impersonal Newcomb
problem of Section~\ref{sec-impn}, except that there the Participant
knows that $q$ is either $R$ or $1\!-\!R$.  As for the impersonal and
standard Newcomb problems, the complexity of the causal relationships
linking the decisions of the two participants in a realistic
Prisoner's Dilemma problem may lead to a lack of clarity of thought,
which might not be present if the causal situation were simpler, as
for a smoking lesion problem. When there is great uncertainty about
what causal factors affect your decision, you may be tempted to reason
according to EDT, even though with \textit{any} more detailed
knowledge of causal relationships you would see that CDT should be
applied.

Nevertheless, one might feel uncomfortable with the CDT recommendation
to defect, since if both Alice and Sue follow this recommendation they
end up much worse off than if they had both followed the EDT
recommendation to cooperate.  This is analogous to the ``Why ain'cha
rich?''  argument against taking two boxes in Newcomb's problem.
I don't regard this argument as valid for realistic Newcomb problems,
but it is at least somewhat troubling, and for fantastic Newcomb problems
involving thought simulation, taking two boxes is actually wrong.
However, as I argue in Section~\ref{sec-impn}, ``Why ain'cha rich?'' is
not a compelling argument against taking two boxes in the impersonal
Newcomb problem. 

And similarly, the fact that Alice and Sue end up
with only \$1K each if they both defect is not a good argument against
defecting in this Prisoner's Dilemma situation.
Suppose would-be embezzlers from some population are randomly paired
up, with each pair obtaining employment at some vulnerable company,
and then ending up in a Prisoner's Dilemma situation like that of
Alice and Sue.  Some fraction $q$ of these people cooperate (leave the
\$1M payment to the other unchanged). Do they do better than the
defectors? No, they do not. They end up \$1K worse off on average.

To see an advantage for cooperation, we would need to suppose that
there are multiple groups of would-be embezzlers, with the
fraction who are cooperators varying from one group to 
another. If each embezzler is paired with another embezzler from the
same group, it's possible that we \textit{would} see that
embezzlers who cooperate do better on average than those who defect,
since those who cooperate are more likely to come from a group
for which the fraction of cooperators is high, and hence are more
likely to be paired with a cooperator.

CDT still recommends defecting in this situation, given the
assumptions regarding Prisoner's Dilemmas listed earlier. And this
seems correct, since (we assume) you have no control over which group
you are a member of. The feeling that there is something wrong with
this seems to me to be a \textit{moral} intuition, that rejects the
psychopathic assumptions without which CDT need not recommend
defecting.

\subsection{Deciding whether to vote}\vspace{-8pt}

Ahmed (2014, p.~117--118) considers the decision whether to vote in a
large election to be analogous to a Prisoner's Dilemma problem (and
hence also a Newcomb problem, in his view), though with many players
rather than two.  I will argue here that, as for Prisoner's Dilemma,
the better analogy is to an impersonal Newcomb problem rather than a
standard Newcomb problem. Indeed, the connection of voting decisions
to the impersonal Newcomb problem is even more direct than for a
two-person Prisoner's Dilemma.

That many people vote in large elections might seem puzzling, given
that voting has a non-negligible cost (in time, at least) and has a
negligible chance of changing the result, so one might think that the
expected benefit is tiny. Possible explanations that have been
advanced include that people enjoy voting, or that they wish to signal
to others that they are responsible citizens. But let's assume that
those explanations are inadequate. If so, why does anyone vote?

Edlin, Gelman, and Kaplan (2007) demonstrate that if people have
altruistic motivations, there is no puzzle.  Voters may believe that
their side wining the election will benefit all (or at least most)
voters, with the benefit to others being something that they
themselves value.  The probability that one vote will change the
election result decreases in proportion to the number of voters, but
the total benefit increases in the same proportion, so the product of
these two --- the expected benefit of voting --- can be non-negligible
even for an election with many voters, perhaps outweighing the cost.
So, to make this an interesting problem in decision theory, let us
assume that voters have only selfish motivations. The benefit of their
side winning the election is then bounded, and hence the expected
benefit will approach zero in a large election.

Ahmed considers the possibility that voters in this situation are
applying EDT, taking their own decision to vote to be an indication that
other people with their political views will also vote. Deciding to
vote would thus be good news for them regarding the result of the election,
which is a reason to do it, according to EDT.  CDT would of course not
countenance this reasoning.

It is simplest to consider voting in a referendum on whether to enact
some policy, rather than an election for some office.  Suppose that
you favour the policy, and that the rules of the referendum are that
the policy is enacted only if more than half the people
\textit{eligible} to vote vote ``yes''. (So opponents of the
policy needn't bother to vote, since not voting has the same effect as
voting ``no''.) Some unknown fraction, $q$, of those eligible will
vote ``yes''. The referendum will pass if $q$ is greater than 1/2.

If we ignore the negligible chance of your single vote tipping the result,
this is exactly the same as an impersonal Newcomb problem, except
there you also know that $q$ is equal to either $R$ or $1\!-\!R$.  In
an impersonal Newcomb problem, the opaque box contains \$1M if
$q>1/2$; in the voting problem, the referendum passes if $q>1/2$.  In
both problems, your own choice changes what you believe $q$ is likely
to be, supposing that you do not already know $q$, and that you see
yourself as representative of the population. So EDT will recommend
one-boxing and voting, whereas CDT will recommend two-boxing and not
voting.

I will not again review the arguments for why I think CDT gives the
right decision here. Instead, I will present another voting example
in which EDT may appear much less appealing.

Consider an election for some office, with two candidates: one who is
``liberal'', and one who is ``conservative''.  People eligible to vote
vary in their personality traits, in particular in how ``tolerant''
and ``conscientious'' they are. You know that being tolerant inclines
a voter to be liberal, and being conscientious inclines a voter to be
conservative.  (Recall, though, that we assume voters are motivated by selfish
concerns, not altruistically.)  You yourself are both tolerant and
conscientious (and aware of this), but the former has had more effect
on your politics, so you support the liberal candidate.

You also know that conscientious and non-conscientious people tend to
decide whether to vote in different ways --- the latter more
impulsively than the former.  You have little idea about the fraction
of conscientious people who will decide to vote in this election, and
consider that your own decision whether to bother to vote is indicative
of what other conscientious people will do, but that your decision
says nothing about how many non-conscientious people will vote.  Should
you vote?

EDT says you shouldn't vote. Doing so would increase your estimate of
the fraction of conscientious people who will vote, which is bad news,
since most conscientious people support the candidate you oppose. CDT
likely also recommends not voting, but for what seems like a much more
acceptable reason --- voting has a cost, while its expected benefit is
very small. (And if the cost were sufficiently small, CDT might recommend
voting even while EDT recommends against voting.) The EDT reasoning
seems bizarre, once its recommendation no longer aligns with
intuitions that are tinged with moral judgements.

\section{The Newcomb insurance problem}\vspace{-10pt}\label{sec-insure}

Ahmed (2014, Section~7.4.3) discusses an interesting two-stage variant
of Newcomb's problem, which he refers to as the Newcomb insurance
problem.  Ahmed's primary aim is to defuse the ``You could have done
better'' argument against EDT's recommendation that you should take
one box in the standard Newcomb problem, which notes that whatever the
Predictor did, you could have done better by taking both boxes (a
reflection of two-boxing dominating one-boxing in CDT terms). In the
Newcomb insurance problem, Ahmed claims that following CDT leads to a
decision for which one can also say that ``You could have done
better', and hence that ``You could have done better'' cannot be a valid
argument for CDT over EDT.  Also worrying is that applying CDT
separately at each stage in the Newcomb insurance problem is
apparently inconsistent with using CDT to select a strategy for
both stages.\footnote{\fsp Ahmed (2014) elaborates on this with regard
to a similar problem in his Section~8.3.\vspace{-15pt}}

I will argue here that the Newcomb insurance problem is impossible to
implement in any non-fantastic manner.  If it is implemented by the
fantastic means of highly-accurate thought simulation, applying CDT is
less straightforward than applying EDT, but in the end both CDT and
EDT recommend the sensible decision, for which one cannot say that you
could have done better, and for which there is no inconsistency
between strategic and sequential choices.

\subsection{\hspace*{-8pt}Specification of the Newcomb insurance 
problem}\vspace{-6pt}

The first stage of the Newcomb insurance problem proceeds as for
the standard Newcomb problem, except the contents of the opaque box
are either \$100 or \$0, and the transparent box is always empty.  In
the second stage, conducted before the opaque box is opened, the
Participant must either bet that the Predictor's prediction was right,
or bet that it was wrong. The payoffs for these bets are such that
both bets are fair if the Predictor's reliability is $R=0.75$, and
betting that the Predictor is right is better when $R$ is larger than
this.  Participants know before choosing to take one box or two that
they will be required to bet one way or the other after their choice.

Ahmed's specification of the problem is a bit odd.  First, as he
notes, it is not essential to his points that the first stage be a
degenerate Newcomb problem with an empty transparent box.  If the
transparent box is empty, CDT provides no reason at the first stage to
take it along with the opaque box (though also no reason not to take
it), obscuring any connection to the standard Newcomb problem.
Accordingly, I will also consider a version in which the transparent
box always contains \$$b$, with other payoffs unchanged.

Ahmed also says ``The predictor makes no prediction about stage 2, of
which he neither knows nor cares.'' This is a strange remark, firstly
because it \textit{seems} irrelevant to his subsequent arguments (it's
never explicitly referenced). But more profoundly, the Predictor must
surely know about the second stage, even while making no prediction
about it. If prediction is based on data regarding people's past
choices, the Predictor must ensure that these people knew they would
be making a bet in a second stage, as this will certainly change
some people's decisions. If prediction is by thought simulation, the
simulated person must know about the second stage, so only by
deliberate ignorance could the Predictor not know.  Ahmed seems to
have assumed that the first stage of a
Newcomb insurance problem is the same as a standard Newcomb problem,
so if one believes standard Newcomb problems are possible, then
Newcomb insurance problems will also be possible, and moreover can
be realized in exactly the same way as standard Newcomb problems.

Figure~\ref{fig-insure} summarizes payoffs for both Newcomb insurance
versions. One can note that for the ``Take 1, Bet wrong'' and the
``Take 2, Bet right'' strategies, the total payoff does not depend on
what the Predictor predicted. This is the motivation for the
``insurance'' nomenclature.

\begin{figure}

\hspace*{-3pt}%
\begin{tabular}{lr@{~~}r@{~~}rr@{~~}r@{~~}r@{~~~~~~}lr@{~~}r@{~~}rr@{~~}r@{~~}r}
\multicolumn{1}{r}{\mbox{Prediction:\hspace{-10pt}}}
 & \multicolumn{3}{c}{~Take 1} 
 & \multicolumn{3}{c@{~~~~~~}}{Take 2}
 & \multicolumn{1}{r}{\mbox{Prediction:\hspace{-10pt}}}
 & \multicolumn{3}{c}{~Take 1} 
 & \multicolumn{3}{c}{Take 2} \\[6pt]
Actions: 
& S1 & S2 & Tot
& S1 & S2 & Tot
& Actions:
& S1 & S2 & Tot
& S1 & S2 & Tot \\[6pt]
Take 1, Bet right & 100 & 25 & 125  & 0 & -75 & -75 & 
Take 1, Bet right & 100 & 25 & 125  & 0 & -75 & -75
\\
Take 1, Bet wrong & 100 & -25 & 75  & 0 & 75 & 75 & 
Take 1, Bet wrong & 100 & -25 & 75  & 0 & 75 & 75
\\
Take 2, Bet right & 100 & -75 & 25  & 0 & 25 & 25 & 
Take 2, Bet right & 100$\!+\!b$ & -75 & 25$\!+\!b$  & $b$ & 25 & 25$\!+\!b$
\\
Take 2, Bet wrong & 100 & 75 & 175  & 0 & -25 & -25 & 
Take 2, Bet wrong & 100$\!+\!b$ & 75 & 175$\!+\!b$  & $b$ & -25 & -25$\!+\!b$
\\[4pt]
\multicolumn{7}{c}{(a)} & \multicolumn{7}{c}{(b)}
\end{tabular}

\caption{Payoffs for two versions of the Newcomb insurance problem:
(a) the version of Ahmed (2014) in which the transparent box is
always empty, and (b) a version in which the transparent box always
contains~\$$b$. For each version, the left column lists the four
strategies obtained by combining taking 1 or 2 boxes with betting that
the Predictor is right or wrong.
The payoffs of these strategies in stage~1 (S1)
and stage~2 (S2) along with their sum (Tot) are then listed, for
situations in which the Predictor predicted that you will take 1
box, or take 2 boxes.}\label{fig-insure}

\end{figure}

\subsection{\hspace*{-8pt}Strategies for the Newcomb insurance 
problem and their analysis}\vspace{-6pt}

We might analyse the Newcomb insurance problem in terms of strategies,
which specify the choices to make at both stages, assuming that
whatever strategy is chosen before the first stage will be followed
without revision in both the first and second stages. From this
viewpoint, one can see that the ``Take~1, Bet right'' strategy is
dominated by ``Take~2, Bet wrong''. In Ahmed's version, if the
Predictor predicted you will take 1~box, the latter strategy pays
\$175, versus \$125 for the former; if instead the Predictor predicted
you will take 2~boxes, it pays $-$\$25 versus $-$\$75. (The
differences are \$$b$ greater if the transparent box contains $b$
dollars.) One can also see that ``Take~2, Bet right'' is dominated by
``Take~1, Bet wrong'', with the latter paying \$75, versus \$25 for the
former, regardless of what the Predictor predicted. ``Take~2, Bet
right'' remains dominated by ``Take~1, Bet wrong'' if the transparent
box contains $b$ dollars, provided $b<50$.

Within this analysis in terms of strategies, CDT will never recommend
either of the dominated strategies. However, unlike the standard Newcomb
problem, there are two non-dominated choices. Deciding between them
using CDT requires using your subjective probabilities of the two
situations (that the Predictor predicted you would take 1~box, or
predicted that you would take~2 boxes). But this will prove problematic for
Newcomb insurance.

Imagine actually trying to choose between the two non-dominated
strategies of ``Take~1, Bet wrong'' and ``Take~2, Bet wrong'', knowing
that the Predictor's reliability is $R=0.8$. Suppose you initially
believe that the prediction was probably that you would take 1 box,
and are therefore tentatively inclined towards ``Take~2, Bet wrong'',
which on that belief will likely have a payoff of \$175 (in Ahmed's
version).  As you firm up this decision, however, you will shift to
believing that the prediction was that you would take 2 boxes (with
probability approaching $R=0.8$ as you get closer to acting on this
decision).  But if so, ``Take~2, Bet wrong'' will have a payoff of
$-$\$25, whereas ``Take 1, Bet wrong'' will have a payoff of \$75. So
you change your mind, and tentatively favour ``Take 1, Bet wrong''.
But this leads you to change your belief again, back to thinking the
prediction was probably that you would take 1 box, which shifts you
back to favouring ``Take~2, Bet wrong''.  And so forth.  There is no
stable state of belief and tentative decision that you can settle on.

This contrasts with CDT applied to the standard Newcomb problem, in
which your initial belief that the Predictor probably predicted you
will take 1 box will decline as your deliberations lead you towards
taking 2 boxes, but this belief change does not change the action that
CDT recommends, since taking 2 boxes dominates taking 1 box.

Next, let us consider a sequential analysis of the problem.  Working
backwards, suppose you have made some choice at stage~1, either taking
1 box or 2 boxes, and must now decide whether to bet that the
Predictor correctly predicted your choice, or did not. If 
condition~(k) of Section~\ref{sec-uncont} holds for this 
Newcomb-like problem, you should believe that the Predictor
was right with the advertised probability $R$. If $R$ is greater than
0.75, your decision (by either CDT or EDT) should be to bet that the
Predictor was right --- regardless of your choice in stage~1, the
expected payoff for this bet is $R\times25\, -\, (1\!-\!R)\times75\, =\,
100\times R\, -\, 75$, which is greater than zero if $R>0.75$, whereas
betting that the Predictor was wrong has expected payoff
$(1\!-\!R)\times75\, -\, R\times25\, =\,
75\, -\, 100\times R$, which is less than zero if $R>0.75$.

So we now see another difficulty --- the choice that a sequential CDT
decision maker will make in stage~2 is the second part of either the
``Take 1, Bet right'' strategy or the ``Take 2, Bet right'' strategy,
both of which we have seen are dominated, and hence inadmissible for a
CDT decision maker!

Leaving that issue aside, there remains the question of what a CDT
sequential analysis should recommend at stage~1. Ahmed considers both
``myopic'' decisions and ``sophisticated'' decisions. The myopic
procedure is to select one of the four strategies, and then do the
first-stage action of that strategy. Since the non-dominated
strategies all have ``Bet wrong'' as the second action, and we have
seen that CDT will recommend ``Bet right'' at stage~2, it seems that
the selected strategy will not actually be carried out in full. The
sophisticated procedure instead accounts for what will be done at
stage~2, and selects only among those strategies for which the second
action is ``Bet right''.

The sophisticated strategy apparently leads to a stable decision for
Newcomb insurance. If you initially think the prediction is likely to
be that you would take 1 box, the sophisticated procedure will favour
taking 1 box, since if that is the prediction, ``Take 1, Bet right''
has payoff (in Ahmed's version) of \$125, versus \$25 for ``Take 2,
Bet right''. Having tentatively decided this, your confidence that the
prediction was that you would take 1 box increases, reinforcing this
decision. On the other hand, suppose you initially think it likely
that the prediction was that you would take 2 boxes. Then you would
tentatively decide to take 2 boxes, since if that was the prediction,
``Take 2, Bet right'' has payoff \$25, versus $-$\$75 for ``Take 1,
Bet right''. Again, this tentative decision increases your confidence
that the prediction was that you would take 2 boxes, and hence
reinforces your tentative decision.  So, as Ahmed concludes, your
decision is determined by your initial belief that the prediction was
that you would take 1 box, with ``Take 1, Bet right'' favoured if this
is greater than 0.5.  Your initial belief about the prediction will be
influenced by various subjective factors, but such subjectivity is not
an issue for either EDT or CDT.\footnote{\fsp One might wonder whether
this procedure is actually stable, though, since if you reflect on
your deliberations, you will realize that your decision depends
crucially on your initial beliefs, which may be based on little
consideration of the matter. If one takes
deliberation to be a fundamental aspect of human decision-making, 
initial impressions before deliberation 
should not, and likely will not, determine the conclusion in this way. 
I will not pursue this further, though, since my conclusion will be that the
scenario is impossible in any case.}

This result --- that a sophisticated sequential application of CDT
leads to either ``Take 1, Bet right'' or ``Take 2, Bet right'' --- is
the basis of Ahmed's claim that if you follow CDT in this situation,
one can say ``You could have done better'' --- since as we've
seen, these strategies are dominated by ``Take 2, Bet wrong'' and
``Take 1, Bet wrong'', respectively, which would have resulted in a
\$50 higher payoff, regardless of what the Predictor predicted.\footnote{\fsp
As Ahmed acknowledges, EDT is also subject to the ``You could have done better''
criticism. His point is to defuse this by showing that this criticism 
does not distinguish EDT from CDT.\vspace{-20pt}}

Ahmed also says that if you apply the myopic procedure, your stage~1
decision will be the opposite --- taking 1 box if you initially think
it likely that the prediction is that you would take 2 boxes, and
taking 2 boxes if the reverse.  And if you assume that your
initial belief remains unchanged, this is true --- ``Take~1, Bet
wrong'' has payoff \$75 if the prediction was that you will take 2
boxes, versus $-$\$25 for ``Take 2, Bet wrong''. However, tentatively
deciding to take 1 box will change your beliefs, moving them towards
thinking you are likely to take 1 box, after which the preferred
decision will be to take 2 boxes, since ``Take 2, Bet wrong'' then has
higher payoff.  This is the same instability seen above when
considering choosing between strategies in a non-sequential manner.

The discussion in Ahmed (2014) does not address this instability 
issue.\footnote{\fsp In
his footnote 34, Ahmed (2014) argues that, for a problem
similar to Newcomb insurance, the sophisticated procedure produces a
stable decision at stage~1.  He implicitly assumes elsewhere that this
is also true for the myopic procedure applied to Newcomb insurance,
but this is not true.} Decision instability has been discussed by others,
however.  Spencer and Wells (2019), for example, try to address instability
by disallowing decision rules based on quantities that the decider does
not have ``stable access'' to:\vspace{-8pt}
\begin{quotation}\noindent
An agent has \textit{stable access} to $Q$ if and only if there is an
option $A\in\cal A$ such that (i) the agent is in a position to know
of $A$ that it maximizes $Q$, and (ii) conditional on $A$, the agent still
is in a position to know of $A$ that it maximizes $Q$.\vspace{-8pt}
\end{quotation}
Causal expected utility (the $U$ function of Section~\ref{sec-recommend})
is not stably accessible in the Newcomb insurance problem when CDT is applied
either to strategies, or to the choice at stage~1 when 
using the myopic procedure.

I will discuss decision instability further when looking at fantastic
realizations of Newcomb insurance using thought simulation.  First,
though, I will discuss whether instability renders the Newcomb
insurance problem impossible to realize by non-fantastic means.

\subsection{\hspace*{-8pt}Imagining a non-fantastic Newcomb insurance 
problem}\vspace{-6pt}

The Predictor for a Newcomb insurance problem could make predictions
based on categories in the same way as for the standard Newcomb
problem, as described in Section~\ref{sec-pr}.  Despite the
potential for instability, Participants (or purported
Participants, when the Predictor is deceptively gathering data) will
in the end choose either one box or two boxes (as well as bet one way or
the other, though the Predictor won't predict this). And just as for the
standard Newcomb problem, common sense tells us that the proportion of
Participants who choose one box is likely to be different in different
categories (keeping in mind that not all Participants will be inclined
to use an unstable decision method), providing scope for successful
prediction.

For the Newcomb insurance problem, $R$ must be greater than $0.75$
--- otherwise betting the prediction is right at stage~2 would
not be favoured.  If the observed proportions of one-boxers in each
category (the $q_i$) do not allow for such an $R$, the payoffs could
be changed to allow a lower $R$, and the Predictor could again gather
data on potential Participants' behaviour, when they are told they
face this modified problem.  It's possible that no suitable $R$ can be
found, with this process just converging on $R=0.5$. Whether this
happens is an empirical question, for which the answer may well be
different for different populations of Participants.  I think, though,
that one can easily imagine that the population of participants allows
for this, which is enough for the Newcomb insurance problem to seem
relevant.

There thus seems to be no fundamental barrier to realizing a Newcomb
insurance problem that satisfies condition~(i) of
Section~\ref{sec-uncont} --- that the Predictor's overall accuracy be
$R$ for both one-boxers and two-boxers.  But what about condition~(k)?
This requires that, after their choice but before boxes are opened,
all Participants (or at least you, and others who are employing
somewhat sensible procedures that we are interested in) believe with
probability $R$ that the Predictor predicted their action correctly,
even given their private knowledge, including knowledge of their own
deliberations.

As discussed above in Section~\ref{sec-bel}, it is plausible, but
not guaranteed, that condition~(k) could hold for the standard Newcomb
problem when prediction is done using categories. One might doubt this
for Participants who were quite uncertain before finally choosing ---
might their choice be effectively random, and hence
unpredictable? For the standard Newcomb problem, though, such
vacillation is likely to be fairly rare. Tentatively deciding to take
2 boxes does not affect the typical argument for taking 2 boxes (if
anything, it reinforces it). Tentatively deciding to take 1 box might
lead one to think taking 2 boxes is better (you're more sure that the
opaque box contains \$1M), but presumably most one-boxers are immune
to this, and in any case this would just change the decision, not produce
uncertainty.

But for the Newcomb insurance problem, the instability of CDT for
choosing among strategies or for myopic sequential choice is a
fundamental issue.  It is to be expected that a
substantial fraction of Participants will favour a decision method
that is unstable, and one cannot dismiss these Participants as not
being of interest, since elucidating whether decision procedures such
as CDT are valid is the motivation for discussing the problem.

So. Suppose that you are a Participant in a Newcomb insurance problem.
You have been presented with convincing evidence that the Predictor
has accuracy, $R$, greater than 0.75, and that you should rationally
believe that the prediction made is correct with this probability,
even conditional on your private information regarding your 
deliberations.\footnote{\fsp
Perhaps you were informed by someone you consider reliable that
past Participants were interviewed afterwards about their
deliberations (whether they knew about CDT and EDT, whether they
tried to apply CDT or EDT, whether they
were at some point puzzled about what to do, whether they almost made
a decision but then changed their mind, and so forth), the
fractions of accurate predictions were computed for various categories
based on these responses, and all these fractions were found to be
close to $R$.} Suppose you decide that the rational approach is to
choose one of the non-dominated strategies --- either ``Take 1, Bet wrong''
or ``Take 2, Bet wrong'' --- with the feeling that surely your future
self will follow your lead.\footnote{\fsp Such a feeling seems sufficient,
and not implausible.
Actually binding your future self is not necessary for this scenario, 
and would be considered by Ahmed (2014, p.~209) to make the problem no longer
a Newcomb insurance problem.}
You think that which of these strategies is best depends on the 
probability that the Predictor predicted you will take one box.
You realize that this leads to the instability described above.

You might think that randomizing your choice between the two non-dominated
strategies would be a reasonable action at this point. But let us assume that
explicit randomization (for instance, by a coin flip) is not 
allowed.\footnote{\fsp Randomization is not discussed by Ahmed (2014),
but is disallowed for a related problem by Spencer and Wells (2019):
``We assume that the agent facing \textit{The Frustrator} cannot play
a mixed strategy. Perhaps the agent is unable to randomize her choice,
or perhaps it is simply unwise to play a mixed strategy, since the Frustrator
is very good at detecting whether an agent is playing a mixed strategy
and punishes the agent severely for doing so.''} Somewhat more fantastically,
suppose that mental randomization (discussed in Section~\ref{sec-rand})
is also disallowed.\footnote{\fsp Without assuming fully-fantastic
levels of brain observation, perhaps such mental processes produce a distinctive
pattern of brain waves, which with current technology 
can be detected by EEG in a fairly non-invasive manner. This is probably
not true, but can be imagined to be true without straining credulity too
much.\vspace{-15pt}} So how do you ever come to a decision?

One must assume that there is some mechanism for ensuring a decision
within a finite time. Here is one possibility. There is a timer
showing a countdown from 10~hours at the start of deliberations to
zero at decision time, a display showing ``1'' or ``2'' which when the
timer reaches zero indicates the number of boxes that will be taken,
and a button under the display that when pressed will change ``1'' to
``2'' or ``2'' to ``1''. You can tentatively indicate a decision by
pressing the button as needed for the display to show the desired
number of boxes. Pressing this button again before the timer reaches
zero will change this tentative decision.

If you are applying an unstable decision procedure with this setup,
the instability will manifest as you pressing the button again and
again, as each tentative decision leads you to conclude that the other
choice is better.  Of course, it's ridiculous to think that an actual
Participant would press the button once every second or so for 10~hours
straight.  Sheer exhaustion would prevent this, if nothing else did.
But also it will take only a few such reversals for you to realize
what's happening, and start reasoning in terms of what the end effect
of such reversals might be.  You might think that you will press the
button many times in the minutes and seconds just before the timer
reaches zero.  If so, you may think that the final decision will be
effectively random, depending on the exact timing of the final button
presses.

Rapid button presses just before the timer reaches zero can be seen as
violating the requirement in condition~(f) of Section~\ref{sec-uncont}
that the Participant have ample time to consider their decision.  We
can modify the procedure to avoid this ---  we remove the timer's
display of how much time is left, and instead have it ring a bell when
only ten minutes remain. The Participant can press the button in these
last minutes to switch between choosing to take 1~box or to take
2~boxes, but they will not know precisely when the timer will reach
zero, at which time the currently-selected choice becomes
final. (I assume that a Participant's internal sense of
time is not able to judge when ten minutes have passed to within a
few seconds.)

What will you think during these final minutes? Your thoughts will
change over time, as your knowledge, $B$, which includes memory of
past thoughts, changes. At any time, you will not be sure
how many boxes you will end up taking. Let $q$ be your subjective
probability that you will end up taking one box, $\prs(A_1\,|\,B)$. 
You will also not be
sure whether the Predictor predicted that you will take one box. Let
$p$ be your subjective probability that you were predicted to take one
box, $\prs(S_1\,|\,B)$. If you have good
evidence that the Predictor has accuracy $R$, and that condition~(k)
of Section~\ref{sec-uncont} holds, then as $p$ and $q$ change, they 
will at each time be rationally related by\vspace{-2pt}
\beq
  p \ \ =\ \ \prs(S_1\,|\,B) & = & \prs(S_1\,|\,A_1\and B)\prs(A_1\,|\,B)\ +\ 
          \prs(S_1\,|\,A_2\and B)\prs(A_2\,|\,B)
  \nonumber\\[4pt]
  & = & Rq\ +\ (1\!-\!R)(1\!-\!q) \ \ =\ \ (2R\!-\!1)q\ +\ (1\!-\!R)
\label{eq-pfromq}
\eeq
Note that $R>1/2$, so $(2R\!-\!1)>0$, and hence an increase in $q$ leads to
an increase in $p$.

Suppose that at some time after the bell sounds your current tentative
decision is to take 1~box (the display above the button shows
``1''). You have subjective probabilities $p$ and $q$ that are related
according to equation~(\ref{eq-pfromq}). What will you do at that moment
(say, in the next second)? Will you press the button so that your
tentative decision switches to taking 2~boxes?

If you are selecting between strategies using CDT, you should decide whether
to press the button based on the expected causal effect of
the ``Take 1, Bet wrong'' strategy versus ``Take 2, Bet wrong''. I
will compute these expected payoffs for the version of the problem in which the
transparent box contains \$$b$ dollars (with $b=0$ giving Ahmed's
version).  The ``Take 1, Bet wrong'' strategy has expected payoff 75 regardless
of $p$. ``Take 2, Bet wrong'' has expected payoff\vspace{-3pt}
\beq
  (175\!+\!b)p\ +\ (-25\!+\!b)(1\!-\!p) & = & 200p - 25+b 
  \label{eq-ppay} \\[-20pt]\nonumber
\eeq
If this is greater than 75, you will have reason to press the button,
switching your tentative decision to ``2'' (at least for a
few seconds); if it is less than 75, you will have reason to not press
the button, leaving the tentative decision as ``1''. 
You will be indifferent to pressing the button if these payoffs are equal --- 
that is, if\vspace{-4pt}
\beq
   75 & = & 200p - 25+b \ \ =\ \ 200\,((2R\!-\!1)q + (1\!-\!R))\ -\ 25\ +\ b
\\[-20pt]\nonumber
\eeq
Here, $p$ has been replaced using equation~(\ref{eq-pfromq}). Solving for
$q$ gives an equilibrium value of
\beq
  q^* & = & {1 \over 2}\,{R-1/2-b/200 \over R-1/2} \label{eq-equilibq}
\eeq
Note that $q^* < 1/2$ when $b>0$, and $q^*=1/2$ when $b=0$. The corresponding
equilibrium value for $p$, found using 
equation~(\ref{eq-pfromq}) (or alternatively by equating (\ref{eq-ppay})
to 75), is\vspace{-4pt}
\beq
  p^* & = & (2R\!-\!1)q^*\ +\ (1\!-\!R)
  \ \ = \ \ 1/2 - b/200 \\[-20pt]\nonumber
\eeq
So another way of seeing whether pressing the button is favoured is to
compare your current $q$ or $p$ (either one) to its equilibrium value.

If the timer reaches zero during a brief time following this decision
whether to press the button or not, whatever your tentative decision
is will become final. Otherwise, you will need to consider how
these events (now part of $B$) will have changed your subjective
probabilities $q$ and $p$.

If you decided not to press the button, leaving ``1'' as your
tentative decision, and the timer does not go to zero immediately,
your $q$ and $p$ probabilities should increase --- first, because the
tentative decision is still ``1'' and the passage of time increases
the chance that the timer will go to zero in the near future, and
second, because not having pressed the button when ``1'' was displayed
should increase your feeling that you will let ``1'' remain in future,
hence increasing your estimate of the relative amount of time your
tentative decision will be ``1'' rather than ``2'', and
therefore increasing the chance that the final decision will be to take 1~box.

On the other hand, if you pressed the button to switch to a tentative
decision of ``2'', your $q$ and $p$ probabilities should decrease. If the
timer goes to zero in the near future, the final decision will now be to
take 2~boxes. And switching to ``2'' in this instance should increase
your estimate of how inclined you will be to do so in future.

So, if your $q$ and $p$ probabilities were below their equilibrium
values, and accordingly you left the tentative decision as ``1'', these
probabilities will increase, perhaps moving above their equilibrium
values. And if $q$ and $p$ were above their equilibrium values,
prompting you to press the button to switch to a tentative decision of
``2'', these values will decrease, perhaps to below their equilibrium
values.  The combined effect of these changes (and corresponding ones
after you have decided whether to switch from ``2'') is to keep $q$
and $p$ near their equilibrium values. There will be small deviations
from this equilibrium, whose magnitude will be smaller if you more
rapidly update your assessments of $q$ and $p$ (deciding to press the
button or not after each re-assessment).

After a while, you may realize that the end effect of these
moment-to-moment decisions is simply to press the button fairly
frequently, switching between ``1'' and ``2'', while spending a
fraction $q^*$ of the time with ``1'' as the tentative decision and a
fraction $1\!-\!q^*$ of the time with ``2''. Since the timer goes to
zero at some time that is unknown to you (within a range of a number of
seconds), the final decision is from your point of view random, with
probability $q^*$ of taking 1~box.

Choosing among non-dominated strategies using CDT has turned out to be
a mental randomization procedure. Randomization was supposed to be
prohibited. But to disallow randomization in this context is
unreasonable, since CDT is one of the decision methods whose
properties we are trying to elucidate using the Newcomb insurance
problem, and reasoning on the basis of CDT leads to effective randomization,
even with no explicit decision to randomize.

We could modify the conditions for the Newcomb insurance problem to
not require that condition~(k) of Section~\ref{sec-uncont} be 
satisfied, but instead impose only the weaker condition~(k$'$) of
Section~\ref{sec-rand}. Equation~(\ref{eq-pfromq}) is then no longer
valid. Instead, we can find a relationship between $p$ and $q$ by
assuming that, as above, you will continually change your tentative decision,
and hence your choice is effectively the result of an unpredictable
mental randomization procedure. Your choice will then be
independent (in your subjective estimation) of the Predictor's 
prediction.\footnote{\fsp
The independence of whether you choose
1~box from whether the Predictor predicted you will choose 1~box
may not hold early in your deliberations, but only after you have
decided to use CDT to choose among strategies, and you have then realized
that you are effectively randomizing your choice by continually
changing your tentative decision.} 
This lets us conclude that\vspace{-4pt}
\beq
  R & = & pq\ +\ (1\!-\!p)(1\!-\!q) \\[-20pt]\nonumber
\eeq
from which it follows that\vspace{2pt}
\beq
  p \ = \ {(1\!-\!q)-R \over 1-2q}
    \ = \ {1\over2}+{1\over2}{R-1/2 \over q-1/2}
  \ \ \ \ \ \mbox{and}\ \ \ \ \
  q \ = \ {(1\!-\!p)-R \over 1-2p} 
    \ = \ {1\over2}+{1\over2}{R-1/2 \over p-1/2}
\label{eq-pfromq2}
\eeq
These imply that either $p$ and $q$ are both in $[0,1\!-\!R]$ or
both are in $[R,1]$.\footnote{\fsp
To avoid division by zero, $q$ must not be $1/2$. If $q<1/2$, then
$p \ge 0$ implies $(R-1/2) \le -(q-1/2)$, and hence $q \le 1-R$.
Also, $q<1/2$ implies that $p<1/2$.
If $q>1/2$, then $p \le 1$ implies $(R-1/2) \le (q-1/2)$, and hence $q \ge R$.
Also, $q>1/2$ implies that $p>1/2$. The same statements hold with $q$ 
and $p$ swapped. In combination, these imply the claim made above.
}

You may suspect that, as above, you might frequently press the button to
change your tentative decision based on deviations from equilibrium
values for $q$ and $p$. The equilibrium value for $p$ can be found by
equating the expected payoff from ``Take 1, Bet wrong'' (75) to that
of ``Take 2, Bet wrong'' ($200p-25+b$), which gives  $p^* \ =\
1/2-b/200$, the same as when we assumed condition~(k).

In the present scenario, however, we have the constraint that either
$p^* \ge R$ or $p^* \le 1\!-\!R$. If $p^* \ge R$, then\vspace{-1pt}
\beq
1/2-b/200 \ \ge\ R, \ \ \ \mbox{which implies}\ \ \ b\ \le\ 100-200R
\eeq
whereas if $p^* \le 1\!-\!R$, \vspace{-4pt}
\beq
1/2-b/200 \ \le\ 1\!-\!R, \ \ \ \mbox{which implies}\ \ \ b\ \ge\ 200R-100
\eeq
Recall that for the odds to favour betting that the Predictor was right,
we must have $R>3/4$. The above conditions then imply that either
$b<-50$ or $b>50$. 

So when mental randomization is considered unpredictable, with 
condition~(k) replaced by condition~(k$'$), Ahmed's version of the Newcomb 
insurance problem (with $b=0$) is impossible to realize, as is any similar
problem with $0<b<50$.
If we set $b<-50$, the transparent box will
contain negative money, which makes little sense for a Newcomb-like
problem.\footnote{\fsp
In detail, when $b<-50$, ``Take 1, Bet right'' dominates ``Take 2, Bet wrong'',
and ``Take 1, Bet wrong'' dominates ``Take 2, Bet right'', so CDT will
recommend that you take 1~box (not randomize).  Having realized that you will 
take 1~box, your subjective probability that the Predictor predicted that
you will take 1~box should be $R$, which is assumed to be greater than $3/4$.
The expected payoff from ``Take 1, Bet right'' will then be $125R-75(1\!-\!R)$,
which is greater than the payoff of 75 for
``Take 1, Bet wrong''. After choosing to take 1~box, your probability that
the Predictor predicted taking 1~box will be $R$, and
CDT will recommend betting that the Predictor was right, consistent with
the ``Take 1, Bet right'' strategy.  EDT would 
also recommend taking 1~box and betting that the Predictor was right.
}
And if we set $b>50$, the ``Take 1, Bet wrong'' strategy no longer dominates 
``Take 2, Bet right'', which also changes the problem to one that is no longer
of much interest.\footnote{\fsp
When $b>50$, ``Take 2, Bet right'' dominates ``Take 1, Bet wrong'', rather than
the reverse, and ``Take 2, Bet wrong'' still dominates ``Take 1, Bet right''.
So CDT recommends taking 2~boxes (not randomizing).  After realizing you
will take 2~boxes, your
subjective probability that the Predictor correctly predicted that you will
take 2~boxes should be $R$. The expected payoff of ``Take 2, Bet wrong''
will be $(-25\!+\!b)R+(175\!+\!b)(1\!-\!R)=175-200R+b$, which is less
than the $25+b$ payoff of ``Take 2, Bet right'' when $R>3/4$.  After your 
decision to take 2~boxes, your probability that the Predictor was right will 
be $R$, so CDT will recommend betting that the Predictor was right, consistent 
with the ``Take 2, Bet right'' strategy. EDT would recommend taking 1~box and
betting the Predictor was right, a difference from CDT analogous
to that for the standard Newcomb problem.
}

So the only versions of Newcomb insurance that might be realizable are
those in which those Participants who follow CDT, and hence seem to be
engaging in mental randomization, make final choices that are actually
predictable, with condition~(k) holding. Might this be possible using
a non-fantastic method, such as prediction based on categories? I
think not. The exact times of button presses will be affected by many
aspects of a Participant's thoughts and environment --- by
distracting thoughts about what to eat for dinner, for example, or by
fidgeting in an uncomfortable chair.  Due to the instability inherent
in this problem, when it is analysed with CDT, prediction will necessarily
involve the whole person (as discussed previously in
Section~\ref{sec-simul}), and can only be imagined if the Participant's
thoughts are simulated in detail. Note that this is so even for a
modest level of accuracy such as $R=0.8$.

\subsection{\hspace*{-8pt}Newcomb insurance with 
 thought simulation}\vspace{-6pt}

Suppose that you are a Participant in a Newcomb insurance problem in
which the Predictor operates by simulating your thoughts in detail, as
discussed in Section~\ref{sec-simul}.  I will assume here that the
Predictor is highly reliable (though not perfect), that you know the
predictions are obtained by simulating your thoughts, and that you
believe that this creates another conscious being that is not
distinguishable (internally) from the ``real'' you.

Figure~\ref{fig-nits} shows the payoffs in this situation for the four
possible strategies of what to do at stage~1 (take one or two boxes)
and later at stage~2 (bet that the predictor was right or
wrong). These payoffs are the same as the totals (for Ahmed's version)
in Figure~\ref{fig-insure}, noting that the prediction is what the
simulated you did (Take~1 or Take~2), and the payoff for the simulated
you is assumed to be whatever the real you gets.  As a convenience, I have
assumed that the simulation continues through the simulated you 
choosing how to bet, but this choice is ignored.\footnote{\fsp
Note that the problem is phrased here in terms of a choice of strategy
for both stages, made before stage~1. Simulated you can make this
choice (not knowing whether they are simulated or real) even if they 
don't actually continue to exist for stage~2.}

\begin{figure}
\begin{center}
\begin{tabular}{l|rrrr|rrrr}
&\multicolumn{4}{c|}{$\!\!$You are real, other is simulated$\!$} 
&\multicolumn{4}{c}{$\!\!$You are simulated, other is real$\!$}\\[2pt]
   & $\!\!$T1,Br & $\!\!$T1,Bw & T2,Br & $\!\!$T2,Bw$\!$ 
   & $\!\!$T1,Br & $\!\!$T1,Bw & T2,Br & $\!\!$T2,Bw$\!$ \\
\hline
\rule{-5pt}{12pt}T1,Br
  & 125~ & 125~ & $-$75~ & $-$75~ & 125~ & 75~ & 25~ & 175~ \\
\rule{-5pt}{12pt}T1,Bw
  & 75~  & 75~  &  75~ &  75~ & 125~ & 75~ & 25~ & 175~ \\
\rule{-5pt}{12pt}T2,Br
  & 25~  & 25~  &  25~ &  25~ & $-$75~ & 75~ & 25~ & $-$25~ \\
\rule{-5pt}{12pt}T2,Bw
  & 175~ & 175~ & $-$25~ & $-$25~ & $-$75~ & 75~ & 25~ & $-$25~
\end{tabular}~~\rule[-43pt]{2pt}{92pt}~~%
\begin{tabular}{rrrr}
\multicolumn{4}{c}{$\!$Average of real and simulated$\!$}\\[2pt]
   $\!\!$T1,Br & $\!\!$T1,Bw & T2,Br & $\!\!$T2,Bw$\!\!$ \\
\hline
\rule{-6pt}{12pt}125~ & 100~ & $-$25~ &  50~ \\
\rule{-6pt}{12pt}100~ &  75~ &  50~ & 125~ \\
\rule{-6pt}{12pt}$-$25~ & 50~  &  25~ &   0~ \\
\rule{-6pt}{12pt}50~ & 125~ &   0~ & $-$25~ \\
\end{tabular}
\end{center}

\caption{Payoffs for Newcomb insurance with thought simulation.
In the table on the left, your four possible strategies are in the
leftmost column, with ``T1,Br'' meaning ``Take 1, Bet right'', and
similarly for the others, while the eight columns show the possible 
states of the world, combining the possibility that you are 
the real or the simulated Participant with the four possible
strategies that the other you may have chosen. 
The table on the right shows the average payoffs,
assuming that you being real or simulated are equally likely.\label{fig-nits}}

\end{figure}

The table on the right in the figure shows the average payoff from
the ``real'' and ``simulated'' parts of the table on the left.
This average can be used to compute expected payoffs on the assumption
that you are equally likely to be the real you or the simulated you,
and that your subjective probabilities for the strategy of the other you
are the same whether you are real or simulated.

Choosing a strategy using EDT in this situation is straightforward.  To
evaluate $V(\mbox{T1,Br})$, you note that if you adopt the ``Take~1,
Bet right'' strategy, then by assumption the other you (whether real
or simulated) is highly likely to adopt the same
strategy.\footnote{\fsp The Predictor is said to predict only box
choice, not the bet made, but if the prediction is done by simulation,
the actions of the simulated you are also predictive of the bet.}
So, consulting the table with average payoffs in
Figure~\ref{fig-nits}, we see that $V(\mbox{T1,Br})$ is close to
125. Similarly, we see that the expected payoffs for the other
strategies, found using EDT, are close to 75, 25, and $-25$, so ``Take~1, Bet
right'' is optimal.  And since stage~2 for this strategy is to bet
that the prediction is correct, there is no reason to doubt that you will
follow stage~2 of the strategy when the time comes.

Applying CDT to this problem is not as straightforward. Looking at the
average payoffs on the right in the figure, one can see that T2,Br is
dominated by T1,Bw, but none of the other strategies are dominated.
It's therefore necessary to compute the causal expected utility of the
strategies, and then pick the one that maximizes this expected
utility.  This requires probabilities for which strategy is or will be
picked by the other you. After making a tentative choice of strategy,
you should update these probabilities to account for the other you
being very likely to make the same choice.  This can lead to instability
if with these updated probabilities you would make a different choice.

If the tentatively chosen strategy still maximizes causal expected
utility with these updated probabilities, the strategy can be said to
be ``ratified''. Ratification is discussed by Egan (2007), who
criticises its general usefulness.

In the Newcomb insurance problem, one can verify that T1,Br is
ratifiable, with an expected utility close to 125, once probabilities
are updated to reflect that the strategy chosen by the other you is
almost certainly also T1,Br. One can also check that none of the other
strategies are ratifiable. Crucially, moreover, none of the other
strategies can possibly have an expected utility greater than 125, so
there seems to be no reason not to just choose the ratifiable strategy
of ``Take 1, Bet right'', the same choice as EDT.

However, in the more general version of Newcomb insurance with $b>0$,
the T1,Bw/T2,Bw combination and the T2,Bw/T1,Bw combination have
average payoff of $125+b/2$, so you might think it conceivable that
you could do better than the expected utility of $125$ obtained with
T1,Br. However, T1,Bw and T2,Bw are not ratifiable --- if you 
have almost decided to do one, the other will seem better.

Rather than look at ratifiability, you could instead look for an
equilibrium randomized (mixed) strategy, choosing a strategy from
T1,Br, T1,Bw, and T2Bw with probabilities $q_{1,r}$, $q_{1,w}$, and
$q_{2,w}$ (note that T2,Br is dominated).  This approach has been
advocated by Arntzenius (2008) and Wallace (2010); see also Plommer
(2016) and Armendt (2019). Whether you are able to randomize in this
problem has not been specified, but in the end this will turn out not to matter.

A characteristic of an equilibrium mixed strategy is that the expected
causal utility is the same for all strategies that are chosen with
non-zero probabilities. (Otherwise, you would not be willing to
randomize, but would instead just choose the apparently best strategy.)  
Using the average payoffs in the right of Figure~\ref{fig-nits}, with
$b$ or $b/2$ added as appropriate for the generalization with \$$b$ in 
the transparent box, the expected causal utilities of 
the non-dominated strategies are\vspace{-3pt}
\beq
 U(\mbox{T1,Br}) & = & 125\,q_{1,r}\ +\ 100\,q_{1,w}\ +\ (50+b/2)\,q_{2,w}
 \label{eq-UT1Br}\\[3pt]
 U(\mbox{T1,Bw}) & = & 100\,q_{1,r}\ +\ 75\,q_{1,w}\ +\ (125+b/2)\,q_{2,w}
 \label{eq-UT1Bw}\\[3pt]
 U(\mbox{T2,Bw}) & = & (50+b/2)\,q_{1,r}\ +\ (125+b/2)\,q_{1,w}\ +\ 
 (-25+b)\,q_{2,w}
 \label{eq-UT2Bw}
\eeq
Here, the probability of the other you choosing each strategy is
taken to be the same as the probability that you choose that strategy,
since the Predictor is assumed to be highly accurate.

\begin{figure}[t]

\begin{center}
\includegraphics[scale=0.9]{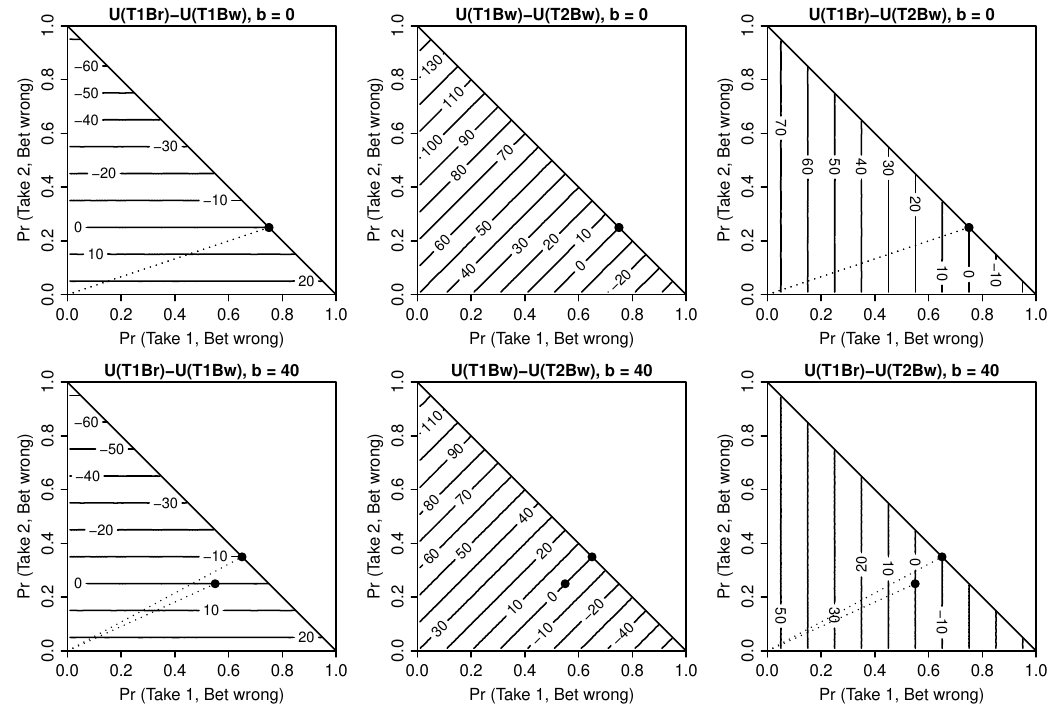}\vspace{-8pt}

\end{center}

\caption{Contour plots of differences in causal expected utilities of
strategies for different mixing proportions over strategies versus
$q_{1,w}$ and $q_{2,w}$, for $b=0$ (top) and $b=40$ (bottom). The dots
show the two equilibria in which more than one strategy has non-zero
probability (which coincide when $b=0$). The dotted lines show a
path in which $q_{1,r}$ is increased by reducing $q_{1,w}$ and 
$q_{2,w}$}\label{fig-newins}

\end{figure}

The equilibria are found by setting some subset (perhaps empty) of
$q_{1,r}$, $q_{1,w}$, and $q_{2,w}$ to zero, and then solving the equations
equating all the expected utilities (as given above) for the
strategies whose probabilities haven't been fixed to zero, with the
constraint that the probabilities sum to one and be greater than zero
(which will eliminate some potential
equilibria). Figure~\ref{fig-newins} shows contour plots of
differences in causal expected utilities for each pair of strategies,
for possible values of $q_{1w}$ and $q_{2,w}$ (with $q_{1,r}$ being
$1-q_{1w}-q_{2,w}$), both with $b=0$ and $b=40$. The zero contours
in these plots indicates values for $q_{1w}$ and $q_{2,w}$ where
the expected utilities of two of the strategies are equal.

\pagebreak

For this problem, the equilibria, and the expected causal utility for
each, are as listed below:\vspace{-2pt}
\beq
\begin{array}{cccc@{~~~~~}l}
 q_{1,r} & q_{1,w} & q_{2,w} & U & \mbox{Stable?} \\[4pt]
   1     &     0   &    0    & 125  & \mbox{Yes} \\[2pt]
   0     &     1   &    0    & 75   & \mbox{No} \\[2pt]
   0     &     0   &    1    & -25+b & \mbox{No} \\[4pt]
   0  & \displaystyle {150-b/2 \over 200}\ & \displaystyle {50+b/2 \over 200} & 
   \displaystyle\ \ {175\over2}+{b\over4}+{b^2\over 800} 
   & \mbox{Yes if $b>0$, No if $b=0$}
   \\[13pt]
  \displaystyle {b\over200} & \displaystyle {150-b\over200} & 
  \displaystyle {1\over4} 
   & \displaystyle {175\over2}+{b\over4}
   & \mbox{No}
\end{array}\nonumber
\eeq
Note that by my definition all the pure strategies (where one of 
the strategy probabilities equals one) are equilibria.
The last two equilibria above are shown as dots in
Figure~\ref{fig-newins}. The first of these is on the diagonal line
where $q_{1,w}+q_{2,w}=1$, and hence $q_{1,r}=0$, and is on the 
zero contour in the middle plots, where $U(\mbox{T1Bw})=U(\mbox{T2Bw})$. 
The second reduces to the one above when $b=0$, and is on the zero contour
in all plots, reflecting that at an equilibrium the utilities of all 
(non-dominated) strategies are the same when mixing over all of them.

I will assume that $b$ is non-negative and is less than $50$, since otherwise
T2,Br is no longer dominated by T1,Bw. The expected causal
utility ($U$) of the first equilibrium strategy (with $q_{1,r}=1$) will
then be greater than that of all the other equilibrium strategies,
which seems like a good reason to adopt it. Note that this strategy
chooses ``Take 1, Bet right'' with probability one. So the end result
of this process of considering randomized decisions is that it is best
to not randomize. We can therefore avoid the issue of whether
randomization is actually possible in this scenario.

Another consideration is that it may be desirable to exclude from
consideration equilibria that are unstable --- for which a slight
change to the probabilities of strategies (including possibly changing
a zero probability to a small non-zero probability) leads to the
strategy whose probability was increased now having higher expected
utility than the others, which would lead you to increase its
probability yet more. For a stable equilibrium, any small change in
probabilities should instead lead you to correct back towards the
equilibrium.

The equilibria above with $q_{1,w}=1$ and $q_{2,w}=1$ are unstable.
After decreasing one of these probabilities to $1-\epsilon$ while
setting $q_{1,r}=\epsilon$ the expected utility of T1,Br will be
higher than that of the other strategy, so its probability should be
increased further.  In contrast, the equilibrium with $q_{1,r}=1$ is
stable.  

The stability of the two other equilibria above can be inferred (for
$b=0$ and $b=40$) from the plots in Figure~\ref{fig-newins}, where the
dotted lines show the direction of a change to the probabilities that
increases $q_{1,r}$. One can see that with $b=0$, a small increase in
$q_{1,r}$ leads to a region where $U(\mbox{T1Br})$ is greater than the
other expected utilities, indicating instability. For $b=40$, this is
also true for the last equilibrium above (for which $q_{1,r}$ is
already non-zero), but the equilibrium with $q_{1,r}=0$ is stable,
since a small reduction in $q_{1,w}$ and $q_{2,w}$ to allow a small
increase in $q_{1,r}$ leads to a region where $U(\mbox{T1Br})$ is less
than the other expected utilities, so you will correct back to the
equilibrium. 

Hence, with $b>0$ there are two stable equilibria. Following
Arntzenius (2008), it seems reasonable to choose the one with higher
expected utility, which in this case is the one with $q_{1,r}=1$.

So, these less-than-straightforward considerations lead in the end to
thinking that when Newcomb insurance is realized using thought
simulation, CDT will recommend the ``Take 1, Bet right'' strategy, the
same as EDT. There should be no inconsistent behaviour at stage 2,
since betting that the Predictor is right will seem correct then.

Although the end result of this analysis seems sensible, I question
whether it's actually reasonable to hypothesize the level of fantasy
involved in assuming near perfect simulation in situations of
potential decision instability, which would seem to follow a
highly-unpredictable course that could amplify even the tiniest
inaccuracy in a simulation.\footnote{\fsp
The simulation would need to be of such high
fidelity that real and simulated Participants cannot effectively
perform unpredictable mental randomization (as in the previous
section). One might then question whether, although the simulation
is a conscious being, it might not be distinct from the
real Participant.  If so, the
prediction does not precede the choice of the real Participant (in
some sense). 
Such accurate simulation also requires the real Participant to
be isolated from the world and presented with a prepared environment
identical to that of the simulation, but one might wonder whether
consciousness inherently requires some connection with the real
world.\vspace{-15pt}}

\section{Discussion}\vspace{-10pt}

One aim of this paper has been to show in a concrete way how Newcomb's
problem can quite plausibly be realized in the real world.  Any viable
theory of how to make decisions should work for such actual problems,
that involve no fantastic elements.  A viable theory must also agree
with well-considered intuitions on such real-life problems --- not
always with one's initial intuitive feelings, but with the intuitions
that survive after examining both the problem to be addressed and
related problems that may provide additional insight.

\pagebreak

In order to aid such insight, I have looked at a succession of 
problems:\vspace{-8pt}
\begin{enumerate}
\item[1)] The smoking lesion problem, with the lesion affecting your 
          desire to smoke.\vspace{-2pt}
\item[2)] The smoking lesion problem, with the lesion affecting the decision
          rule you use.\vspace{-2pt}
\item[3)] The impersonal Newcomb problem, where the prediction is the
          same for all Participants.\vspace{-2pt}
\item[4)] The standard Newcomb problem.\vspace{-8pt}
\end{enumerate}
I claim that all these problems can exist in real life, or at least
be readily imagined as existing without upsetting our conception of
how the world works.  I also claim that CDT gives the right decision in
all these problems --- to smoke or to take two boxes.

I consider the primary support for use of CDT in these problems to be
our intuitions regarding how causality operates in the world. It seems
to be universally agreed that the decision CDT recommends for the
smoking lesion affecting desire problem (1) is correct.  My hope is
that looking at the problems above in succession will carry the intuition
behind this judgement forward to Newcomb's problem. If EDT or some
other theory disagrees with CDT for Newcomb's problem (4), we can ask
at what point in the path from (1) through (2) and (3) to (4) does the
disagreement with CDT begin.

To me, it seems that EDT disagrees with CDT for all of (2), (3), and
(4). If so, it would seem productive to focus debate on the smoking
lesion problem affecting decision rule problem (2), which is simpler
than (3) and (4).  But perhaps EDT advocates could argue that EDT
would recommend smoking for (2), or like Ahmed (2014, p.~93) think
that this problems is not possible.

The Functional Decision Theory (FDT) of Yudkowsky and Soares (2018) is
an interesting alternative to both CDT and EDT. FDT advocates viewing
a decision as choosing what action the decision function you are using
should output, keeping in mind that this same function may also be
used elsewhere and elsewhen. Yudkowsky and Soares say that FDT
recommends smoking in the smoking lesion affecting desire
problem~(1).\footnote{\fsp They refer to it simply as ``smoking
lesion'', but describe ``an arterial lesion that causes those
afflicted with it to love smoking''.} In my understanding, FDT would
think that making your decision function output ``smoking'', which you
enjoy, does not change the lesion, and hence does not change your risk
of cancer, and if you should perhaps imagine that someone else uses
the same function, the change in output for them is irrelevant to you.

It seems that FDT should recommend smoking in a smoking lesion
affecting decision rule problem~(2) as well, since again the decisions
of any other people whom you might view as sharing your decision
function are irrelevant to you.  These recommendations for smoking
lesion problems both agree with CDT, since the distinction between
choosing an action (CDT) and choosing a function that outputs an
action (FDT) makes no difference when there is only one relevant
instance of the function.

But what of the impersonal (3) and standard (4) Newcomb problems, when
they are realized using the concrete, non-fantastic methods
described in Sections~\ref{sec-impn} and~\ref{sec-pr}? 

Yudkowsky and
Soares say that FDT recommends taking one box in the standard Newcomb
problem, but they appear to assume that the Predictor uses a method
much different from prediction using categories as I have described. They
say (pp.~4--5):\vspace{-8pt}
\begin{quotation}\noindent
\ldots Newcomb's predictor builds an accurate model of the agent
and reasons about her behavior \ldots When the FDT agent imagines a
world where she two-boxes, she visualizes a world where the predictor
did not fill the box; when she imagines a world where she one-boxes,
she imagines a world where the box is full. She bases her decision solely
on the appeal of these hypothetical worlds, and one-boxes.
\end{quotation}
So, they consider that with FDT, the Predictor's accurate model incorporates 
the same decision function as the agent, and thus imagining this function
producing different outputs requires imagining that the result of 
the Predictor's model, and hence of the box contents, also change.\footnote{\fsp
Note that a CDT agent might also be described as imagining
hypothetical worlds, but those worlds differ only in the \textit{future},
whereas FDT hypothetical worlds may also differ in the past.}

But a Predictor who uses categories does not employ an accurate (or
even not-so-accurate) model of each Participant --- they simply
determine which category the Participant is in, perhaps based on
ordinary observations, and make a (possibly randomized) prediction
on that basis. 

So perhaps FDT recommends two-boxing when the Predictor uses categories.
If so, however, a major argument for FDT is lost --- ``FDT works where
CDT and EDT fail'' (Yudkowsky and Soares 2018, p.~33).  In so far as
the ``Why ain'cha Rich?" argument has any force against CDT, it would
equally tell against FDT in this concrete Newcomb problem.

Or perhaps FDT does recommend one-boxing even when the Predictor uses
categories.  Although the Predictor does not model your decision
function in this scenario, the Predictor's predictions are based on
actions of past Participants, who did have some decision function.  In
imagining that your own decision function outputs one-boxing or
two-boxing, you might see this as going with different actions for these
previous Participants, and hence for what the Predictor predicts that you
will do. This seems to me to be a rather extravagant hypothetical, but
I will not consider the matter further here.

Fantastic Newcomb scenarios involving thought simulation would seem
to be where FDT is most clearly applicable, with the Predictor's model
of your decision function being extremely accurate.  Yet if we assume
a functionalist view of consciousness, this very accuracy renders FDT
superfluous, since CDT then also recommends one-boxing, as I discuss
in Section~\ref{sec-simdecision}. Lacking any adequate understanding
of consciousness, it is hard to say whether there is any intermediate
ground where FDT wins --- where the simulation is too accurate to be
seen as prediction based on categories (where I argue that two-boxing
as recommended by CDT is correct) but not accurate enough to be a
conscious being.\footnote{\fsp Another intermediate situation is
where the simulation is accurate enough to be a conscious being, but 
sufficiently inaccurate that this being can recognize
inconsistencies that point to it being a simulation. Then you should
one-box if you are the simulation and two-box if you are the real you.
Of course, the Predictor will then not achieve a high accuracy.}

Those skeptical of CDT have advanced scenarios other than Newcomb's
problem in which CDT is said to give the wrong result. These often
involve the sort of decision instability that I discuss in relation to
the Newcomb insurance problem in Section~\ref{sec-insure}. An example
is the ``adversarial offer'' scenario of Oesterheld and Conitzer
(2021), in which a buyer may choose to purchase one of two boxes, A
and B, for \$1, or to decline to purchase either, knowing that the
seller predicted the buyer's decision, and has put \$3 in every box
that is predicted to not be purchased. As soon as you decide to
purchase box A, or box B, or to make no purchase, you will immediately
regret this decision.  Randomization would solve this dilemma, so it
is specified that it is in some way prohibited.\footnote{\fsp
Egan (2007) and Spencer and Wells (2019) present several similar examples.}

For the Newcomb insurance problem, which exhibits similar instability,
I argued that in any non-fantastic instance of this problem actual
reasoning using CDT would result in effective mental randomization,
even if the decision maker had no initial inclination to randomize, so
prohibiting randomization is equivalent to simply prohibiting use of
CDT, rendering the problem uninteresting. It would be interesting to
see whether this argument generalizes to showing that the adversarial
offer problem, and others that have been seen as posing difficulties
for CDT, are also unrealizable.

\section*{Appendix:\ \ Proofs of claims}\vspace{-10pt}

Recall the conditions for a Predictor who uses categories $G_1,\ldots,G_K$
to have accuracy $R$, from equation~(\ref{acc-eq-R}):

\hfill $\displaystyle
  R_1\ =\ 
  {\sum\limits_{i=1}^K p_i q_i w_i \over \sum\limits_{i=1}^K q_i w_i} 
   \ =\ R\ \ \ \ \ \mbox{and}\ \ \ \ \ \,
  R_2\ =\ 
  {\sum\limits_{i=1}^K (1\!-\!p_i) (1\!-\!q_i) w_i \over 
   \sum\limits_{i=1}^K (1\!-\!q_i) w_i} \ =\ R
$ \hfill (\ref{acc-eq-R})

\noindent
For brevity, let $w$ denote $w_1,\ldots,w_K$, let $q$ denote
$q_1,\ldots,q_K$, and let $p$ denote $p_1,\ldots,p_K$. 

The subjective prior probability density that a Participant has for
values of $q$, before learning the Predictor's accuracy, will be
denoted by $f(q)$.  The probability density for the distribution
obtained from $f$ by restricting to values of $q$ for which accuracy $R$
is possible according to~(\ref{acc-eq-R}), with given $w$,
will be denoted by $f_{R,w}(q)$. Note that $f_{R,w}$ restricts to
values of $q$ that (in combination with $w$) are compatible with 
accuracy $R$ for \textit{some}~$p$, on the assumption that the
Predictor will be able to find such a $p$, regardless of how
small a volume the $p$ that allow this occupy.

Given an assumed value for $R$, define $e_i(w,q)$ to be the
expectation of $p_i$ given $q$, where the expectation is with respect
to the Predictor selecting $p$ uniformly from those values that result
in predictions having accuracy $R$ --- that is, that satisfy the
equations of~(\ref{acc-eq-R}).  Note that the dependence on $R$ is not
explicit in the notation, and that $e_i(w,q)$ is undefined if $q$ and
$w$ do not allow the Predictor to have accuracy $R$ for any $p$.

\vspace{8pt} 

A probability density $h(q_1,\ldots,q_K)$ will be called \textit{shift
symmetric} if for all $q$ and any $t$ in $\{1,\ldots,K\}$, 
\beq
  h(q_t,\ldots,q_K,q_1,\ldots,q_{t-1}) & = & h(q_1,\ldots,q_K)
\eeq
If $q$ denotes $q_1,\ldots,q_K$ and $q^{\shft t}$ denotes
$q_{t},\ldots,q_K,q_1,\ldots,q_{t-1}$, then this can be written as
$h(q^{\shft t})\,=\,h(q)$ for all $t$ in $\{1,\ldots,K\}$.

Note that shift symmetry is a weaker condition than exchangeability
(invariance of $h$ under any permutation of the $q_i$) which is in
turn weaker than the $q_i$ being independent and identically
distributed under $h$ (so that $h(q_1,\ldots,q_K)=\prod_i g(q_i)$ for
some univariate density $g$).

\noindent\textbf{Lemma 1:}\ \ For any $q$ and $i$, 
$q_i = q^{\shft i}_1$; similarly for any other shifted quantity, such as $p$.

\noindent\textbf{Proof:}\ \ By definition, the first element of
$q^{\shft i}$ is $q_i$, and similarly for other shifted
quantities. \QED

Note that when all $w_i=1/K$, the conditions 
of~(\ref{acc-eq-R}) reduce to
\beq
  R_1\ =\ 
  {\sum\limits_{i=1}^K p_i q_i \over \sum\limits_{i=1}^K q_i} 
   \ =\ R\ \ \ \ \ \mbox{and}\ \ \ \ \ \,
  R_2\ =\ 
  {\sum\limits_{i=1}^K (1\!-\!p_i) (1\!-\!q_i) \over 
   \sum\limits_{i=1}^K (1\!-\!q_i)} \ =\ R
\label{acc-eq-R-uw}
\eeq
This allows us to show the following Lemmas leading to Claim A.

\noindent\textbf{Lemma 2:}\ \ If $w_i=1/K$ for all $i$, then for all $R$
and $f$,
if $f(q)$ is shift symmetric then $f_{R,w}(q)$ is also shift symmetric.

\noindent\textbf{Proof:}\ \ 
When all $w_i=1/K$, 
observe that $q$ and $p$ will satisfy the conditions of~(\ref{acc-eq-R-uw})
if and only if $q^{\shft t}$ and $p^{\shft t}$ satisfy them, 
since the terms in the sums above
have the same form for all $i$.  So either $q$ and $q^{\shft t}$
both cannot satisfy these conditions for any $p$, in which case
$f_{R,w}(q^{\shft t})=f_{R,w}(q)=0$, or both do satisfy these conditions
(for some $p$ and $p^{\shft t}$), in which case we will have
$f_{R,w}(q^{\shft t})\,=\,cf(q^{\shft t})\,=\,cf(q)\,=\,f_{R,w}(q)$,
where $c$ is the constant needed to normalize $f_{R,w}(q)$ to integrate to one 
over the region of feasible $q$. \QED

\noindent\textbf{Lemma 3:}\ \ If $w_i=1/K$ for all $i$, then for all $R$
and $s$, $e_s(w,q)\,=\,e_1(w,q^{\shft s})$.

\noindent\textbf{Proof:}\ \ If the conditions of~(\ref{acc-eq-R-uw})
hold when all $w_i=1/K$, they will also hold after shifting the $q_i$
and $p_i$.  So, applying Lemma~1, the set of values for $p_s$ that
satisfy~(\ref{acc-eq-R-uw}) with $q$ is the same as the set of values
for $p_1$ that satisfy (\ref{acc-eq-R-uw}) with $q^{\shft s}$, and
hence the expectation over this set is both $e_s(w,q)$ and
$e_1(w,q^{\shft s})$.  \QED

\noindent\textbf{Claim A:}\ \ Suppose a Participant knows they are in
category $G_s$, and they know that all $w_i=1/K$, but they do not know
$q$, and apart from $R$ they know nothing else of relevance.  Suppose 
also that their
prior for $q$ before accounting for the Predictor's accuracy, $f(q)$,
is shift symmetric, and their uncertainty about how the Predictor
chooses $p$ given $q$ is expressed by a uniform distribution.  Then
after choosing one box or two (but before looking in the opaque box)
their posterior probability that the Predictor correctly predicted
their action will equal the Predictor's accuracy, $R$.

\noindent\textbf{Proof:}\ \ A Participant's 
posterior probability that
the Predictor was correct after they take one box will be:
\beq
 \prs(C\giv A_1\and G_s\and B)\ \ =\ \ 
 \prs(S_1\giv A_1\and G_s\and B)\ \ =\ \
 {\prs(S_1\and A_1\giv G_s\and B) \over \rule{0pt}{10pt}\prs(A_1\giv G_s\and B)}
\eeq
where $B$ is background information, such as that $w_i=1/K$. The numerator
and denominator above can be written in terms of expectations over
$q$ and $p$, giving:\vspace{-6pt}
\beq
 \prs(C\giv A_1\and G_s\and B) & = &
 {\displaystyle \rule[-8pt]{0pt}{1pt}\int e_s(w,q)\, q_s\, f_{R,w}(q)\, dq 
  \over \rule{0pt}{17pt}\displaystyle \int q_s\, f_{R,w}(q)\, dq}
\eeq
Using Lemmas 1, 2, and 3, this can be rewritten as
\beq
 \prs(C\giv A_1\and G_s\and B) & = &
 {\displaystyle \rule[-8pt]{0pt}{1pt}\int e_1(w,q^{\shft s})
   \, q_1^{\shft s}\, f_{R,w}(q^{\shft s})\, dq 
  \over \rule{0pt}{17pt}\displaystyle 
   \int q_1^{\shft s}\, f_{R,w}(q^{\shft s})\, dq}
 \ \ = \ \
 {\displaystyle \rule[-8pt]{0pt}{1pt}\int e_1(w,q)\, q_1\, f_{R,w}(q)\, dq 
  \over \rule{0pt}{17pt}\displaystyle \int q_1\, f_{R,w}(q)\, dq}\ \ 
\eeq
where the last step simply changes the order of integration over the components
of $q$.
Since the right side above does not depend on $s$, it must be that
$\prs(C\giv A_1\and G_i\and B)$ is the same for all $i$, and since\vspace{-4pt}
\beq
  R \ \ =\ \ \prs(C\giv A_1\and B)\ \ =\ \ 
  \sum_{i=1}^K \prs(C\giv A_1\and G_i\and B)\,\prs(G_i\giv A_1\and B)
\eeq
it must be that all $\prs(C\giv A_1\and G_i\and B)\ =\ R$, and in particular
that $\prs(C\giv A_1\and G_s\and B)\ =\ R$.

Similarly,\vspace{-6pt}
\beq
 \prs(C\giv A_2\and G_s\and B) & = &
 {\displaystyle \rule[-8pt]{0pt}{1pt}\int (1\!-\!e_s(w,q))\, (1\!-\!q_s)\, 
                                          f_{R,w}(q)\, dq 
  \over \rule{0pt}{17pt}\displaystyle \int (1\!-\!q_s)\, f_{R,w}(q)\, dq}
\\[4pt] & = &
 {\displaystyle \rule[-8pt]{0pt}{1pt}\int (1\!-\!e_1(w,q^{\shft s}))
   \, (1\!-\!q_1^{\shft s})\, f_{R,w}(q^{\shft s})\, dq 
  \over \rule{0pt}{17pt}\displaystyle 
   \int (1\!-\!q_1^{\shft s})\, f_{R,w}(q^{\shft s})\, dq}
\\[4pt] & = &
 {\displaystyle \rule[-8pt]{0pt}{1pt}\int (1\!-\!e_1(w,q))\, (1\!-\!q_1)\, 
                                           f_{R,w}(q)\, dq 
  \over \rule{0pt}{17pt}\displaystyle \int (1\!-\!q_1)\, f_{R,w}(q)\, dq}\ \ 
\eeq
and since this does not depend on $s$, it follows in the same way as above
that $\prs(C\giv A_2\and G_s\and B)\ =\ R$.
\QED

\vspace{8pt} 

Let $1\!-\!p$ refer to $1\!-\!p_1,\ldots,1\!-\!p_K$, and similarly let
$1\!-\!q$ refer to $1\!-\!q_1,\ldots,1\!-\!q_K$.  A probability
density $h(q)$ will be called \textit{inversion symmetric} if
$h(q)=h(1\!-\!q)$ for all $q$.

\noindent\textbf{Lemma 4:}\ \ For any $w$ and $R$, $q$ and $p$
satisfy~(\ref{acc-eq-R}) if and only if $1\!-\!q$ and $1\!-\!p$
satisfy~(\ref{acc-eq-R}).

\noindent\textbf{Proof:}\ \ Substituting $1\!-\!p$ for $p$ and
$1\!-\!q$ for $q$ in the first equation converts it to the second,
and vice versa, so if these two equations are satisfied by $q$ and $p$, they
will also be satisfied by $1\!-\!q$ and $1\!-\!p$, and similarly in
the other direction. \QED

\noindent\textbf{Lemma 5:}\ \ For any $w$ and $R$, $e_i(w,1\!-\!q) \,=\,
1-e_i(w,q)$ for all $i$ in $\{1,\ldots,K\}$.

\noindent\textbf{Proof:}\ \ By Lemma~4, if $p$
satisfies~(\ref{acc-eq-R}) with $q$, then $1\!-\!p$
satisfies~(\ref{acc-eq-R}) with $1\!-\!q$, and vice versa.
Since the Jacobian of the
mapping from $p$ to $p' = 1\!-\!p$ has absolute value one, a uniform
distribution on the $p$ that satisfy~(\ref{acc-eq-R}) with $q$ maps to
a uniform distribution on the $p'$ that satisfy~(\ref{acc-eq-R}) with
$1\!-\!q$. Hence $e_i(w,1\!-\!q) \,=\, E(1\!-\!p_i|q,w) \,=\, 
1-E(p_i|q,w) \,=\, 1-e_i(w,q)$. \QED

\noindent\textbf{Lemma 6:}\ \ For all $w$ and $R$, if $f(q)$ is inversion 
symmetric then $f_{R,w}(q)$ is also inversion symmetric.

\noindent\textbf{Proof:}\ \ By Lemma 4, there is a $p$ with which $q$
satisfies~(\ref{acc-eq-R}) if and only if there is a $p$ with which $1\!-\!q$
satisfies~(\ref{acc-eq-R}). Hence, either $f_{R,w}(q)$ and $f_{R,w}(1\!-\!q)$ 
are both zero because neither $q$ nor $1\!-\!q$ can satisfy~(\ref{acc-eq-R}) for
any $p$, or $f_{R,w}(q)\,=\,cf(q)\,=\,cf(1\!-\!q)\,=\,f_{R,w}(1\!-\!q)$, where 
$c$ is the constant needed to normalize $f_{R,w}(q)$ to integrate to one over
the region of feasible $q$. \QED

\noindent\textbf{Claim B:}\ \ Consider a Participant who knows they are in
category $G_s$, and knows the values of the $w_i$, and apart from $R$
knows nothing else of relevance.  Suppose that the joint prior for $q$
(before conditioning on $R$), $f(q)$, is inversion symmetric. Then
\beq
  \prs(C\giv A_1\and G_s\and B)\ =\ \prs(C\giv A_2\and G_s\and B) 
\eeq
where $B$ is background information, such as the $w_i$.

\noindent\textbf{Proof:}\ \ 
A Participant's posterior probability that
the Predictor was correct after they take one box is\vspace{-2pt}
\beq
 \prs(C\giv A_1\and G_s\and B)\ = \ 
 \prs(S_1\giv A_1\and G_s\and B)\ =\
 {\prs(S_1\and A_1\giv G_s\and B) \over \rule{0pt}{10pt}\prs(A_1\giv G_s\and B)}
 \ =\ 
 {\displaystyle \rule[-8pt]{0pt}{1pt}\int e_s(w,q)\, q_s\, f_{R,w}(q)\, dq 
  \over \rule{0pt}{17pt}\displaystyle \int q_s\, f_{R,w}(q)\, dq}\ \ \ \
\label{eq-B1}
\eeq
Their posterior probability that the Predictor is correct after they
take two boxes is
\beq
 \prs(C\giv A_2\and G_s\and B)\ \ =\ \ 
 {\displaystyle \rule[-8pt]{0pt}{1pt}\int 
  (1\!-\!e_s(w,q))\, (1\!-\!q_s)\, f_{R,w}(q)\, dq 
  \over \rule{0pt}{17pt}\displaystyle \int (1\!-\!q_s)\, f_{R,w}(q)\, dq}
  \ \ =\ \ 
 {\displaystyle \rule[-8pt]{0pt}{1pt}\int 
  e_s(w,1\!-\!q)\, (1\!-\!q_s)\, f_{R,w}(1\!-\!q)\, dq 
  \over \rule{0pt}{17pt}\displaystyle \int (1\!-\!q_s)\, f_{R,w}(1\!-\!q)\, dq}
\eeq
where the second equality follows from Lemmas 5 and 6.  After substituting 
$1\!-\!q$ for $q$ in the integrals in the numerator and denominator above they 
are identical to the integrals in equation~(\ref{eq-B1}), and hence
$\prs(C\giv A_1\and G_s\and B)\ =\ \prs(C\giv A_2\and G_s\and B)$.
\QED

\vspace{8pt} 

If $w$ is not known, a Participant's inference must be based on some
prior distribution for it, considered before the Participant learns
their own category.  In general, a Participant might have prior
beliefs in which $w$ and $q$ are dependent, with density $g(w,q)$
(with respect to the uniform distribution over the simplex of valid $w$
and of $q$ on $[0,1]^K$).  The result of restricting this density to
values of $w$ and $q$ that allow the Predictor to have accuracy $R$
will be denoted as $g_R(w,q)$.  Note that even if $w$ and $q$ are
independent in $g(w,q)$, they will in general be dependent in
$g_R(w,q)$.

Claim~C below can be seen as a generalization of Claim~A to when
$g(w,q)$ is shift symmetric in both arguments --- that is, $g(w^{\shft
t},q^{\shft t})=g(w,q)$ for all $t$ in $\{1,\ldots,K\}$. It is proved
using the generalizations of Lemmas~2 and~3 below:

\noindent\textbf{Lemma 7:}\ \ For all $R$ and $g$,
if $g(w,q)$ is shift symmetric then 
$g_R(w,q)$ is also shift symmetric.

\noindent\textbf{Proof:}\ \ Observe that $w$, $q$, and $p$ will
satisfy the conditions of~(\ref{acc-eq-R}) if and only if $w^{\shft
t}$, $q^{\shft t}$, and $p^{\shft t}$ satisfy them, since the terms in
the sums above have the same form for all $i$.  So either $q$ with $w$
and $q^{\shft t}$ with $w^{\shft t}$ both cannot satisfy these
conditions for any $p$, in which case $g_R(w^{\shft t},q^{\shft
t})=g_R(w,q)=0$, or both do satisfy these conditions (for some $p$ and
$p^{\shft t}$), in which case we will have $g_R(w^{\shft t},q^{\shft
t})\,=\,cg(w^{\shft t},q^{\shft t})\,=\,cg(w,q)\,=\,g_R(w,q)$, where $c$ is the
constant needed to normalize $g_R(w,q)$ to integrate to one over the
region of feasible $w$ and $q$. \QED

\noindent\textbf{Lemma 8:}\ \ For all $R$, $w$, $q$,
and $s$, $e_s(w,q)\,=\,e_1(w^{\shft s},q^{\shft s})$.

\noindent\textbf{Proof:}\ \ If the conditions of~(\ref{acc-eq-R})
hold, they will also hold after shifting the $w_i$, $q_i$, and
$p_i$.  So, applying Lemma~1, the set of values for $p_s$ that
satisfy~(\ref{acc-eq-R}) with $w$ and $q$ is the same as the set of values
for $p_1$ that satisfy (\ref{acc-eq-R}) with $w^{\shft s}$ and
$q^{\shft s}$, and hence the expectation over this set is both $e_s(w,q)$ 
and $e_1(w^{\shft s},q^{\shft s})$. \QED

\noindent\textbf{Claim C:}\ \ Suppose a Participant knows they are in
category $G_s$, but do not know $w$ and $q$, and apart from $R$ they
know nothing else of relevance.  Suppose also that their prior
distribution for the unknown $w$ and $q$ (before accounting for the
Predictor's accuracy), denoted $g(w,q)$, is shift symmetric for $w$
and $q$, and their uncertainty about how the Predictor chooses $p$
given $w$ and $q$ is expressed by a uniform distribution.  Then after
choosing one box or two (but before looking in the opaque box) their
posterior probability that the Predictor correctly predicted their
action will equal the Predictor's accuracy,~$R$.

\noindent\textbf{Proof:}\ \ A Participant's 
posterior probability that
the Predictor was correct after they take one box will be:
\beq
 \prs(C\giv A_1\and G_s\and B)\ \ =\ \ 
 \prs(S_1\giv A_1\and G_s\and B)\ \ =\ \
 {\prs(S_1\and A_1\giv G_s\and B) \over \rule{0pt}{10pt}\prs(A_1\giv G_s\and B)}
\eeq
where $B$ is background information. The numerator
and denominator above can be written in terms of expectations over
$w$, $q$, and $p$, giving:\vspace{-6pt}
\beq
 \prs(C\giv A_1\and G_s\and B) & = &
{\displaystyle \rule[-8pt]{0pt}{1pt}\int\!\int e_s(w,q)\,q_s\,
        w_s\,g_R(w,q)\,dq\,dw
  \over \rule{0pt}{17pt}\displaystyle \int\!\int q_s\,
        w_s\,g_R(w,q)\,dq\,dw}
\eeq
Here, the factor $w_s$ accounts for the Participant knowing they are in $G_s$.
Using Lemmas 1, 7, and 8, this can be rewritten as\vspace{-6pt}
\beq
 \prs(C\giv A_1\and G_s\and B)
 & = &
 {\displaystyle \rule[-8pt]{0pt}{1pt}\int\!\int 
   e_1(w^{\shft s},q^{\shft s})
   \, q_1^{\shft s}\,w_1^{\shft s}\,g_R(w^{\shft s},q^{\shft s})\, dq\,dw
  \over \rule{0pt}{17pt}\displaystyle 
   \int\!\int 
   q_1^{\shft s}\,w_1^{\shft s}\,g_R(w^{\shft s},q^{\shft s})\,dq\,dw}
 \\[7pt]
 & = &
 {\displaystyle \rule[-8pt]{0pt}{1pt}
  \int\!\int e_1(w,q)\, q_1\,w_1\,g_R(w,q)\,dq\,dw
  \over \rule{0pt}{17pt}\displaystyle \int\!\int q_1\,w_1\,g_R(w,q)\,dq\,dw }
\eeq
where the last step simply changes the order of integration over the components
of $w$ and $q$.
Since the right side above does not depend on $s$, it must be that
$\prs(C\giv A_1\and G_i\and B)$ is the same for all $i$, and since\vspace{-4pt}
\beq
  R \ \ =\ \ \prs(C\giv A_1\and B)\ \ =\ \ 
  \sum_{i=1}^K \prs(C\giv A_1\and G_i\and B)\,\prs(G_i\giv A_1\and B)
\eeq
it must be that all $\prs(C\giv A_1\and G_i\and B)\ =\ R$, and in particular
that $\prs(C\giv A_1\and G_s\and B)\ =\ R$. In similar fashion, one can show
that $\prs(C\giv A_2\and G_s\and B)\ =\ R$.
\QED

Note that though Claim C is phrased in terms of a density function for 
a continuous $w$, it extends trivially to prior distributions
in which $w$ is discrete, as in the examples of footnote~\ref{foot-unk-w}.

\vspace{8pt} 

Suppose a Participant has knowledge of which of a second set of
categories, $H_1,\ldots,H_L$, they are in, and knows that membership
in one of these categories was not used by the Predictor. Let the
prevalence of $H_j$ for Participants in $G_i$ be denoted by $u_{i,j}$,
so the fraction of all Participants who are in both $G_i$ and $H_j$ is
$w_i u_{i,j}$. (If $u_{i,j}$ is the same for all $i$, the notation can
be abbreviated to $u_j$, as it was in Section~\ref{sec-additional}.)
Suppose that the Participant does not know the $w_i$ or $u_{i,j}$.

The Predictor makes predictions having accuracy $R$ based on the
one-boxing probabilities, $q_i$, for the categories $G_1,\ldots,G_K$.
Suppose that the Participant does not know these $q_i$, and also does
not know the one-boxing probabilities, $r_{i,j}$, for Participants in
$G_i$ and $H_j$, from which the $q_i$ could be derived as $q_i=\sum_j
u_{i,j}r_{i,j}$.

The Participant will have some joint prior distribution for these
unknown $w_i$, $u_{i,j}$, and $r_{i,j}$, before
knowing which of these categories they themselves are in.  Let the
joint density of this prior
(with respect to the uniform distributions over simplexes for $w$ and $u$
and $[0,1]^{KL}$ for $r$)
be written as $h(w,u,r)$.  Define $h$ to be \textit{simultaneously shift
symmetric} if for all $s$ in $\{1,\ldots,K\}$ and $t$ 
in $\{1,\ldots,L\}$,\vspace{-8pt}
\beq
  h(w,u,r) & = & h\Big(w^{\shft s},u\upst,r\upst\Big)
\eeq
where\vspace{-8pt}
\beq
  \Big[u\upst\Big]_{i,j}
  & = & u_{\,1+(s+i-2\,\mbox{mod}\, K),\ 1+(t+j-2\,\mbox{mod}\, L)}
  \\[-15pt]\nonumber
\eeq
and similarly for $r\upst$.

After learning that the Predictor has accuracy $R$, a Participant will
update their prior $h(w,u,r)$ to $h_R(w,u,r)$ by restricting it to
values of $w$ and $q_i=\sum_j u_{i,j}r_{i,j}$ for which accuracy $R$
is possible. We have the following Lemma:

\noindent\textbf{Lemma 9:}\ \ For all $R$ and $h$,
if $h(w,u,r)$ is simultaneously shift symmetric then 
$h_R(w,u,r)$ is also simultaneously shift symmetric.

\vspace{-10pt}

\noindent\textbf{Proof:}\ \ A shift of $u$ and $r$ to $u\upst$ and $r\upst$
affect whether accuracy $R$ is achievable only through the
$q_i=\sum_j u_{i,j}r_{i,j}$. Shifting the $j$ indexes by $t$ has no
effect on this sum. That shift symmetry is preserved by the shift of
the $i$ indexes by $s$ follows in the same way as for Lemma~7.
\QED

\noindent\textbf{Claim D:}\ \ Suppose a Participant knows they are in
category $G_s$, and also in category $H_t$ (which is not known, or at
least not used, by the Predictor), but do not know $w$, $u$, $r$, or
$q$, and apart from $R$ they know nothing else of relevance.  Suppose
also that their prior distribution for the unknown $w$, $u$, and $r$,
denoted $h(w,u,r)$, is simultaneously shift symmetric, and their
uncertainty about how the Predictor chooses $p$ given the $q_i=\sum_j
u_{i,j}r_{i,j}$ is expressed
by a uniform distribution.  Then after choosing one box or two (but
before looking in the opaque box) their posterior probability that the
Predictor correctly predicted their action will equal the Predictor's
accuracy,~$R$.

\noindent\textbf{Proof:}\ \ 
This claim can be seen as a generalization of Claims~A and~C, which are
special cases when $L=1$, and can be proved in similar but more general fashion.

A Participant's 
posterior probability that
the Predictor was correct after they take one box will be:
\beq
 \prs(C\giv A_1\and G_s\and H_t\and B)\ \ =\ \ 
 \prs(S_1\giv A_1\and G_s\and H_t\and B)\ \ =\ \
 {\prs(S_1\and A_1\giv G_s\and H_t\and B) 
   \over \rule{0pt}{10pt}\prs(A_1\giv G_s\and H_t\and B)}
\eeq
where $B$ is background information. The numerator
and denominator above can be written in terms of expectations over
$w$, $u$, $r$, and $p$, giving:\vspace{-2pt}
\beq
 \prs(C\giv A_1\and G_s\and H_t\and B) & = &
{\displaystyle \rule[-8pt]{0pt}{1pt}
 \int\!\int\!\int 
  e_s(w,q(r,u))\,r_{s,t}\, w_s\, u_{s,t}\,h_R(w,u,r)\,dr\,dw\,du
  \over \rule{0pt}{17pt}\displaystyle 
  \int\!\int\!\int r_{s,t}\, w_s\, u_{s,t}\,h_R(w,u,r)\,dr\,dw\,du}
\eeq
where $[q(r,u)]_i\,=\,\sum_j u_{i,j}r_{i,j}$.  Note that the factor
$w_s\, u_{s,t}$ accounts for the Participant knowing they are in $G_s$
and $H_t$.  The above can be rewritten as\vspace{-10pt}
\beq
 \prs(C\giv A_1\and G_s\and H_t\and B)
 & \!=\!\! & {\displaystyle \rule[-8pt]{0pt}{1pt}\int\!\int\!\int
   e_1\big(w^{\shft s}\!,q\big(r\upst,u\upst\big)\big)
   \, r\upst_{1,1}
      \, w_1^{\shft s}\,u_{1,1}\upst\,
       h_R\big(w^{\shft s},u\upst,r\upst\big)\, dr\,dw\,du
  \over \rule{0pt}{17pt}\displaystyle 
   \int\!\int\!\int\!
   \, r\upst_{1,1}
      \,w_1^{\shft s}\,u_{1,1}\upst\, 
       h_R\big(w^{\shft s},u\upst,r\upst\big)\, dr\,dw\,du}
\ \ \ \ \ \ \ \\[10pt]
 & \!=\!\! &
{\displaystyle \rule[-8pt]{0pt}{1pt}
 \int\!\int\!\int 
  e_1(w,q(r,u))\,r_{1,1}\,w_1\, u_{1,1}\,h_R(w,u,r)\,dr\,dw\,du
  \over \rule{0pt}{17pt}\displaystyle 
  \int\!\int\!\int r_{1,1}\,w_1\, u_{1,1}\,h_R(w,u,r)\,dr\,dw\,du}
\eeq
where the last step simply changes the order of integration over the components
of $w$, $u$, and $r$.
Since the right side above does not depend on $s$ or $t$, 
$\prs(C\giv A_1\and G_i\and H_j\and B)$ must be the same for all $i$ and $j$,
and hence $\prs(C\giv A_1\and G_s\and H_t\and B)\ =\ R$. 
In similar fashion, one can show
that $\prs(C\giv A_2\and G_s\and H_t\and B)\ =\ R$.~\QED

\section*{References}\vspace{-10pt}

\leftmargini 0.2in

\begin{description}

\item[\hspace{-5pt}]
Aaronson, S.\ (2005) ``Dude, it’s like you read my mind'',
  \texttt{scottaaronson.blog/?p=30}

\item[\hspace{-5pt}]
Aaronson, S.\ (2013) \textit{Quantum Computing since Democritus},
  Cambridge University Press.

\item[\hspace{-5pt}]
Agresti, A.\ (2013) \textit{Categorical Data Analysis}, 3rd edition,
 Wiley.

\item[\hspace{-5pt}]
Ahmed, A.\ (2014) \textit{Evidence, Decision and Causality}, Cambridge
  University Press.

\item[\hspace{-5pt}]
Ahmed, A.\ (2018) ``Introduction'', in A.~Ahmed (editor) 
  \textit{Newcomb's Problem}, Cambridge University Press, pp.~1--18.

\item[\hspace{-5pt}]
Ahmed, A.\ and Price, H.\ (2018) ``Arntzenius on `Why ain'cha rich?'{''},
  Erkenntnis, vol.~77, pp.~15--30.

\item[\hspace{-5pt}]
Armendt, B.\ (2019) ``Causal Decision Theory and decision instability'',
  \textit{The Journal of Philosophy}, vol.~116, pp.~263--277.

\item[\hspace{-5pt}]
Arntzenius, F.\ (2008) ``No regrets, or: Edith Piaf revamps decision theory'',
  Erkenntnis, vol.~68, pp.~277--297.

\item[\hspace{-5pt}]
Berm\'udez, J.~L.\ (2018) ``Does Newcomb's problem actually exist'', in
  A.\ Ahmed (editor) \textit{Newcomb's Problem}, Cambridge University Press.


\item[\hspace{-5pt}] 
Edlin, A., Gelman, A., and Kaplan, N.\ (2007) ``Voting as a rational choice:
 Why and how people vote\linebreak to improve the well-being of others'',
 NBER Working Paper 13562, \texttt{www.nber.org/papers/w13562}

\item[\hspace{-5pt}]
Eells, E.\ (1981) ``Causality, utility, and decision'', \textit{Synthese},
  vol.~48, pp.~295--329.

\item[\hspace{-5pt}]
Eells, E.\ (1982) \textit{Rational Decision and Causality}, Cambridge
University Press.

\item[\hspace{-5pt}]
Egan, A.\ (2007) ``Some counter-examples to Causal Decision Theory'',
  \textit{Philosophical Review}, vol.~116, pp.~93--114.

\item[\hspace{-5pt}]
Gallow, J.~D.\ (2021) ``Riches and rationality'',
  \textit{Australian Journal of Philosophy}, vol.~99, pp.~114--129.

\item[\hspace{-5pt}]
Gibbard, A.\ and Harper, W.~L.\ (1976) ``Counterfactuals and two kinds
  of expected utility'', Discussion Paper no.~194, 
  \texttt{hdl.handle.net/10419/220553}

\item[\hspace{-5pt}]
Joyce, J.~M.\ (1999) \textit{The Foundations of Causal Decision Theory},
  Cambridge University Press.

\item[\hspace{-5pt}]
Joyce, J.~M.\ (2018) ``Deliberation and stability in Newcomb problems and
  pseudo-Newcomb problems'',
  in A.\ Ahmed (editor) \textit{Newcomb's Problem}, Cambridge University Press.

\item[\hspace{-5pt}] Kirk, R.\ (2023), ``Zombies'', \textit{The
  Stanford Encyclopedia of Philosophy (Fall 2023 Ed.)}, E..~N.~Zalta and
  U.~Nodelman (editors), 
  \texttt{plato.stanford.edu/archives/fall2023/entries/zombies}

\item[\hspace{-5pt}]
Levi, J.\ (1975) ``Newcomb's many problems'', \textit{Theory and Decision},
  vol.~6, pp.~161--175.

\item[\hspace{-5pt}]
Levin, J.\ (2023) ``Functionalism'', \textit{The Stanford Encyclopedia of 
  Philosophy (Summer 2023 Ed.)}, E.~N.~Zalta and U.~Nodelman (editors),
  \texttt{plato.stanford.edu/archives/sum2023/entries/functionalism}

\item[\hspace{-5pt}]
Lewis, D.\ (1979) ``Prisoners' Dilemma is a Newcomb Problem'', 
 \textit{Philosophy \& Public Affairs}, vol.~8, pp.~235--240.

\item[\hspace{-5pt}]
Lewis, D.\ (1981) ``Causal decision theory'',
 \textit{Australian Journal of Philosophy}, vol.~59, pp.~5--30.

\item[\hspace{-5pt}]
McKay, P.\ (2004) ``Newcomb's problem: the causalists get rich'',
  \textit{Analysis}, vol.~64, no.~2, pp.~187--189.

\item[\hspace{-5pt}]
Mackie, J.~L.\ (1977) ``Newcomb's paradox and the direction of causation'',
  \textit{Canadian Journal of Philosophy}, vol.~7, no.~2, pp.~213--225.

\item[\hspace{-5pt}]
Neal, R.~M.\ (2006) ``Puzzles of anthropic reasoning resolved using
  full non-indexical conditioning'', 
  \texttt{arxiv.org/abs/math/0608592} or \texttt{philsci-archive.pitt.edu/2888}

\item[\hspace{-5pt}]
Nozick, R.\ (1969) ``Newcomb's problem and two principles of choice'',
  in N.~Rescher (editor) \textit{Essays in Honor of Carl G.~Hempel},
  Boston: Reidel, pp.~114--146.

\item[\hspace{-5pt}]
Oesterheld, C.\ and Conitzer, V.\ (2021) ``Extracting money from causal
  decision theorists'', \textit{The Philosophical Quarterly}, vol.~71,
  pp.~701--716.

\item[\hspace{-5pt}]
Pearl, J.\ (2016) ``The sure-thing principle'', \textit{Journal of Causal
  Inference}, vol.~1, pp.~81--86.


\item[\hspace{-5pt}]
Plommer, B.\ (2016) ``A new problem with mixed decisions, or: You'll
  regret reading this article, but you still should'', \textit{Erkenntnis},
  vol.~81, pp.~349--373.

\item[\hspace{-5pt}]
Pollock, J.~L.\ (2010) ``A resource-bounded agent addresses the Newcomb
  problem'', \textit{Synthese}, vol.~176, pp.~57-82.

\item[\hspace{-5pt}]
Schmidt, J.~J.\ (1998) ``Newcomb's paradox realized with backward causation'',
 \textit{The British Journal for the Philosophy of Science}, vol.~49, no.~1,
 pp.~67--87.

\item[\hspace{-5pt}]
Shafir, E.\ and Tversky, A.\ (1992) ``Thinking through uncertainty:
 Nonconsequential reasoning and choice'', \textit{Cognitive Psychology},
 vol.~24, pp.~449--474.

\item[\hspace{-5pt}]
Silverberg, R. (1965) ``The Sixth Palace'', \textit{Galaxy Magazine},
 February 1965, available at\\
 \texttt{scifiwise.com/the-sixth-palace-robert-silverberg-science-fiction}

\item[\hspace{-5pt}]
Slezak, P.\ (2013) ``Realizing Newcomb's problem'',
 \texttt{philsci-archive.pitt.edu/9634}

\item[\hspace{-5pt}]
Spencer, J.\ and Wells, I.\ (2019) ``Why take both boxes?'', 
 \textit{Philosophy and Phenomenological Research}, vol.~99, pp.~27--48.

\item[\hspace{-5pt}]
Spohn, W.\ (2011) ``Reversing 30 years of discussion: why causal decision
 theorists should one-box'', \textit{Synthese}, vol.~187, pp.~95--122.

\item[\hspace{-5pt}]
Stalnaker, R.\ (2018) ``Game theory and decision theory (causal and 
  evidential)'', in
  A.\ Ahmed (editor) \textit{Newcomb's Problem}, Cambridge University Press.

\item[\hspace{-5pt}]
Viger, C., Hefer, C., and Viger, D.\ (2019) ``The philosopher's paradox:
 How to make a coherent decision in the Newcomb Problem'', \textit{Theoria},
 vol.~34, pp.~407--421.

\item[\hspace{-5pt}]
Wallace, D.\ (2010) ``Diachronic rationality and prediction-based games'',
  \textit{Proceedings of the Aristotelian Society}, New Series, vol.~110,
  pp.~243--266.

\item[\hspace{-5pt}]
Wolpert, D.~H.\ and Benford, G.\ (2011) ``The lesson of Newcomb's paradox'',
  Synthese, vol.~190, pp.~1--10.

\item[\hspace{-5pt}]
Yudkowsky, E.\ and Soares, N.\ (2018) ``Functional Decision Theory: A
  new theory of instrumental rationality'', 
  \texttt{arxiv.org/abs/1710.05060v2}

\end{description}

\end{document}